\documentclass[%
 reprint,%
 amssymb, amsmath, 
pre, aps,
]{revtex4-1}

\usepackage{bm}%
\usepackage{graphicx}
\expandafter\ifx\csname package@font\endcsname\relax\else
 \expandafter\expandafter
 \expandafter\usepackage
 \expandafter\expandafter
 \expandafter{\csname package@font\endcsname}%
\fi
\usepackage{subcaption}
\begin{document}

\title{Escape time, relaxation and sticky states of a softened Henon-Heiles model: low-frequency vibrational modes effects}%
 
\author{J. Quetzalc\'oatl Toledo-Mar\'in}%
\affiliation{Departamento de Sistemas Complejos, Instituto de
F\'{i}sica, Universidad Nacional Aut\'{o}noma de M\'{e}xico (UNAM),
Apartado Postal 20-364, 01000 M\'{e}xico, Distrito Federal,
M\'{e}xico}%

\author{Gerardo G. Naumis}%
\email{naumis@fisica.unam.mx}
\affiliation{Departamento de Sistemas Complejos, Instituto de
F\'{i}sica, Universidad Nacional Aut\'{o}noma de M\'{e}xico (UNAM),
Apartado Postal 20-364, 01000 M\'{e}xico, Distrito Federal,
M\'{e}xico}%

\date{September 2017}%
\revised{?}%

\begin{abstract}
Here we study the relaxation of a chain consisting of $3$ masses joined by non-linear springs and periodic conditions when the stiffness is weakened. This system, when expressed in their normal coordinates, yields a \textit{softened Henon-Heiles} system. By reducing the stiffness of one low-frequency vibrational mode,  a faster relaxation is enabeled. This is due to a reduction of the energy barrier heights along the softened normal mode as well as for a widening of the opening channels of the energy landscape in configurational space. The relaxation is for the most part exponential, and can be explained by a simple flux equation. Yet, for some initial conditions the relaxation follows as a power law and, and in many cases, there is a regime change from exponential  to power law decay. We pin point the initial conditions for the power law decay, finding two regions of sticky states. For such states, quasiperiodic orbits are found since almost for all components of the initial momentum orientation, the system is trapped inside two pockets of configurational space. The softened Henon-Heiles model presented here is intended as the simplest model in order to understand the interplay of rigidity, non-linear interactions and relaxation for non-equilibrium systems like glass-forming melts or soft-matter.    
\end{abstract}


\maketitle

\section{Introduction}
Relaxation processes are abundant in the Universe. In fact, if one were to take a snapshot of the Universe one would see that only a few number of physical systems are to be in equilibrium, while the rest are either in steady states or relaxing towards equlibrium. All the same, it is not clear how relaxation processes depend upon the energy landscape \cite{trachenko2008relationship,zwanzig2001nonequilibrium, berthier2011theoretical, wolynes2012structural, binder2011glassy, naumisbifurcation}. It is of common knowledge that this is due to mainly two contributions, one being the energy landscape complexity itself,  and the other being the intrinsic stochastic behavior given that there are no isolated systems in nature. However, even if one were to exclude this stochastic ingredient and only consider the complexity of the energy landscape, one would find that the system has an intermittent chaotic behavior. In this way, the deterministic feature lacks of meaning, similar to whats happens in stochastic processes \cite{Born1968}. Furthermore, what we call stochastic behavior and chaotic behavior seems to be two faces of the same coin \cite{shlesinger1993strange}. In this sense, Alvaro \textit{et al.} in \cite{diaz2017relating} are able to establish this sort of correspondence for two particular well-known stochastic models.

In order to garantee that a given system under study will tend to an equilibrium state, one usually imposes detailed balance or assumes the conditions are fulfilled for the fluctuation-dissipation theorem or the equipartition theorem. However, this is not obvious when going from a classical mechanical approach to a statistical mechanical one. As it is well known, this is what Fermi, Pasta and Ulam investigated back in 1954, by considering a chain of nonlinear oscillators. What they found is quite long to sumarize here (see \cite{fermi1955studies}). However, as one would expect, this relaxation depends strongly on the mode-coupling, i.e., the non-linear terms. In fact, Ponno in Ref. \cite{ponno2005fermi} presents some estimates on how the energy is transfered from one mode to another, as well as the characteristic relaxation time which is proportional to the number of oscillators. What is even more interesting is that this energy sharing starts in the low vibrational modes due to resonances, which was first pointed out in Ref. \cite{ford1961equipartition}. This is because the dispersion relation for the low vibrational modes is linear and the frequencies are linearly dependent. Then each mode will resonate with their mode-coupling term. Here lies the importance of the low vibrational modes when in comes to relaxation processes and, in particular, glassy dynamics, turbulence and protein folding.

Nonetheless, the relaxation mechanism, in particular, in supercooled liquids has proven to be a very complex one. In fact, the general features of supercooled liquids still lack of a scientific explanation, because of the complex nature of it. On the one hand, the problem is difficult because the harmonic approximation breaks down for the Hamiltonian at long-time scales, which are relevant to describe the relaxation and viscosity properties of glassy melts \cite{trachenko2008relationship}. On the other hand, the glass transition is a non-equilibrium transition problem where the system does not have long range order. These arguments are just the tip of the iceberg that give foundation to why this problem is a very complex one. Despite the amount of research focused on it (see \cite{pedersen2016thermodynamics,albert2016fifth, hansen2017connection,gleim2000relaxation, mezard2012glasses, trachenko2011heat, micoulaut1999glass, Mauro2} and in particular \cite{dyre2006col} and references therein), there is not to much of a consensus and rather different points of view.  To spice things up a little, experiments and simulations have not yet met in the sense that it takes to much computational time to drive the system towards a region near the glass transition temperature. Yet attention should be payed on recent simulations \cite{ninarello2017models}.

One of the questions that arise in this phenomenon is how the glass transition temperature, $T_g$, is related to the composition of the glass former \cite{naumis1998stochastic,kerner2000stochastic}. Rigidity theory \cite{phillips1979topology,thorpe1983continuous,huerta,huerta2002evidence,flores2012mean} gives some insight on this aspect in a qualitative manner, and works quite well in the case of chalcogenide glasses. Another rather interesting feature in supercooled liquids is the viscosity behavior during a quench. As is well known, viscosity is a property that depends upon relaxation, i.e., the time that the system takes in order to leave a basin of the energy landscape and produce a structural relaxation. Depending on this behavior, the supercooled liquid is classified as a strong one if it follows the Arrhenius equation and as a fragile one if it follows the Vogel-Fulcher-Tamman equation \cite{dyre2006col,berthier2011theoretical}. The fragility or non-fragility of a supercooled liquid is related with the glass forming tendency in the sense that strong supercooled liquids have a strong glass forming tendency such that do not require large quenches in comparison with fragile ones which are poor glass formers. Thus, the glass forming tendency is clearly related with the time-relaxation of the system.

It is well known that there is a correlation between the glass transition temperature, $T_g$, and the cooling rate. Quite recently, Lerner \textit{et al.} in Refs. \cite{lerner2016statistics,lerner2017effect} have shown that the statistics and localization of low-frequency vibrational modes depend upon the cooling rate. Thus, there lies a trichotomy, namely, glass transition, relaxation and low-frequency vibrational. In a series of previous papers, we have discussed how these are related in a very natural way \cite{naumis2005energy,naumis,flores,flores2011boson,toledo2016minimal,toledo2017short}. In fact, rigidity theory has 
allowed to rationalize how they are interrelated\cite{naumis2006variation}. In their rigidity theory, Phillips and later Thorpe, consider covalent bonding as a mechanical constraint\cite{phillips1979topology,thorpe1983continuous}. In this sense, one may summarize the main feature of this theory as follows. When the number of bond constraints equals the number of degrees of freedom, the glass forming ability is optimized, i.e., producing glass requires the slowest cooling rate. In this situation, the mean coordination number equals the critical percolation coordination number, i.e., domains of floppy modes (zero frequency modes) and rigid modes coexist. As the mean coordination number decreases, which may be tuned by varying the chemical composition, floppy mode domains grow while rigid mode domains disappear. As floppy modes increase in number, the glass formation is more difficult.

However, there are no general models to deal with the non-linear regime which is the one which is interesting for glass transition.  
With these ideas in mind, here we study and present our findings on the relaxation behavior when we decrease the frequency of the normal modes towards zero, i.e., floppy modes, in the case of a chain formed by three non-linear oscillators, which when expressed in the coordinates that diagonalize the linear part yields the Henon-Heiles potential. 
This potential is a particular case of the Fermi-Pasta-Ulam (FPU) model, in which it is known that low-frequency modes are responsible for relaxation \cite{romero2009thermal,romero2008thermal,ponno2005chaotic,onorato2015route}. In the FPU model,
the system can be roughly described in terms of what happens in turbulence, where energy can only be dissipated at small scales. Thus, the removal or addition of such modes modify in an important 
way the relaxation properties. The advantage of the Henon-Heiles model is that it contains the minimal ingredients to understand the effects of non-linearity. In that sense, here we provide a minimal model to understand how low-frequency modes impact the escape time and relaxation of the system. 

It is worthwhile mentioning that the Henon-Heiles potential has been widely studied \cite{henon1964applicability,fordy1991henon,aguirre2001wada,waite1981mode, toda2012theory,zhao2007threshold}. Concerning the escape dynamics, it has been observed the fact that the phase space escape flow follows an exponential law  connected to chaotic dynamics, whereas in non-chaotic dynamics the phase space escape follows a power law. Using simple arguments, Zhao \textit{et al.} \cite{zhao2007threshold}  obtain the exponential law, which they then compare succesfully with their simulations in the case of chaotic dynamics, with a small threshold energy. Bauer and Bertsch \cite{bauer1990decay} also obtained the exponential law before Zhao \textit{et al}. Furthermore, from a heuristic and restrospective approach they obtain the power law. However, one of the results in the present work is the crossover between exponentially decaying law and power law which is not seen in \cite{zhao2007threshold} because they consider smaller threshold energies and short times.

The article is organized as follows: in  the following section we present the model and its features, in section \ref{sec:results} 
we present the results obtained from the simulations and how the exponential relaxation is affected by low frequency vibrational modes. In section \ref{sec:power}
we study special states that are sticky, which instead present a power law relaxation. Finally, in section \ref{sec:discussion} we discuss these results.

\section{Softened Henon-Heiles model} \label{sec:model}
Let us consider a chain consisting of $3$ masses joined by non-linear springs and periodic conditions. Thus,  the Hamiltonian is
\begin{equation}
H=\sum_{i=1}^3 \frac{1}{2m}\vec{P}_i^2+\frac{1}{2}k_{i+1,i} \left( \Delta Q_{i+1,i} \right)^2 +\frac{\gamma}{3} \left( \Delta Q_{i+1,i} \right)^3\; ,
\end{equation}
where $\Delta Q_{i+1,i}=Q_{i+1}-Q_i$ and $k_{i+1,i}$ is the spring stiffness between masses $i$ and $i+1$ .

Let us ignore for the moment the cubic interaction terms. Using the Hamilton Eqs. with the ansatz $Q_j(t)=q_j e^{-\imath \omega t}$ yields
\begin{equation}
\left(\mathbf{D}-\omega^2 \mathbf{I}\right)\vec{q}=\vec{0} \; ,
\end{equation}
where $\mathbf{D}$ is the dynamic matrix given by
\[
\mathbf{D}=k
\begin{bmatrix}
   1+\beta    & -1 & -\beta   \\
    -1       & 1+\alpha & -\alpha  \\
    -\beta      & -\alpha & \alpha+ \beta	\\
\end{bmatrix} \; ,
\] 
and $k_{21}=k, \; k_{21}\beta=k_{13}, \; $ and $k_{32}=\alpha k_{21}$. Notice that here we introduce
the stiffness in terms of the control parameters $\alpha$ and $\beta$.

One of the eigenvalues of $\mathbf{D}$ is always zero and corresponds to the center of mass motion, while the other two depend upon $\alpha$ and $\beta$. However, as was stated in the introduction, we are interested in studying the low vibrational dynamics relaxation process. In this sense, it happens that $\omega_x$ becomes zero only when $\alpha=\beta=0$ while $\omega_y$ becomes $\sqrt{2}$ (see appendix \ref{app:Eigenvectors} for further details). Thus, here on we consider $\alpha=\beta$ and without loss in generality we assume $k=1$. Therefore, the eigenvalues become
\begin{equation}
\omega_{x}^2 =3\beta \; , \quad 
 \omega_{y}^2 =2+\beta \; , \quad 
 \omega_z^2= 0  \; .
\end{equation} \label{eq:eigenvalues}

Now, let us define as $\mathbf{A}$ the unitary transformation which relates our former coordinates with the normal ones denoted as $(x,y,z)$. Therefore, using the aforementioned matrix we may write the cubic interaction in the normal coordinates which takes on the following form (see appendix \ref{app:Eigenvectors} for further details):
\begin{equation}
 -\frac{3 \gamma}{2^{1/2}}  \left(\frac{1}{3} y^3- x^2 y \right) \; .
\end{equation}
Thus, the Hamiltonian written in of the center of mass and in terms of the normal coordinates and momenta is
\begin{equation}
H=\frac{1}{2}\left(p_x^2 +p_y^2 \right) + \frac{1}{2}\left(\omega_x^2 x^2 + \omega_y^2 y^2 \right) - \frac{3 \gamma}{2^{1/2}}  \left(\frac{1}{3} y^3- x^2 y \right) \; ,
\end{equation}
which corresponds to two particles interacting via a Henon-Heiles-type potential \cite{toda2012theory}.

Furthermore, let us do the following rescalings:
\begin{eqnarray}
y&\rightarrow&\frac{2^{1/2}}{\gamma}y \; , \qquad x\rightarrow \frac{2^{1/2}}{\gamma} x \; , \nonumber \\
t & \rightarrow & t/3^{1/2} \; , \qquad H \rightarrow \frac{6}{\gamma^2} H \; .
\end{eqnarray}
This gives the scaled Hamiltonian:
\begin{equation}
H=\frac{1}{2}\left(p_x^2 +p_y^2  \right) + \frac{1}{6}\left(\omega_x^2 x^2 + \omega_y^2 y^2 \right) -  \left(\frac{1}{3} y^3- x^2 y \right) \; , \label{eq:Hamiltonian}
\end{equation}
and the Hamilton Eqs. are
\begin{eqnarray}
\dot{x}_i(t) &=& p_i(t) \; , \qquad x_i=\lbrace x,y,z \rbrace \nonumber \\
\dot{p}_x(t)&=&-x(t)\left(\beta +2  y(t) \right) \; , \nonumber \\
\dot{p}_y(t)&=& - \frac{1}{3} \left(2+\beta \right) y(t) + \left( y(t) \right)^2 - \left( x(t) \right)^2 \; . \nonumber \\ \label{eq:HamilEq}
\end{eqnarray}

\begin{figure}[hbtp]
\centering
\includegraphics[width=3.3in]{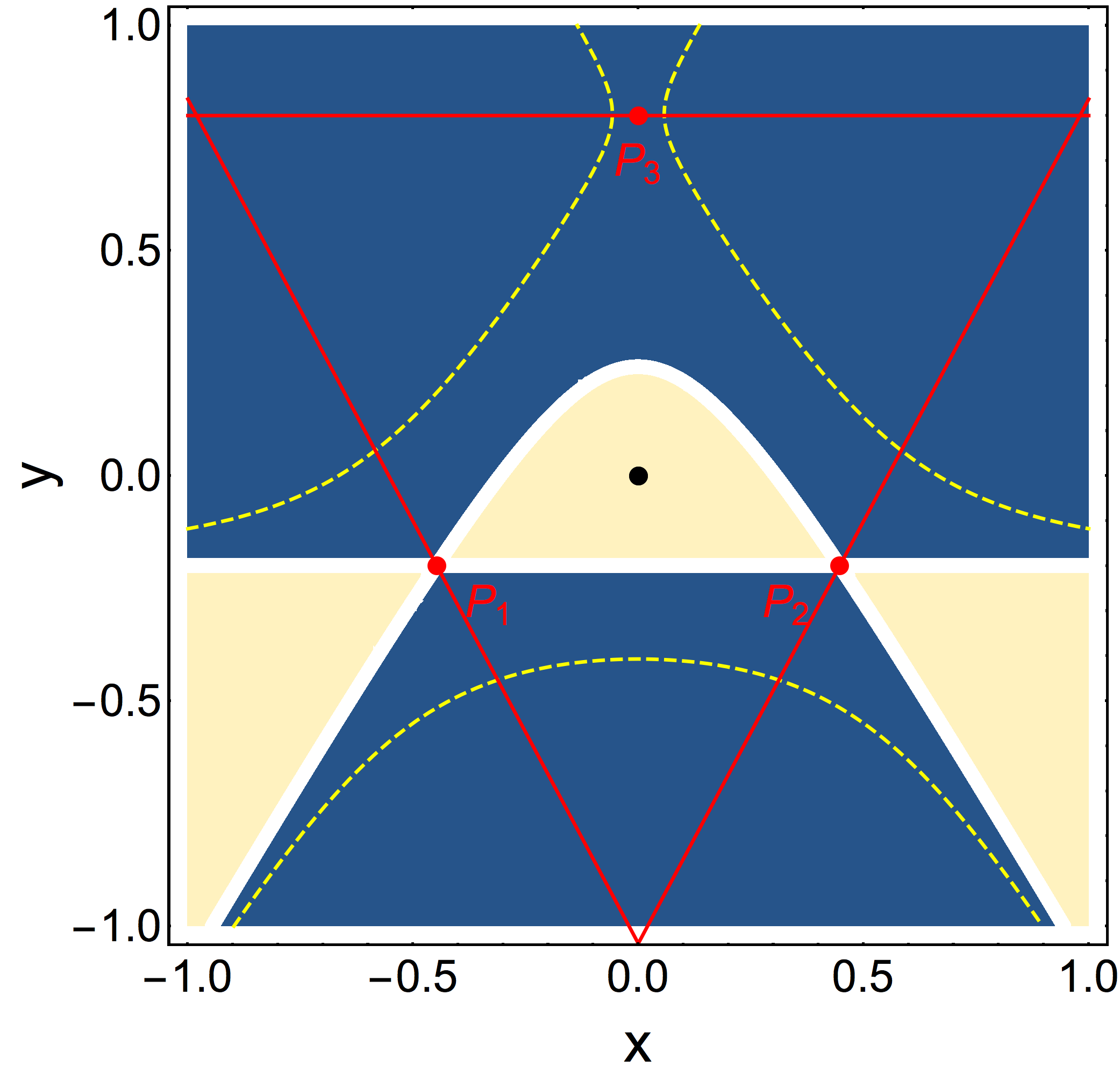}
\includegraphics[width=3.3in]{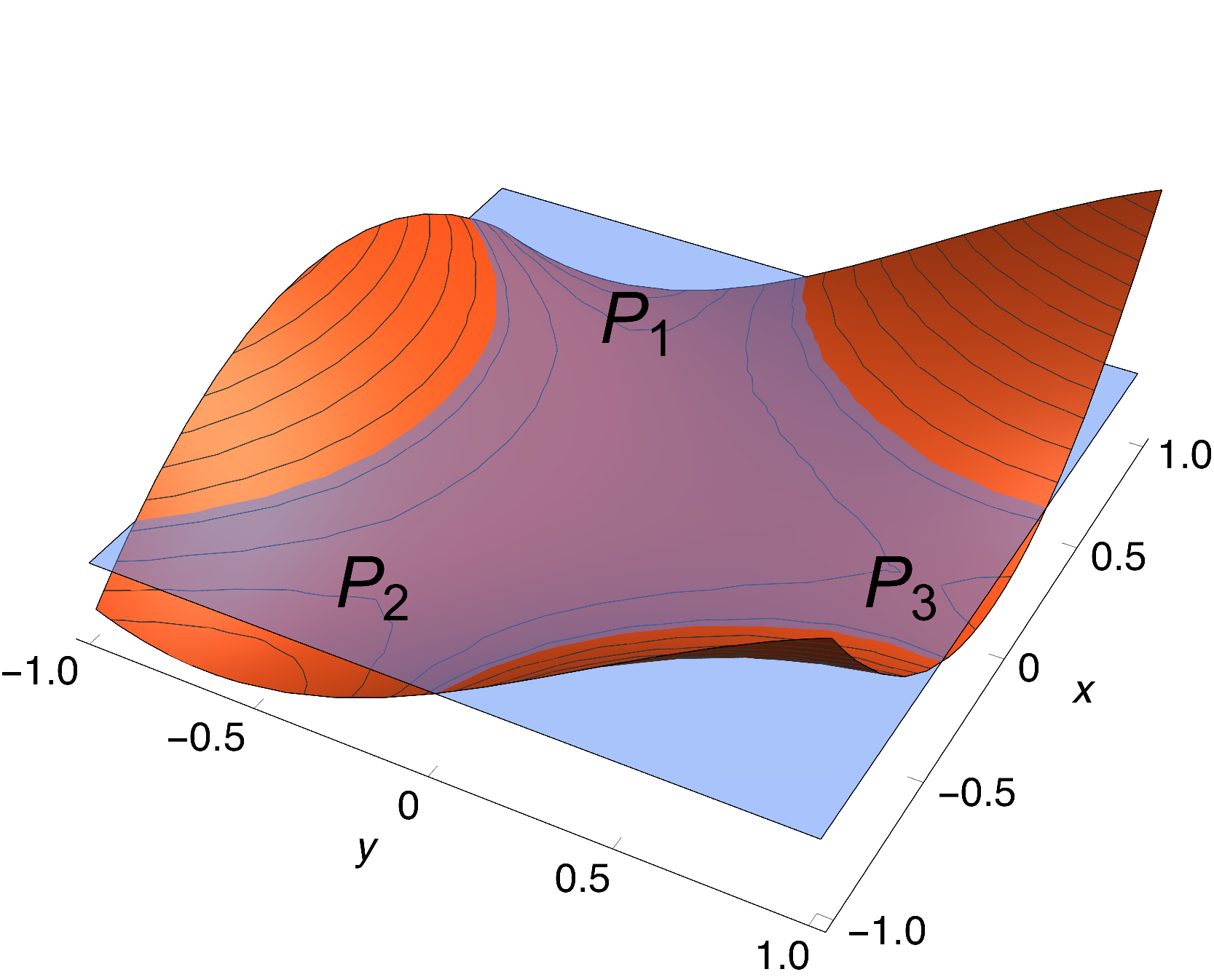}
\caption{{\footnotesize Energy landscape showing an isopotential with $\beta=0.4$. \textbf{Upper panel} The dashed line correspond to the ispotential with energy, $\Delta E=0.07$, above the lower saddle points height (yellow) while the inner soften triangle corresponds to $\Delta E=0$. Each line of the red triangle goes through a saddle point and has the direction of the eigenvector corresponding to the positive eigenvalue of the Hessian matrix in the corresponding saddle point. \textbf{Lower panel} Here we show a 3D plot of the potential. The transversal plane corresponds to $\Delta E=0.07$.}} \label{fig:regionOfInt}
\end{figure}

\begin{figure}[hbtp]
\centering
\includegraphics[width=3.3in]{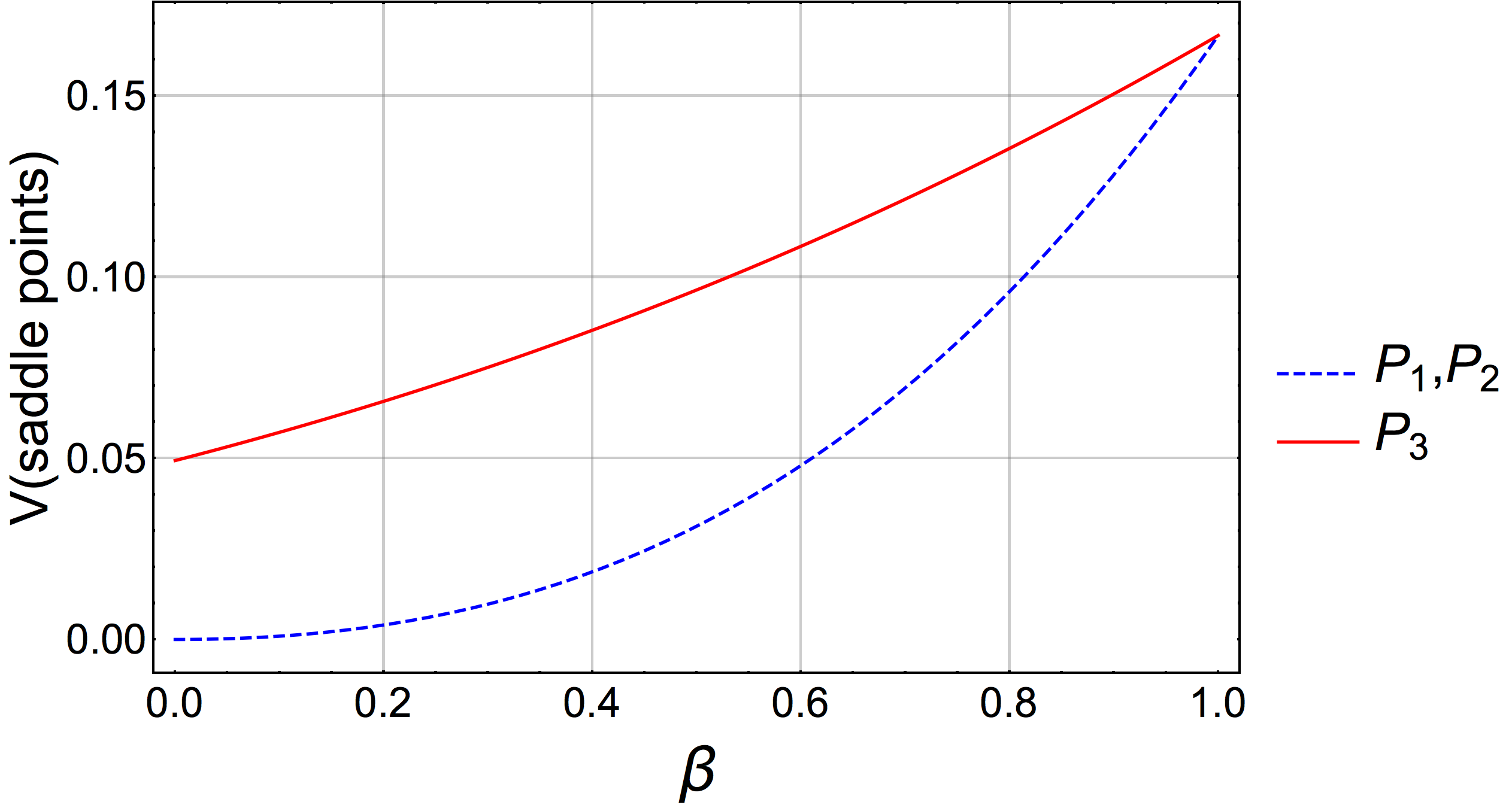}
\caption{{\footnotesize Plot of the saddle point heights as a function of $\beta$. The blue dashed curve corresponds to the saddle point height at $P_1$ and $P_2$, while the red continuous curve corresponds to the saddle point height at $P_3$ (see fig. \ref{fig:regionOfInt}) }} \label{fig:saddlepointH}
\end{figure}

The resulting model is a softened \textit{Henon-Heiles} system, since by making $\beta \rightarrow 0$, $\omega_x$ goes to zero, resulting in a floppy mode. Thus, $\beta$ is a control parameter that allows us to reduce the stiffness of the low frequency vibrational modes. This results in a lowering of two saddle points height.
In the upper panel of figure \ref{fig:regionOfInt} we have depicted the isopotential for a fixed $\beta$ and different energies. We also show the three saddle points (red dots) and the local minimum for the potential (black dot) where the potential energy is zero. In the lower panel of figure \ref{fig:regionOfInt} we may appreciate how the height of the saddle points $P_1$ and $P_2$ are the same and smaller than $P_3$. Moreover, in figure \ref{fig:saddlepointH} we have plotted the saddle points height vs $\beta$. When $\beta=0$ the potential barriers located at $P_1$ and $P_2$ drop to zero, while the other barrier drops to $\simeq 0.05$ (see fig. \ref{fig:saddlepointH}). 

In the case for which $\beta=1$, all saddle points have the same height. This Hamiltonian corresponds to the model used by \textit{H\'enon} and \textit{Heiles} to study the motion of a star in a galaxy with cylindrical symmetry \cite{henon1964applicability}. For a certain choice of parameters it has been proven to be an integrable problem \cite{fordy1991henon}, but it is not in general. Moreover, numerical results suggest that when the energy of the system is $E<1/12$, the system is non-chaotic and non-ergodic, yet above this energy the region with chaotic behavior in phase space increases with the energy up until $E=1/6$ where the whole phase space is chaotic and, which is also, saddle point's height \cite{zhao2007threshold, toda2012theory}.

In this way, it seems that as the energy increases the dynamics become chaotic and ergodicity is established. However, in this work we show that this is not always the case for the softened model. Actually, as the energy increases, there are certain islets in the phase space for which quasiperiodicity is established. This is done in the following section.

\section{Relaxation properties of the softened Henon-Heiles model} \label{sec:results}

\begin{figure}[hbtp]
\centering
\includegraphics[width=3.3in]{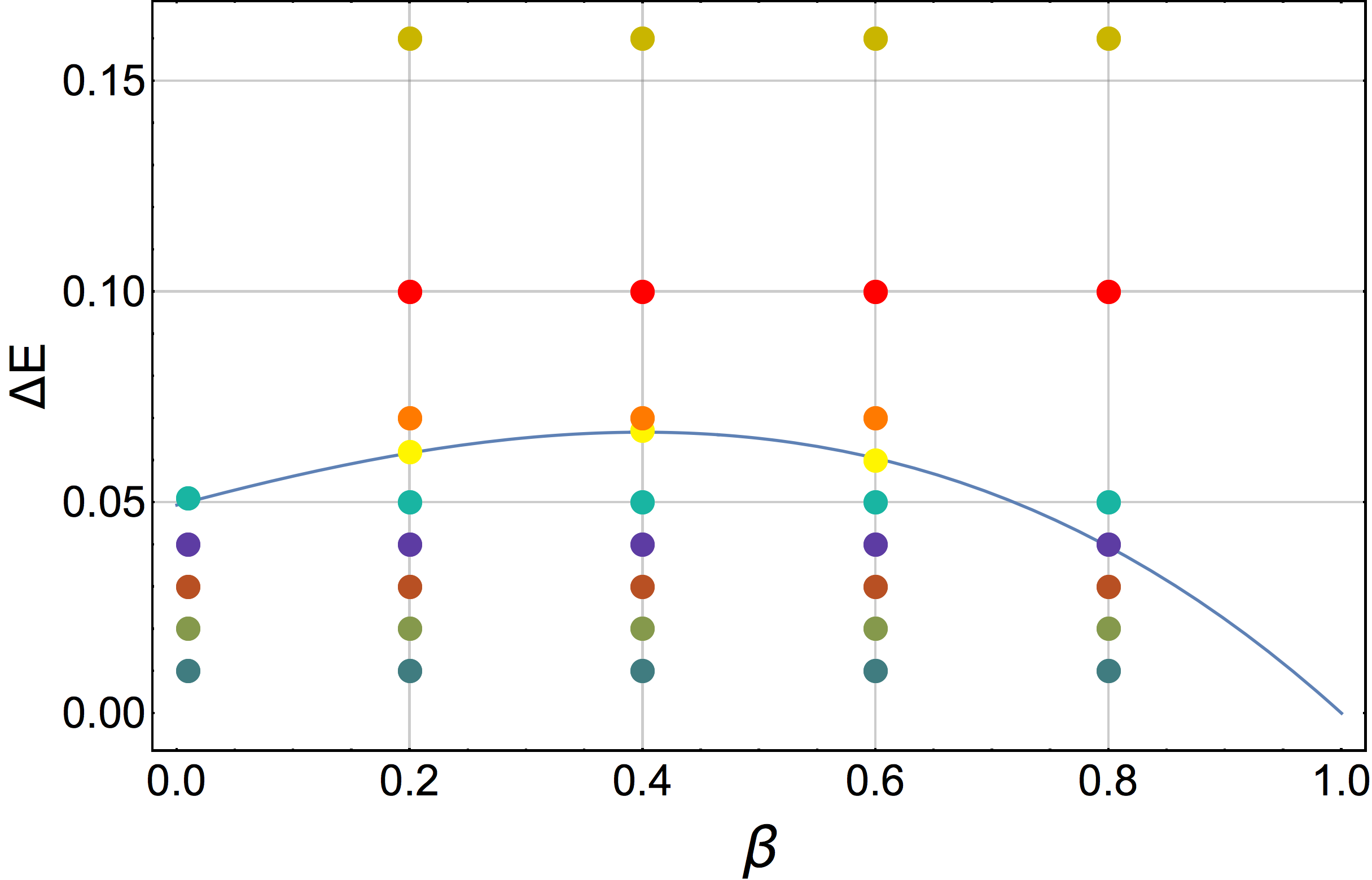}
\caption{{\footnotesize Sets of parameters $(\beta, \Delta E)$ used in our simulations to study relaxation. Points of the same color have the same energy $\Delta E$ above the lowest saddle points.  The blue curve corresponds to the energy difference between the high and low saddle points, i.e., for $\Delta E$ under the blue curve there are two exit channels, namely, $P_1$ and $P_2$ and for $\Delta E$ above the blue curve, there are three exit channels. }} \label{fig:saddle}
\end{figure}

In this section we present the results obtained from solving numerically the Hamiltonian Eqs. (Eqs. (\ref{eq:HamilEq})). We first fixed $\beta$ and the energy,  $\Delta E$, above the lower saddle points height (see fig. \ref{fig:regionOfInt}), then we took $N=16000$ randomly chosen initial conditions, i.e.,  linear moment orientation and position, given the fixed energy. Then we let them evolve and we studied the distribution of their escape time, i.e., the time taken to escape the well through any of the exit channels denoted as $P_1,P_2$ and $P_3$ (see figure \ref{fig:regionOfInt}). We did this for different values of the parameters $\beta$ and $ \Delta E$, which we present in figure \ref{fig:saddle} (the blue line in figure \ref{fig:saddle} corresponds to the energy difference between the upper saddle point and the lower saddle points, i.e., for a given fixed $\beta$ when $\Delta E$ is under the blue curve there are two exit channels, namely, $P_1$ and $P_2$; while when $\Delta E$ is above the blue curve, there are three exit channels).

\begin{figure}[hbtp]
\centering
\includegraphics[width=3.3in]{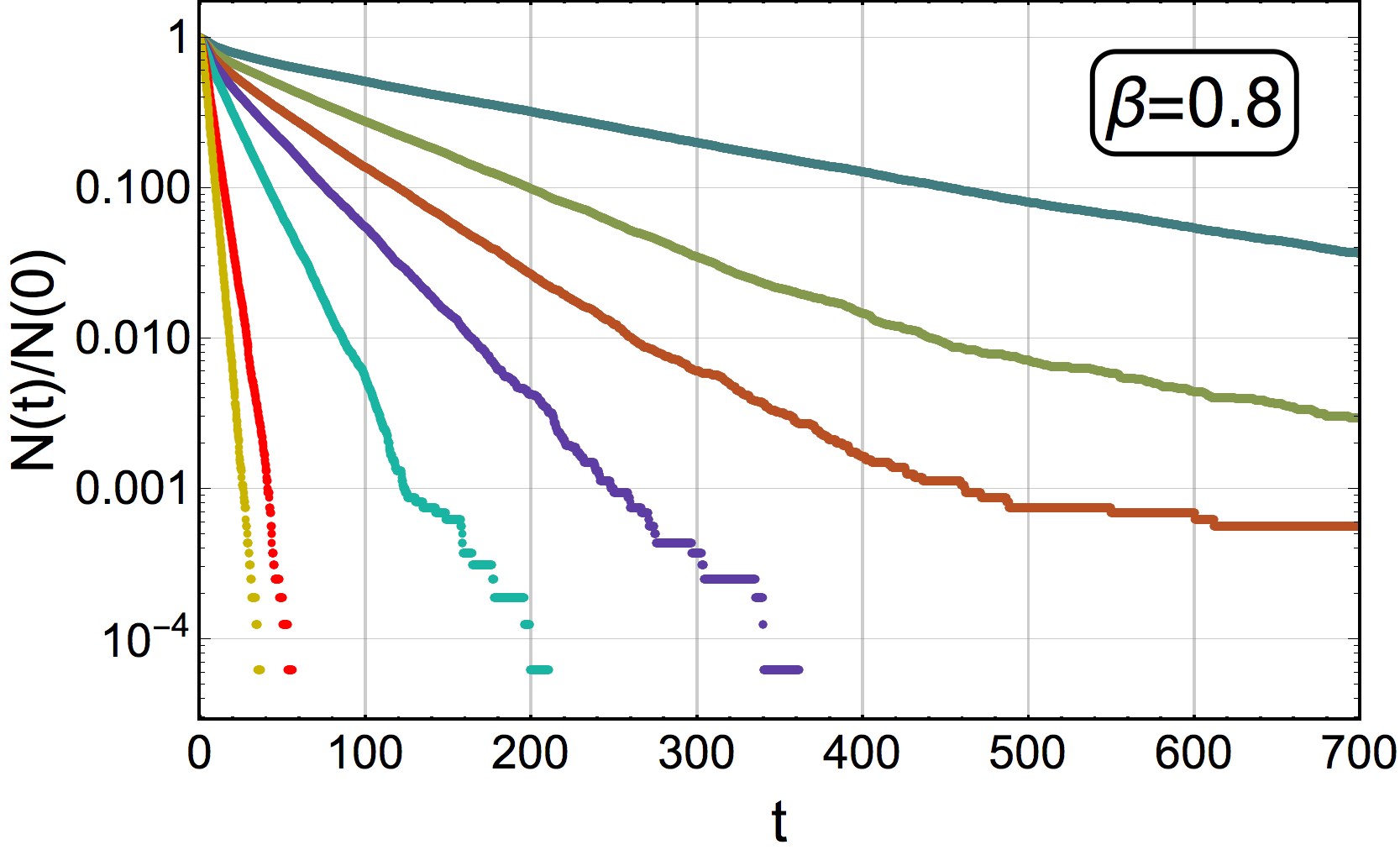}
\includegraphics[width=3.3in]{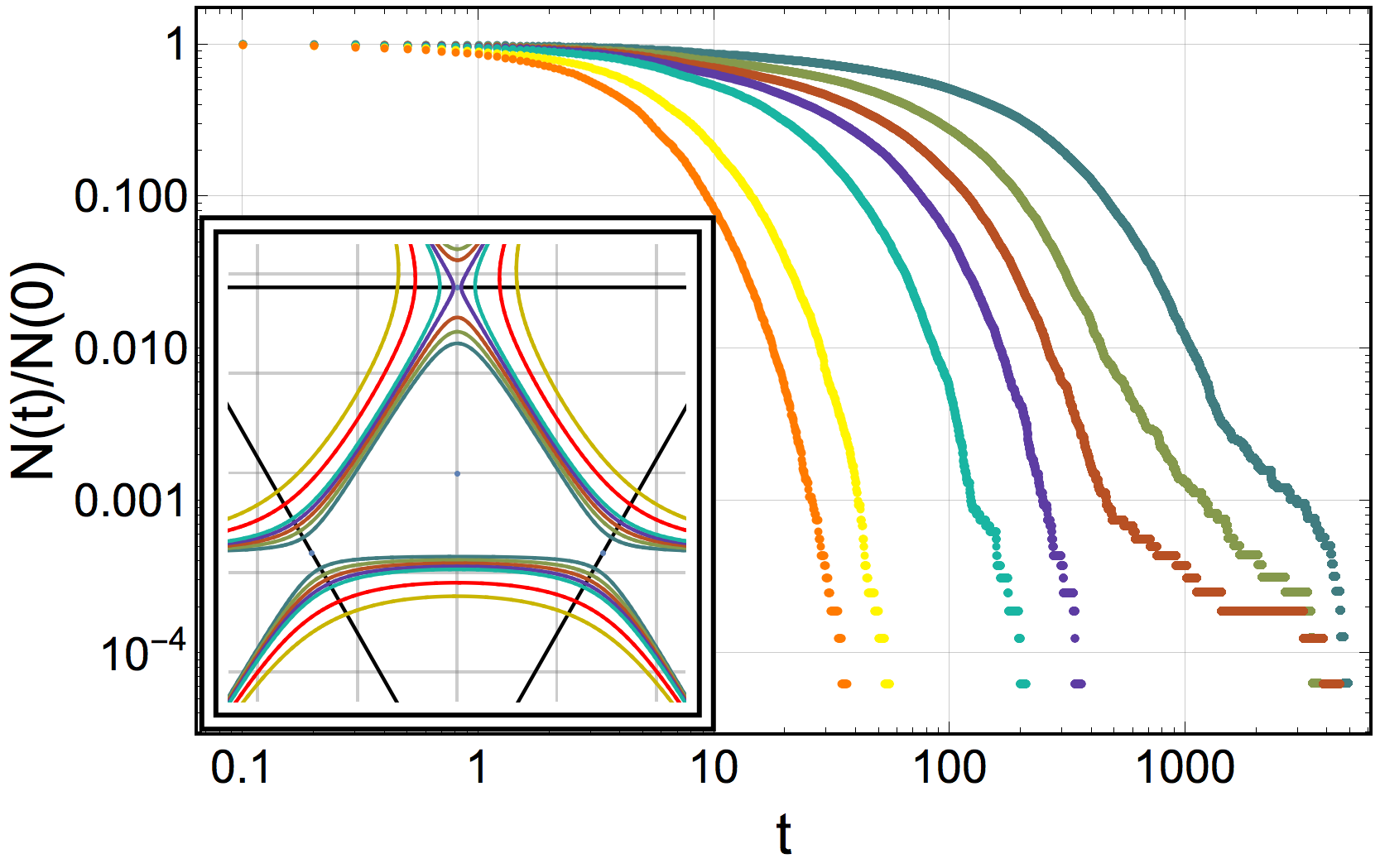}
\caption{{\footnotesize For $\beta=0.8$: (\textbf{Upper panel}) The  population in the potential at time $t$ in a Log-Linear plot, for different energies as indicated in the color code of figure  \ref{fig:saddle} .  (\textbf{Lower panel}) The population in the potential at time $t$ in Log-Log. (\textbf{Inset}) The isopotential for different values of $\Delta E$(see fig. \ref{fig:saddle}).}}  \label{fig:contour08}
\end{figure}

\begin{figure}[hbtp]
\centering
\includegraphics[width=3.3in]{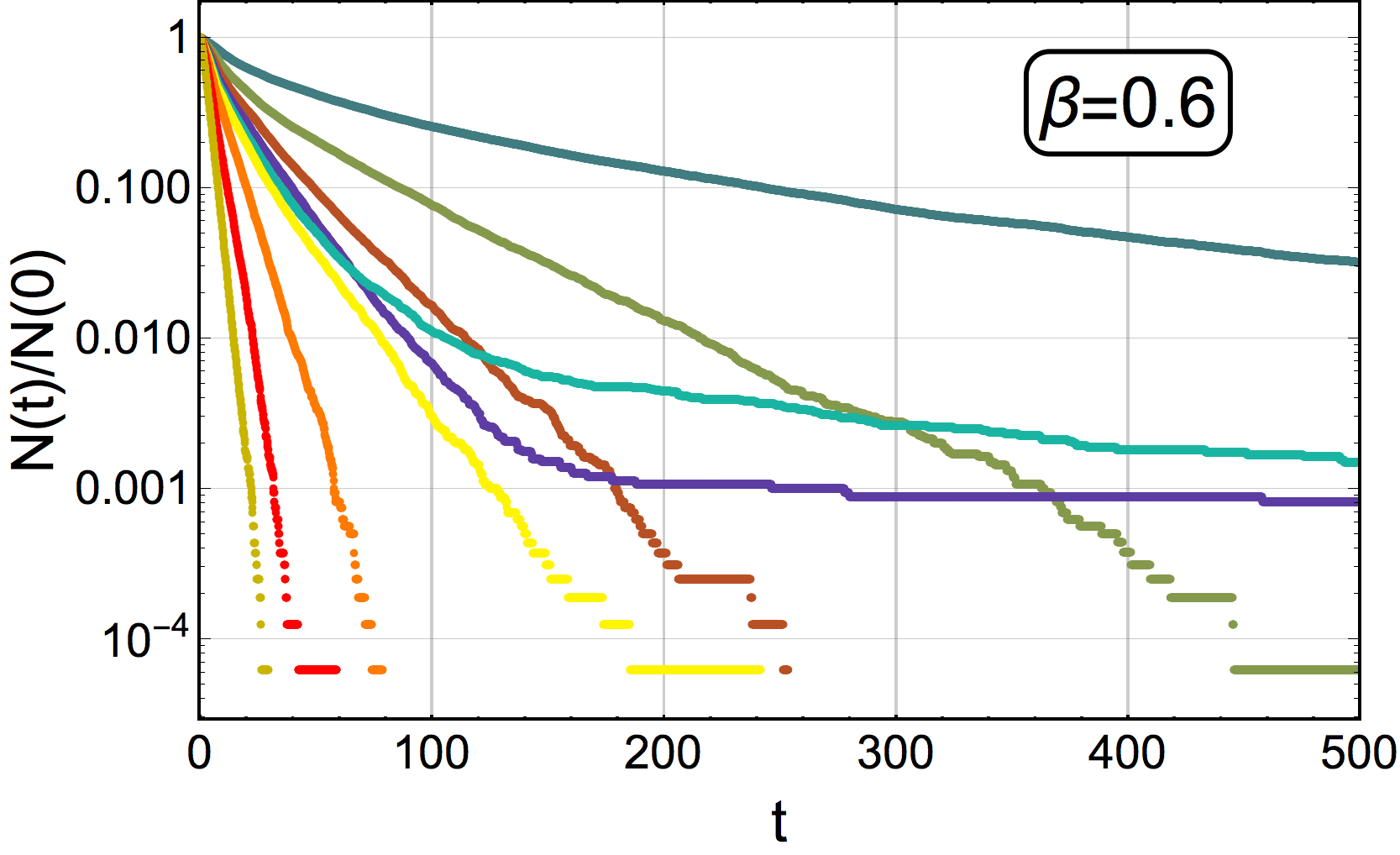}
\includegraphics[width=3.3in]{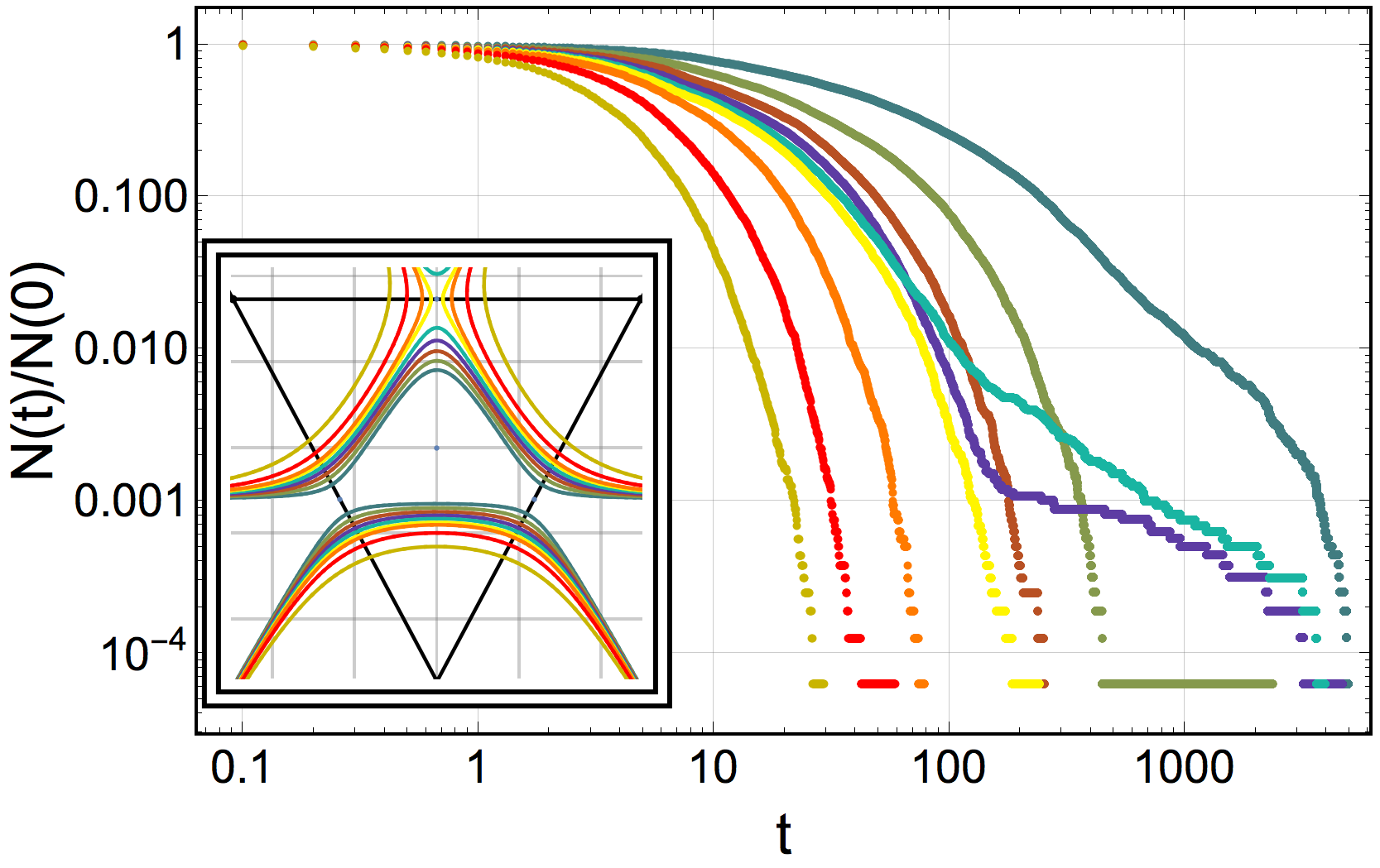}
\caption{{\footnotesize For $\beta=0.6$: (\textbf{Upper panel}) The  population in the potential at time $t$ in Log-Linear,  for different energies as indicated in the color code of figure  \ref{fig:saddle} .  (\textbf{Lower panel}) The population in the potential at time $t$ in Log-Log. (\textbf{Inset}) The isopotential for different values of $\Delta E$(see fig. \ref{fig:saddle}). }}  \label{fig:contour06}
\end{figure}

\begin{figure}[hbtp]
\centering
\includegraphics[width=3.3in]{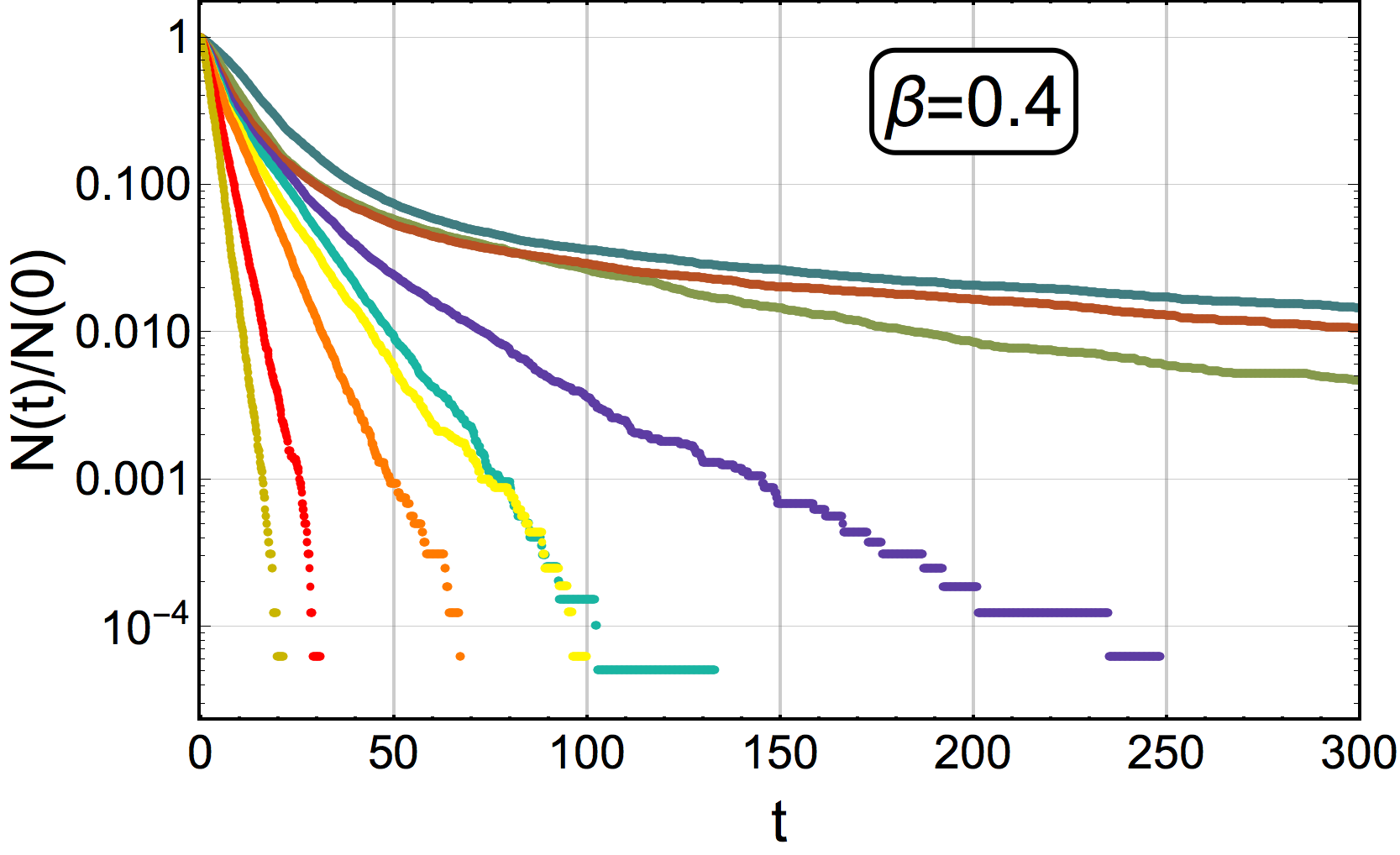}
\includegraphics[width=3.3in]{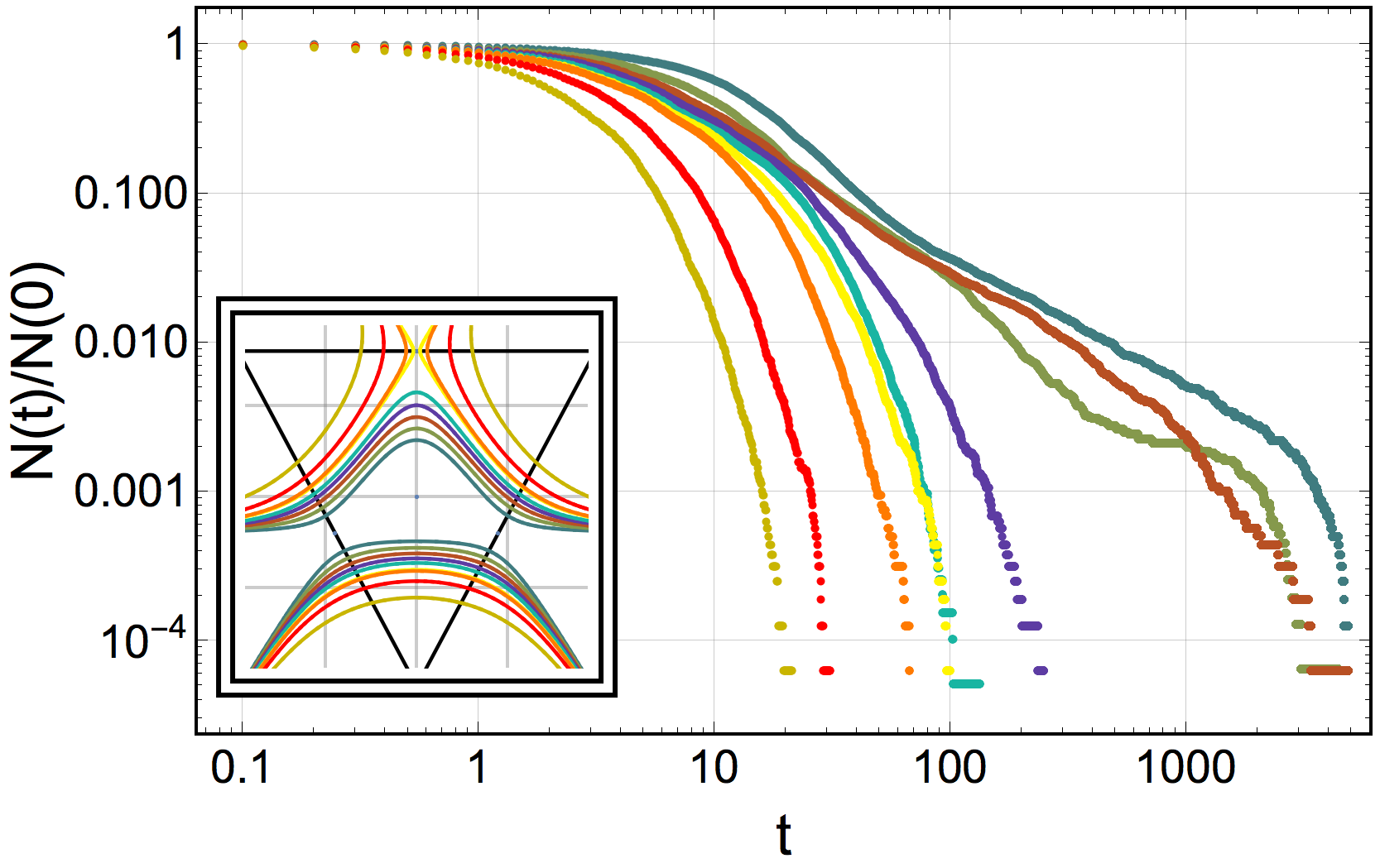}
\caption{{\footnotesize For $\beta=0.4$: (\textbf{Upper panel}) The  population in the potential at time $t$ in Log-Linear,  for different energies as indicated in the color code of figure  \ref{fig:saddle} .  (\textbf{Lower panel}) The population in the potential at time $t$ in Log-Log. (\textbf{Inset}) The isopotential for different values of $\Delta E$(see fig. \ref{fig:saddle}).} } \label{fig:contour04}
\end{figure}

\begin{figure}[hbtp]
\centering
\includegraphics[width=3.3in]{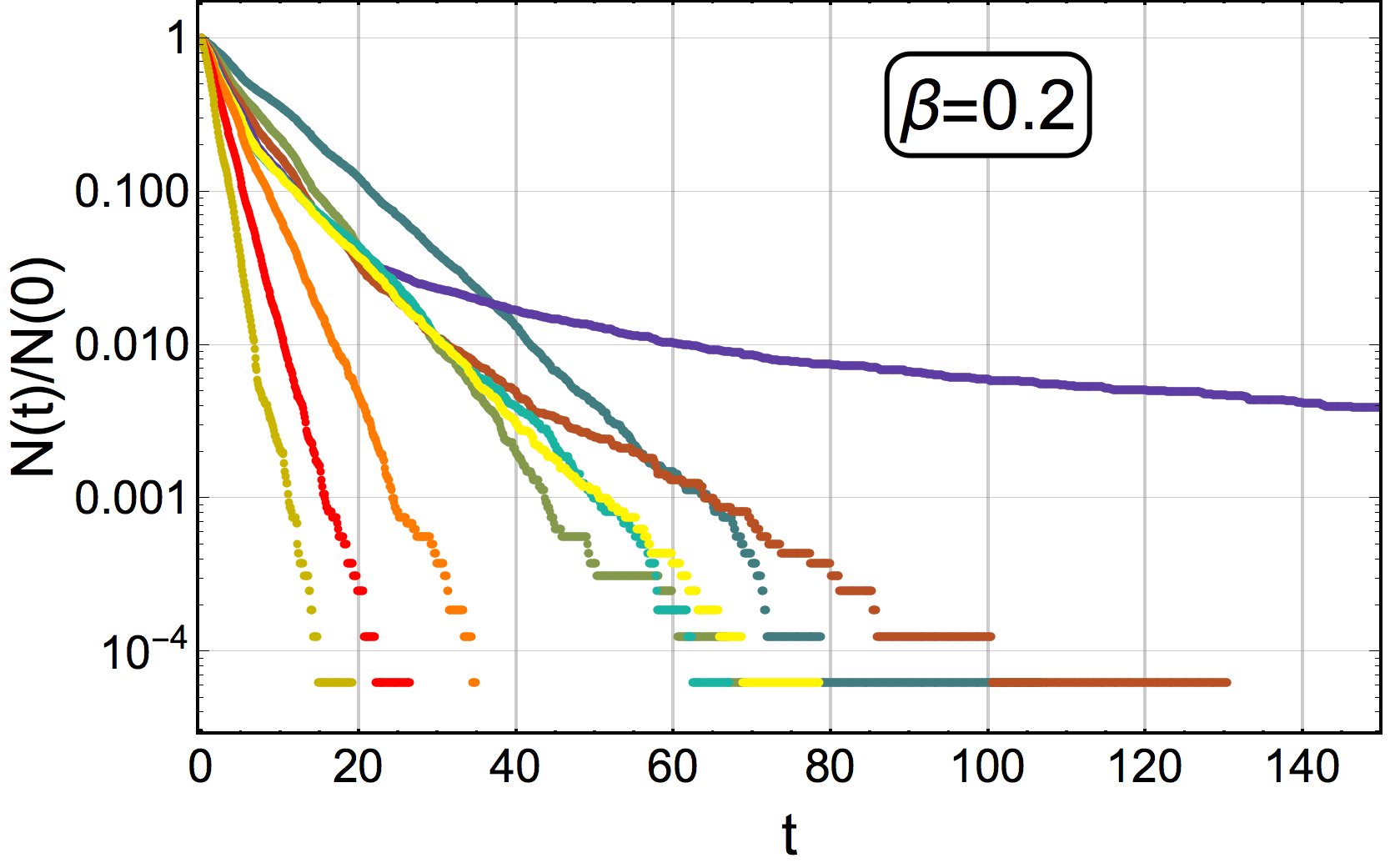}
\includegraphics[width=3.3in]{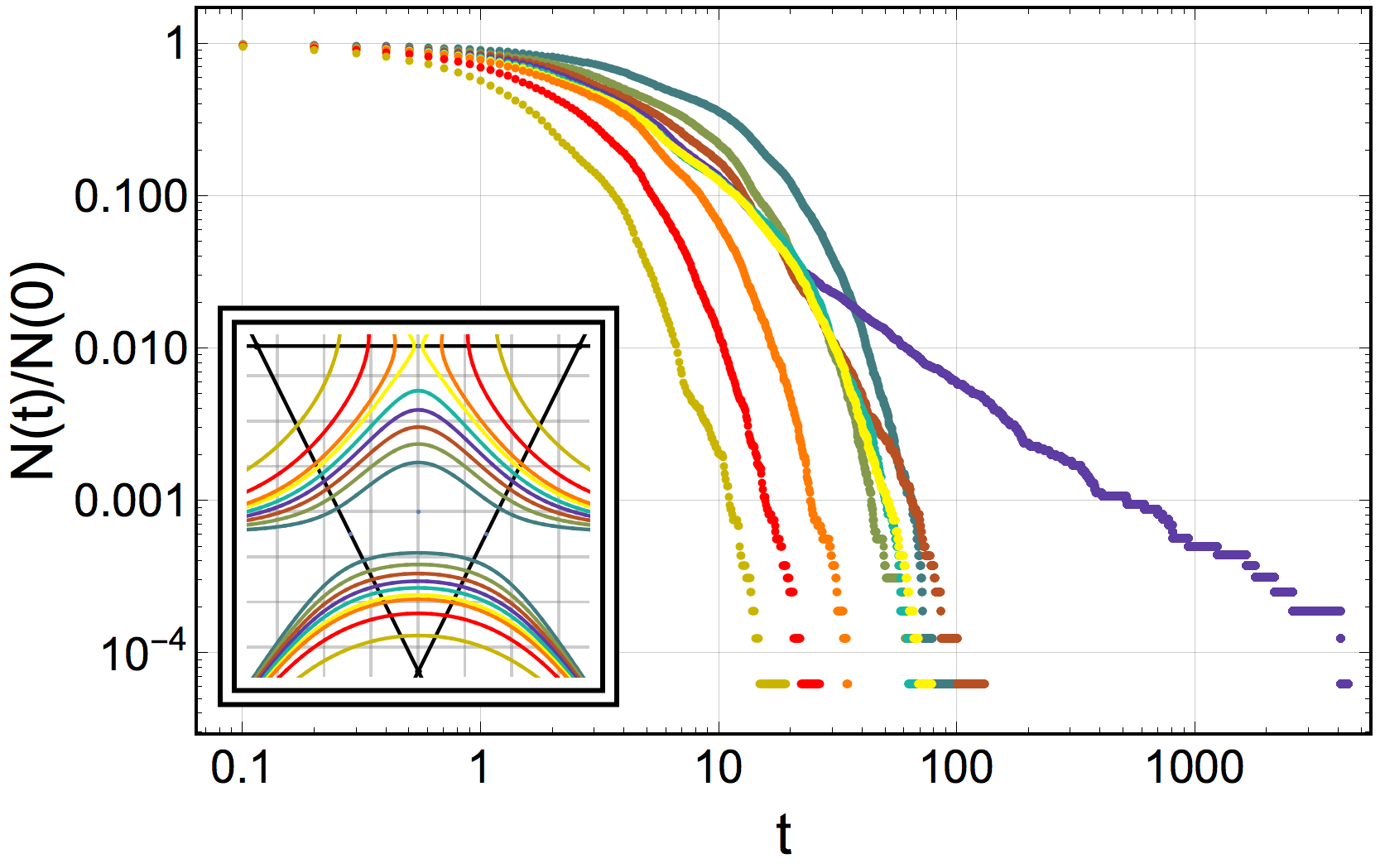}
\caption{{\footnotesize For $\beta=0.2$: (\textbf{Upper panel}) The population in the potential at time $t$ in Log-Linear,  for different energies as indicated in the code of figure  \ref{fig:saddle} .  (\textbf{Lower panel}) The population in the potential at time $t$ in Log-Log. (\textbf{Inset}) The isopotential for different values of $\Delta E$ (see fig. \ref{fig:saddle}).}} \label{fig:contour02}
\end{figure}

In figures \ref{fig:contour08}, \ref{fig:contour06}, \ref{fig:contour04} and \ref{fig:contour02} we present the population $N(t)$ inside the well as a function of time in Linear-Log and Log-Log plots. We also present the potential contour corresponding to each of the values for $\Delta E$ for a fixed $\beta$. Rather than using legends in each of the plots, we used instead the same colors, i.e., the red curves in the potential contour plot and in the \textit{ N(t) vs t} correspond to the same value of $\Delta E$ which we show in figure \ref{fig:saddle}. The first thing one may notice is that, in general, the escape flow at a given time follows an exponential decay. However, there are some values of $\beta$ and $ \Delta E$ for which the escape flow at a given time has a crossover unto a power law decay and \textit{ sticky states} appear, but we will come back to this later.

\begin{figure}[hbtp]
\centering
\includegraphics[width=3.3in]{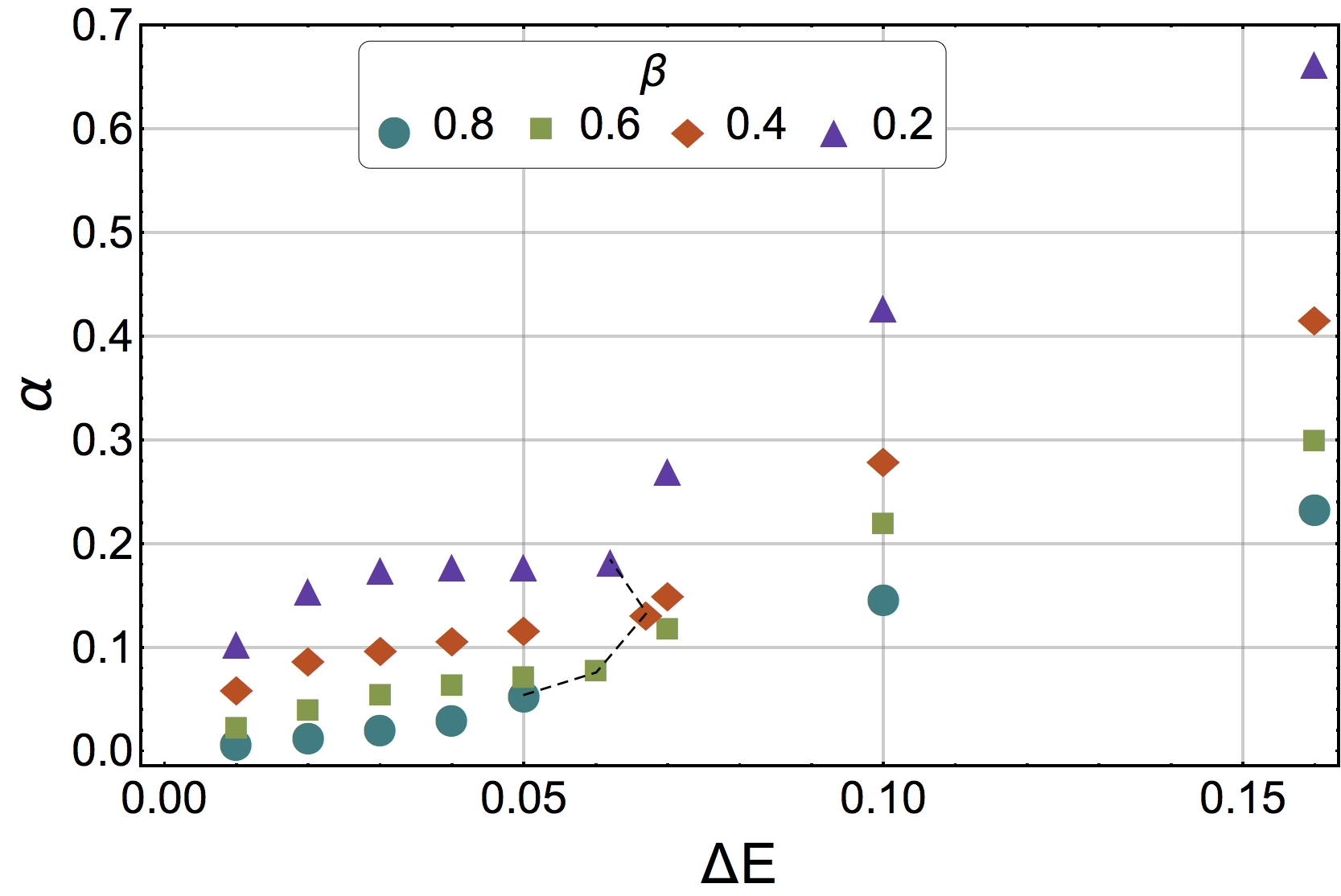}
\caption{{\footnotesize $\alpha(\beta, \Delta E)$ obtained numerically by solving the Hamilton Eqs. (\ref{eq:HamilEq}). The dashed line separates the region where the exit channel corresponding to the saddle point $P_3$ is  forbidden (left) and the region where it is accessible (right).} } \label{fig:alfaNumerically}
\end{figure}

Notice that for small $t$ it seems that $N(t) \sim \exp (-\alpha t)$. In figure \ref{fig:alfaNumerically} we present the values of  $\alpha$ as a function of $\beta$ and $\Delta E$, obtained by fitting the curves in figures \ref{fig:contour08}, \ref{fig:contour06}, \ref{fig:contour04} and \ref{fig:contour02} in the short-time regime (see supplementary material). The dashed line divides the scenario in which there are only two exit channels (which correspond to the two lower saddle points) from the scenario where there are three exit channels (see fig. \ref{fig:regionOfInt}). \\
To understand this exponential decay behavior for the unsoftened Henon-Heiles model ($\beta=1$), first Bauer and Bertsch and then Zhao \textit{et al.} (see \cite{zhao2007threshold, bauer1990decay}) used simple rather clever arguments. They considered that all initial conditions contained in the energy landscape well would flow out and assumed that the population change rate equals the flux with momentum orientation between $-\pi/2$ and $\pi/2$ relative to the normals of the exit channels line. Hence
\begin{equation}
\frac{d N(t)}{d t}=-N(t)\rho \int_{-\pi/2}^{\pi/2} d\theta \int_{r_0}^{r_1} dl \vert \vec{v}(x,y) \vert \cos \theta  \; ,
\label{eq:Zhao}
\end{equation} 
where $\rho=1/2\pi S(\Delta E)$ is the distribution of the variables $(x,y,\theta)$ and $S(\Delta E)$ is the area of the well. The integral goes over the opened exit channel lines and, the points $r_0$ and $r_1$ correspond to the classical return points at these opened exit channel lines. In the case where $\beta=1$, all exit channels are identical. Then, integrating over the exit channel $P_3$ and multiplying by $3$ yields
\begin{equation}
\frac{d N(t)}{d t}=-\frac{\sqrt{3} \Delta E}{S(\Delta E)}N(t) \; .
\end{equation}
Hence, $N(t)\sim e^{-\alpha t}$, with $\alpha =\sqrt{3} \Delta E/S(\Delta E)$.

In the softened Henon-Heiles model, the exit channels $P_1$ and $P_2$ are identical and different from $P_3$. Following this same idea, we determined $\alpha(\beta , \Delta E)$  by estimating the escaping flux numerically from Eq. (\ref{eq:Zhao}), but evaluated for the open channels for a given energy where  
\begin{eqnarray}
\vert v(x,y) \vert &=&\left[ 2\left(\Delta E + \frac{1}{12}\beta^2 (1+\beta) \right. \right.  \label{eq:velocity} \\
 & - & \left. \left.  \frac{1}{6}\left(3\beta x^2+\left(2+\beta \right)y^2 \right)+ \left(\frac{1}{3} y^3-x^2 y \right) \right) \right]^{1/2} \nonumber \; .
\end{eqnarray}
The first two terms on the right hand side of Eq. (\ref{eq:velocity}) are simply the energy above the lower saddle points  and the lower saddle points energy height.The third term in the right hand side of Eq. (\ref{eq:velocity})  is the softened Henon-Heiles potential. The classical returning points as well as the area of the wells, $S(\Delta E)$, were also determined numerically for different values of $\Delta E$.

\begin{figure}[hbtp]
\centering
\includegraphics[width=3.4in]{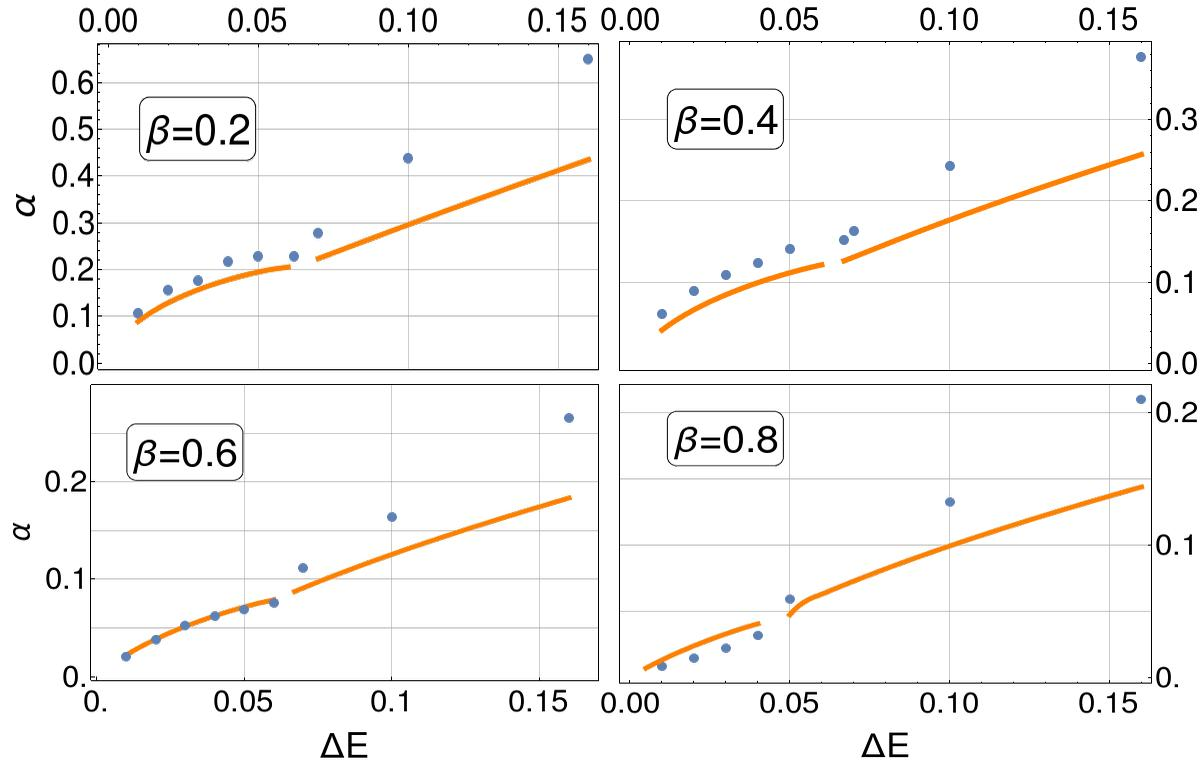}
\caption{{\footnotesize $\alpha(\beta, \Delta E)$ obtained from numerically solving the Hamilton Eqs. (\ref{eq:HamilEq}) (blue dots) and by considering the initial conditions escape flux (see Eq. (\ref{eq:Zhao})) (yellow lines).}} \label{fig:alfaComparing}
\end{figure}

These results can be compared with the actual fitting of $\alpha$ obtained from the numerical results of  $\log N(t)$ for small $t$, as obtained in figures  \ref{fig:contour08}, \ref{fig:contour06}, \ref{fig:contour04} and \ref{fig:contour02}. In figure \ref{fig:alfaComparing} we compare both methods, namely, the blue points correspond to the fitting methods while the orange curves correspond to the numerical  flux estimation method. We should clarify that the orange curves were obtained by interpolating numerical results, i.e., no analytical Eq. was obtained. One may appreciate how this heuristic approach works quite well for small $\Delta E$, not so much for when the flow may come out through the upper channel, i.e., $P_3$. This has to do with the fact that in this regime, not all initial conditions in  configurational space flow out of the basin, as we will show later.

From figures \ref{fig:alfaNumerically} and   \ref{fig:alfaComparing} is clear that the relaxation time is decreased as the stiffness of the model is reduced, since  in general $\alpha$ grows as $\beta$ is reduced for a fixed energy. This is explained in general by two effects. The first is a reduction of the energy barrier heights along the softened normal modes (see fig. \ref{fig:saddlepointH}), and the second is a widening of the opening channels.

\section{Power law relaxation and sticky states}\label{sec:power}

\begin{figure}[hbtp]
\centering
\includegraphics[width=3in]{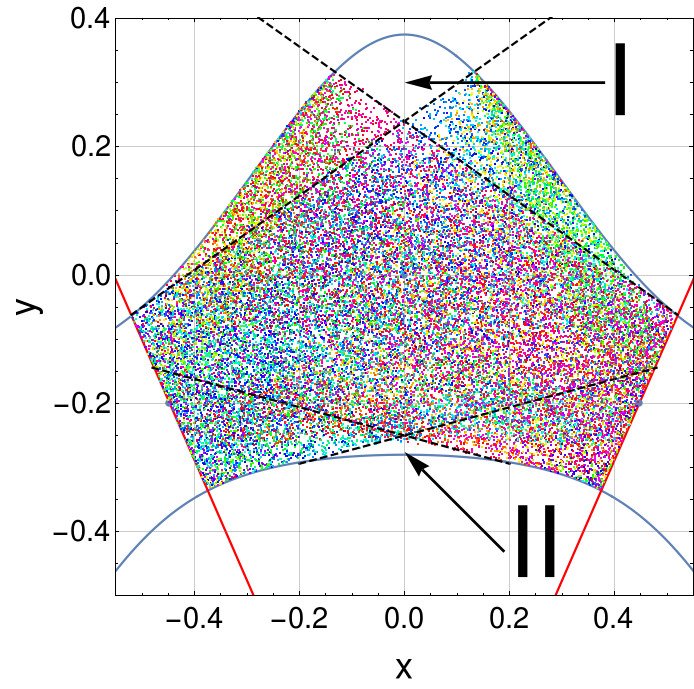}
\includegraphics[width=3.3in]{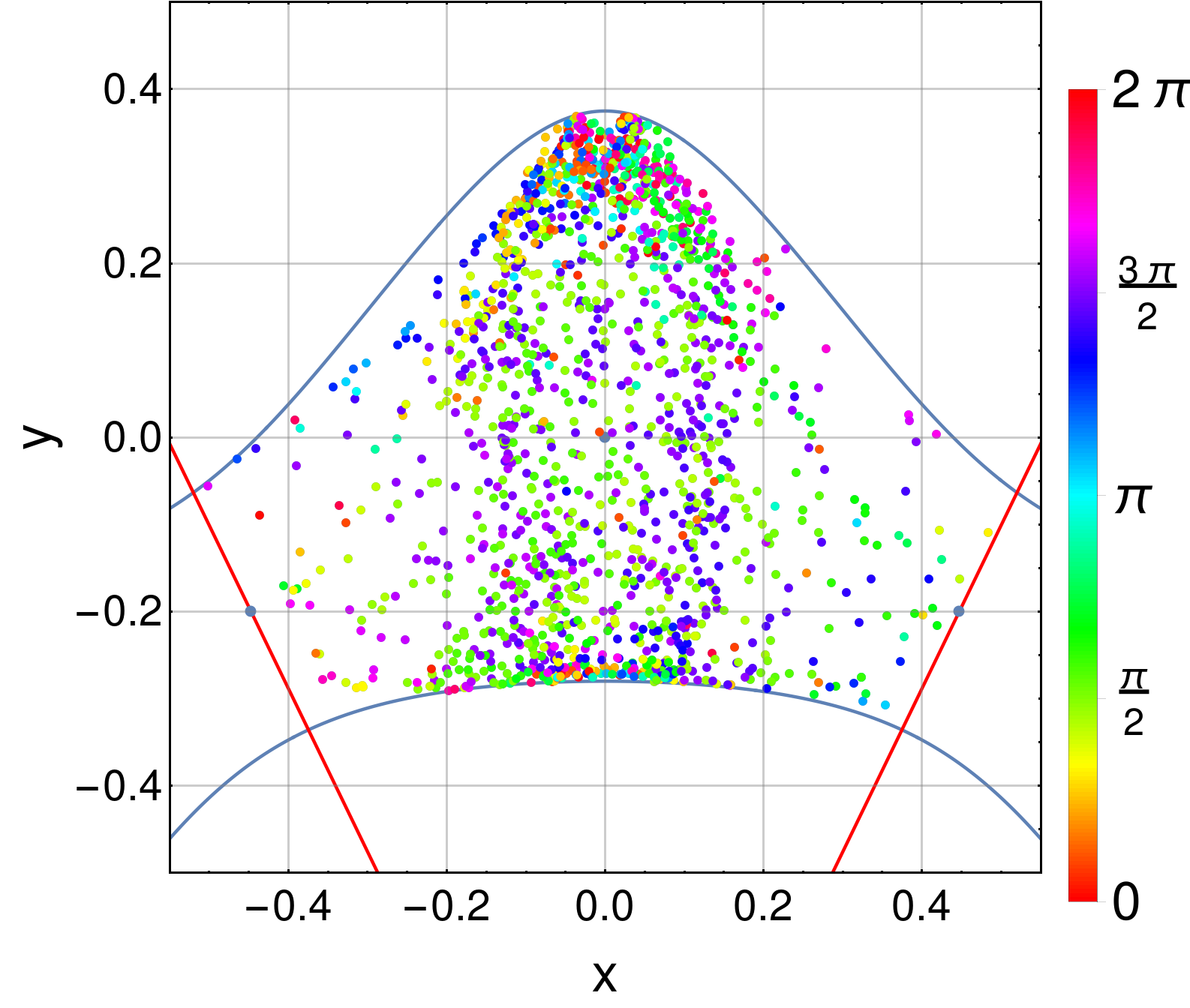}
\caption{{\footnotesize Escape decay regimen as a function of the initial position for $\beta=0.4$ and $\Delta E=0.02$. The coloring corresponds to the initial moment orientation with respect to the horizontal (see legend). (\textbf{Upper panel}) Particles which escape under the exponential decay regime. The black dashed lines are an eye-guidelines to indicate the cone-like regions of particles with initial moment orientation directed towards one of the exits, and also indicate the regions I and II were particles there take longer to escape no matter what their initial moment orientation is. (\textbf{Lower panel}) Particles which escape under the power law regime as a function of the initial positions. The isopotential curve is indicated in blue while the exits of the potential basin is indicated by lines.} } \label{fig:EscapeDiagUnder7}
\end{figure}

To have a better grip and to qualitatively differentiate  which conditions flows out following an exponential decay from the region which flows out following a power law, we fixed $\beta=0.4$ and $\Delta E =0.02$ and solved numerically the Hamilton Eqs. (\ref{eq:HamilEq}) for $N \simeq 5\cdot 10^4$ different initial conditions, i.e., $(x,y,\theta)$. In the upper panel of figure \ref{fig:EscapeDiagUnder7} we have plotted $\sim 50\%$ of the studied initial positions  which  flows out first , i.e., we plot the configurational space region that corresponds to the exponential decay regime of $N(t)$, and we have colored each point according to their initial moment orientation. We also did this for the initial conditions which flow out the slowest, corresponding to the power law regime of $N(t)$, and we show this in the lower panel of figure \ref{fig:EscapeDiagUnder7}. There are several features we may extract from this. First, notice that the conditions that flow out first  are those with an initial moment orientated towards the exits. Yet, there are some initial positions which define two regions, namely, region I and region II (see upper panel in figure \ref{fig:EscapeDiagUnder7}), which does not flow out during the exponential decay regime no matter what their initial moment orientation is. This is further verified in the lower panel of figure \ref{fig:EscapeDiagUnder7} from which we may qualitatively appreciate a density gradient in the $y$ direction for $y>0$ and in the $-y$ direction for $y<0$.

\begin{figure}[hbtp]
\centering
\includegraphics[width=3.3in]{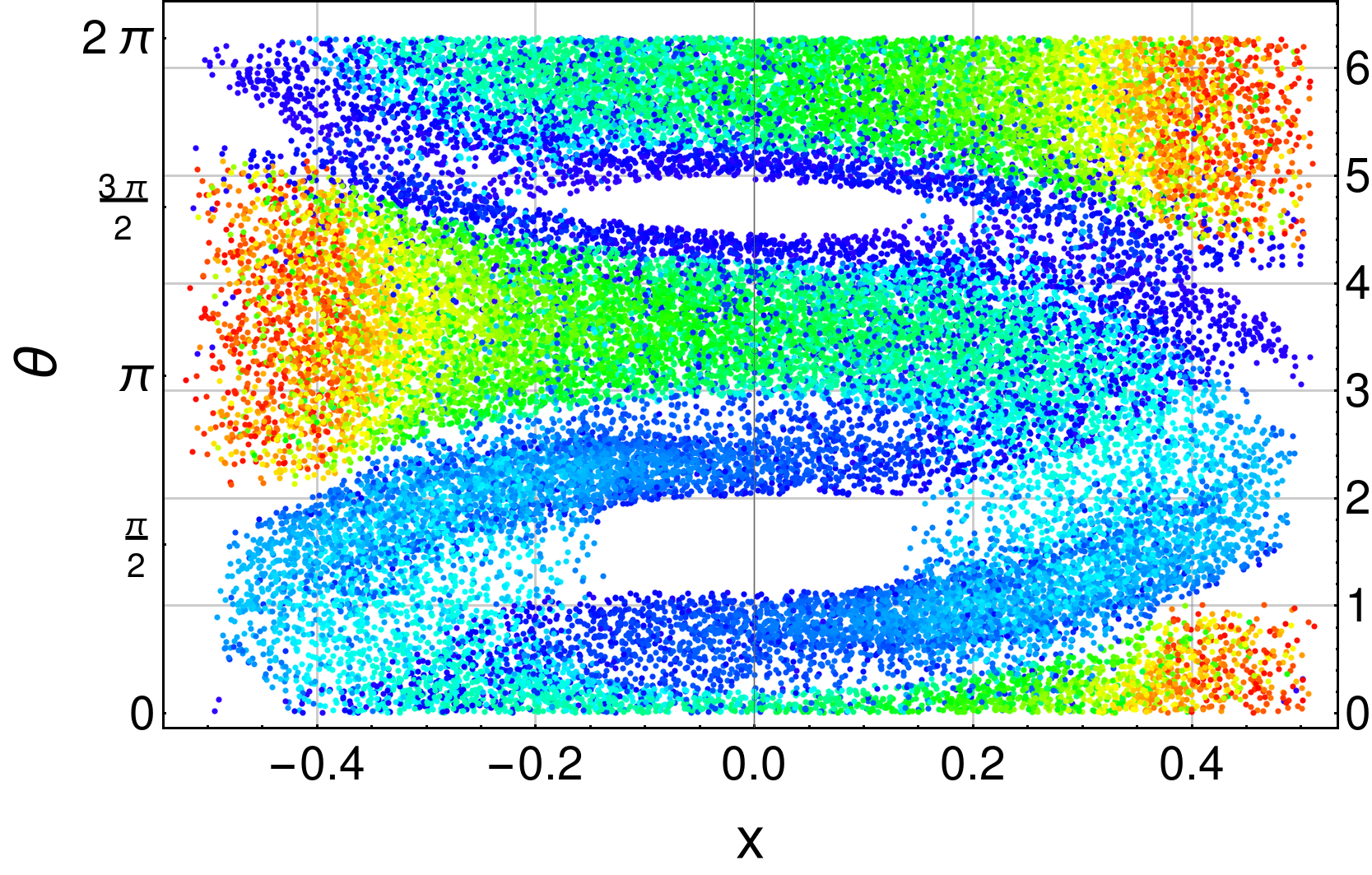}
\includegraphics[width=3.3in]{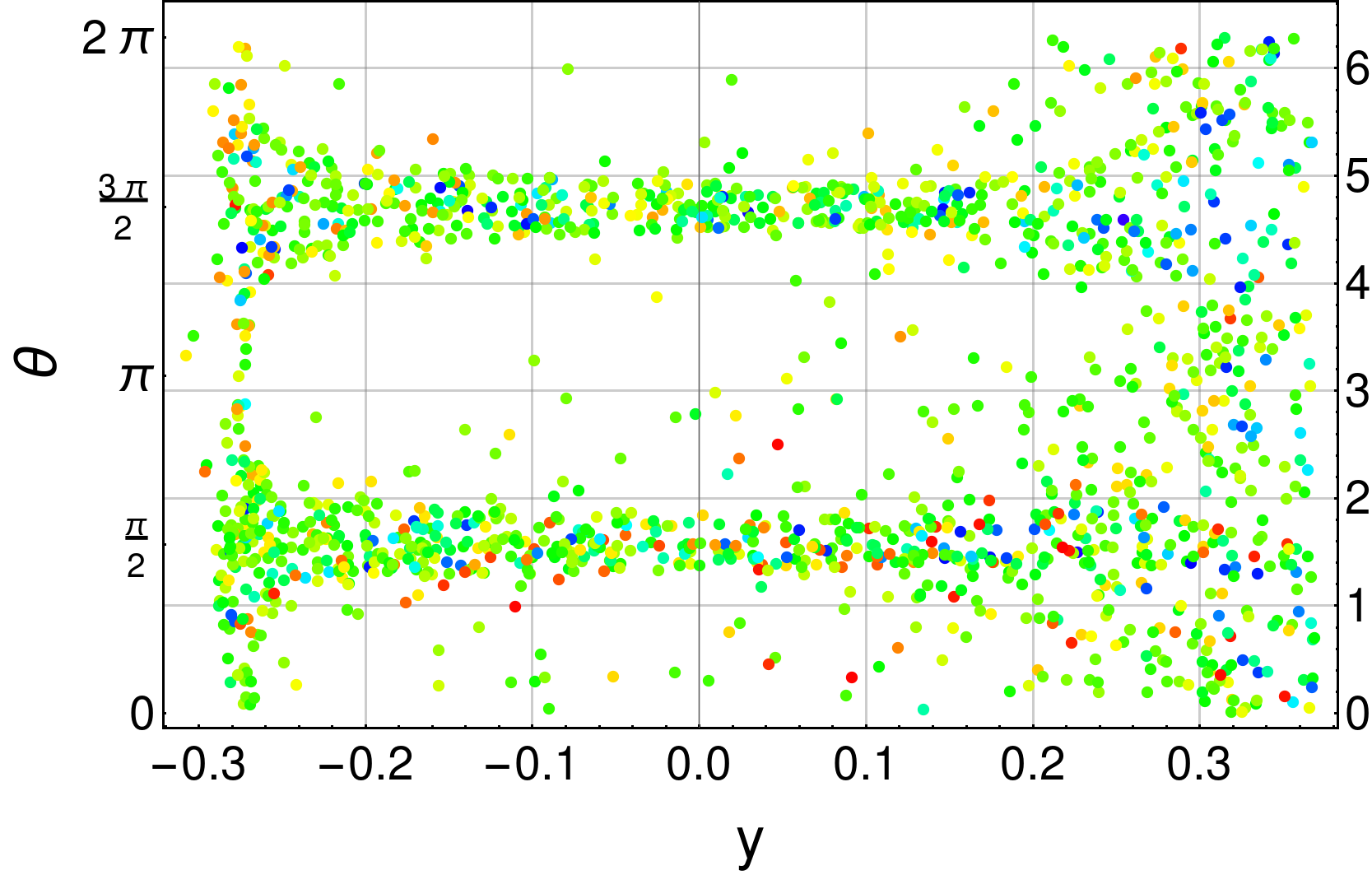}
\caption{{\footnotesize Decay regime as a function of the initial position $x$ and initial moment orientation $\theta$ in the case of $\beta=0.4$ and $\Delta E=0.02$. (\textbf{Upper panel}) Particles which escape under the exponential decay regime. (\textbf{Lower panel}) Particles which escape under the power law regime. The coloring corresponds to the time taken to escape, such that red is short exit times and navy blue is for large exit times (see legend). }} \label{fig:EscapeDiagx}
\end{figure}

Also notice that the bulk of the initial positions which flow out slowly is concentrated in a vicinity of $x\simeq 0$. However, there is clearly an overlap in this vicinity with the initial positions that exit quite fast. For this reason, in figure \ref{fig:EscapeDiagx} we have plotted in the upper panel the initial condition coordinate $x$ and the initial moment orientation $\theta$ for $50\%$ for the initial conditions that flows out first, while in the lower panel of figure \ref{fig:EscapeDiagx} we plotted the initial condition coordinate $y$ and the initial moment orientation $\theta$ of the slowest flowing initial conditions. In both plots the color is a function of the exit time, i.e., red corresponds to the smallest exiting time while navy blue corresponds to the largest exiting time. Notice that from the upper panel of figure \ref{fig:EscapeDiagx} one may appreciate fairly well that the configurational space region that flows out first is the one next to the left (right) exit channel and have initial moment orientation between $2.3$ and $3\pi/2$ ($4.5$ and $2\pi+1$), then the initial conditions region that follows has an initial moment orientation contained in these intervals but with smaller absolute value of the initial position coordinate $x$, in other words, the initial moment orientation still corresponds to that directed towards one of the exits channels. Now, the initial conditions region that follows has an initial moment orientation in the vicinity of $\pi/2$ and is distributed around $x\pm 0.3$. This means that this initial conditions region collides with the potential barrier before being able to flow out. 
 
Now, from  the lower panel in figure \ref{fig:EscapeDiagx} we may appreciate that the initial conditions region which takes the longest to flow out is distributed all over the classically permitted interval in the $y$ axis but with either $\pi/2$ or $3\pi/2$ as the value of the initial moment orientation.  All this suggests that the region that takes the longest to flow out corresponds to oscillating trajectories in the $y$ direction and in a quasiperiodic manner with $x(t)\simeq 0$, i.e., sticky orbits appear.

\begin{figure}[hbtp]
\centering
\includegraphics[width=3.3in]{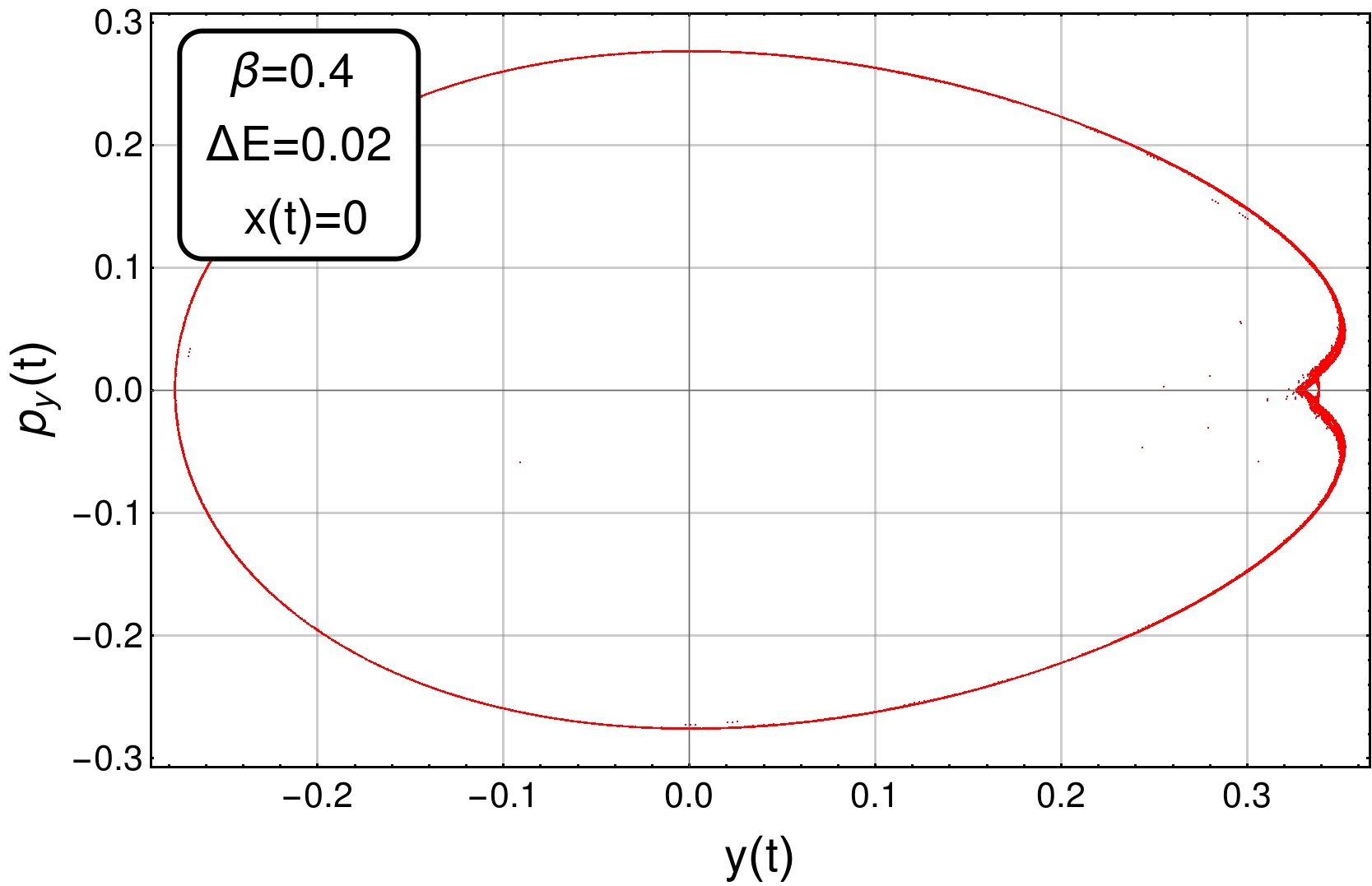}
\caption{{\footnotesize Poincar\'e map for $\beta=0.2, \Delta E=0.04$ and some initial condition for which the particle takes a long time to exit the basin, corresponding to a sticky state.}} \label{fig:PoincareMap}
\end{figure}

Going one step further, in figure \ref{fig:PoincareMap} we show a \textit{Poincar\'e} section for one of the initial conditions which takes a long time to flow out, obtained in the case of $\beta=0.4$ and $\Delta E=0.02$. Cleary, this type of section corresponds to a quasi-periodic trajectory. Eventually, the small deviations amplify and the trajectory escapes the well.

\section{Conclusion} \label{sec:discussion}
In this work we have studied the relaxation process of a chain consisting of $3$ masses joined by non-linear springs, periodic conditions and weakened stiffness. The idea was to explore
how relaxation is modified by changes in the low-frequency vibrational mode region. We found that the relaxation time is, for the most part, exponential, which is in agreement with the idea of using an ergodic description. This was confirmed by using a simple flux balance that relates the accesible area of the basin, the initial conditions and the size of the basin appertures above the saddle points.  We have shown that by reducing the rigidity of the model,i.e. by softening one of the normal modes, the system relaxes faster. There are two reasons for this. One is the decreasing of the energy barriers since two of the saddle points of the landscape are reduced in energy. The other is related with the shape of the basin. This leads to the conclusion that relaxation occurs mainly into directions of soft or floppy modes \cite{naumis2005energy}.    
However, we also found some energies and regions in phase space where sticky states appear. For these states, the relaxation decay follows a power law. In these sticky states, the trajectory is quasiperiodic in the sense that, given the isopotential curvature, small deviations occur. Eventually, these deviations are amplified in such a way that the quasiperiodicity is lost and the trajectory goes out of the well. This can also be understood as energy transfer from one normal mode to another. It is believed that the sticky states phase region diminish as the energy increases. However, we have shown that is not always the case and, in fact, there are certain islets in the phase space for which sticky states appear as energy increases (see figure \ref{fig:contour02}). Furthermore, as the control parameter $\beta$, which measures the stiffness of the model decreases, the sticky states region in phase space diminishes. This is expected since energy sharing (transfer) starts in the low vibrational modes due to resonances. The sticky states regions were clearly identified with initial conditions for which the assumptions made for the simple flux balance are broken, i.e., regions in which a kind of cavity exist. Therein, for most particles is impossible to leave the energy landscape basin without having many reflections, since their momenta do not have components on the direction of the normal to the lines which are the gates of the energy landscape basin.   

\section{Acknowledgments}

This work was partially supported by DGAPA-UNAM project IN102717. J.Q.T.M. acknowledges a doctoral fellowship from CONACyT.

\newpage

\appendix

\section{Eigenvectors} \label{app:Eigenvectors}
In this section we obtain the eigenvectors and eigenvalues of the 
dynamical matrix, in order to write the model in normal modes
coordinates. The eigenvalues of the dynamical matrix $\mathbf{D}$ are,
\begin{equation}
\begin{cases}\omega_0/k= 0  , \\
 \omega_{x}/k =1+\alpha+\beta +\sqrt{1-\alpha + \alpha^2-\beta -\alpha \beta +\beta^2} \rbrace \; , \\
 \omega_{y}/k =1+\alpha+\beta - \sqrt{1-\alpha + \alpha^2-\beta -\alpha \beta +\beta^2} \rbrace
 \end{cases} \; .
\end{equation} \label{eq:eigenvalues2}

Let us determine the eigenvectors. To do so, notice that since there are no fixed endpoints, the first eigenvalue corresponds to the center mass movement. Hence, 
one of the normal modes is,
\begin{equation}
\vert 0 \rangle =\frac{1}{\sqrt{3}} 
\begin{bmatrix}
1 \\ 1 \\ 1\\
\end{bmatrix} \; .
\end{equation}
The remaining two eigenvectors must be orthonormal. Therefore, one way to tackle this is by considering two generic eigenvectors in the $XY$ plane, i.e.,
\begin{equation}
\vert 1, 0 \rangle = 
\begin{bmatrix}
\sin \chi \\ -\cos \chi \\ 0\\
\end{bmatrix} \; .
\end{equation}
\begin{equation}
\vert 2, 0 \rangle = 
\begin{bmatrix}
\cos \chi \\ \sin \chi \\ 0\\
\end{bmatrix} \; .
\end{equation}

We rotate these vectors using the same rotation matrices which transforms vector $(0,0,1)$ into $(1,1,1)/\sqrt{3}$, i.e., first we rotate an angle $\theta_x$ such that $\cos \theta_x=1/\sqrt{3}$ around the $X$ axis clockwise, and then we rotate an angle $\theta_z=\pi/4$ around the $Z$ axis clockwise. Thus
\[
\mathbf{R}_x(\theta_x)=
\begin{bmatrix}
   1 & 0 & 0  \\
    0 & \cos \theta_x & -\sin \theta_x  \\
   0 & \sin \theta_x & \cos \theta_x 	\\
\end{bmatrix} \; ,
\] 
\[
\mathbf{R}_x(\theta_x)=
\begin{bmatrix}
   \cos \theta_z & -\sin \theta_z & 0  \\
   \sin \theta_z & \cos \theta_z & 0 	\\
   0 & 0 & 1 \\
\end{bmatrix} \; ,
\]  
\begin{equation}
\vert 1 \rangle \equiv \mathbf{R}_z(\theta_z)  \mathbf{R}_x(\theta_x) \vert 1, 0 \rangle = 
\begin{bmatrix}
-\frac{\cos \chi}{\sqrt{6}}+\frac{\sin \chi}{\sqrt{2}} \\ 
-\frac{\cos \chi}{\sqrt{6}}-\frac{\sin \chi}{\sqrt{2}} \\ 
\sqrt{\frac{2}{3}} \cos \chi
\end{bmatrix} \; .
\end{equation}
\begin{equation}
\vert 2 \rangle \equiv \mathbf{R}_z(\theta_z)  \mathbf{R}_x(\theta_x) \vert 2,0 \rangle =
\begin{bmatrix}
\frac{\cos \chi}{\sqrt{2}}+\frac{\sin \chi}{\sqrt{6}} \\ 
-\frac{\cos \chi}{\sqrt{2}}+\frac{\sin \chi}{\sqrt{6}} \\ 
-\sqrt{\frac{2}{3}} \sin \chi
\end{bmatrix} \; ,
\end{equation}

Now, notice that both eigenvectors must fulfill the characteristic equation for $\mathbf{D}$,
from which we obtain, using $\vert 2 \rangle$ that
\begin{equation}
\tan \chi =
 \sqrt{3} \frac{2+\alpha -\omega_{-}^2}{3\alpha -\omega_{-}^2} \; .
\label{eq:tantheta}
\end{equation}

Before moving on, notice that when $\alpha=\beta=1$ the eigenvalues become $\lbrace 0, 3 \rbrace$ being the last one degenerate. Also notice that making $\chi \rightarrow \chi-\pi/2$ in $\vert 2 \rangle$ gives $\vert 1 \rangle$.

Now that we have the eigenvectors, we diagonalize $\mathbf{D}$. We define $\vert i c \rangle$ as the canonical base with $i=\lbrace 1, 2, 3 \rbrace$ and the matrix $\mathbf{A}$, which diagonalizes the interaction, i.e., $\mathbf{D}_0=\mathbf{A} \mathbf{D} \mathbf{A}^T$. Then $A$ is
\begin{eqnarray}
\mathbf{A}&=&\vert 1c \rangle \langle 1 \vert +\vert 2c \rangle \langle 2 \vert + \vert 3c  \rangle \langle 0 \vert \\
&=&\begin{bmatrix}
-\frac{\cos \chi}{\sqrt{6}}+\frac{\sin \chi}{\sqrt{2}} & -\frac{\cos \chi}{\sqrt{6}}-\frac{\sin \chi}{\sqrt{2}}  & \sqrt{\frac{2}{3}} \cos \chi \\
 \frac{\cos \chi}{\sqrt{2}}+\frac{\sin \chi}{\sqrt{6}} & -\frac{\cos \chi}{\sqrt{2}}+\frac{\sin \chi}{\sqrt{6}} & -\sqrt{\frac{2}{3}} \sin \chi \\ 
 1/\sqrt{3} & 1/\sqrt{3} & 1/\sqrt{3}  \\ 
\end{bmatrix} \; .
\end{eqnarray}

Therefore, the normal (canonical) coordinates are
\begin{equation}
\vec{x}=\mathbf{A} \vec{Q} \; ,
\end{equation}
\begin{equation}
\vec{p}=\mathbf{A} \vec{P} \; .
\end{equation}
Hence, in our Hamiltonian we make the following substitutions:
\begin{equation}
Q_i=\sum \mathbf{A}_{ij}^t x_j \; ,
\end{equation}
\begin{equation}
P_i=\sum \mathbf{A}_{ij}^t y_j \; .
\end{equation}
Substituting these last Eqs. in the Hamiltonian, we obtain the diagonalized Hamiltonian, i.e., 
\begin{equation}
H=\frac{1}{2}\left(p_x^2+p_y^2+p_z^2 \right)+ \frac{1}{2} \left(\omega_x^2 x^2+\omega_y^2 y^2 \right) \; .
\end{equation}
Notice that $p_z$ is a constant since $z$ is a cyclic coordinate.

Consider the cubic interaction term, i.e.,
\begin{equation}
\frac{\gamma}{3} \left( \left(Q_1-Q_2 \right)^3 + \left(Q_2-Q_3 \right)^3 + \left(Q_3-Q_1 \right)^3 \right) \; ,
\end{equation}
and let us apply the transformation $Q_i \rightarrow x_i$. This last operation yields
\begin{equation}
\frac{3\gamma}{\sqrt{2}} \left(- \xi_2 \left(\frac{\xi_2^2}{3}-\xi_3^2 \right) \cos 3 \chi + \xi_3 \left(\frac{\xi_3^2}{3}-\xi_2^2 \right) \sin 3 \chi \right) \; .
\end{equation}

In the case where $\alpha=\beta$, let us examine what happens to the angle $\theta$. Notice that the matrix $\mathbf{D}$ becomes

\[
\mathbf{D}=
\begin{bmatrix}
    1+\beta      & -1 & -\beta  \\
   -1      & 1+\beta & -\beta  \\
    -\beta & -\beta & 2\beta \\
\end{bmatrix} \; ,
\] 
with eigenvalues $\omega_y^2=2+\beta$ and $\omega_x^2=3\beta$.
When fixing the angle $\theta$ for which the secular Eq. is fulfilled, we may use 
\begin{equation}
-\beta \left( \sqrt{\frac{2}{3}} \sin \chi \right) - \left( 2\beta - \omega_{x,y}^2 \right)\sqrt{\frac{2}{3}} \sin \chi=0 \; , \label{eq:secular}
\end{equation} 
which is the last Eq. of  $\left(\mathbf{D}-\omega_{2,3}^2 \mathbf{I} \right) \vert 2 \rangle =0$. Now, if we plug in the eigenvalue $3\beta$ in Eq. (\ref{eq:secular}) we obtain the trivial solution. On the contrary, when we plug in the eigenvalue $2+\beta$ in Eq. (\ref{eq:secular}), this tells us that $\chi=n \pi$, with $n=0,1,2,...$. We diagonalize the matrix $\mathbf{D}$ by obtaining the following diagonal matrix $\mathbf{D}_0$:
\begin{equation}
\mathbf{D}_0=\mathbf{A} \mathbf{D} \mathbf{A}^T=\begin{bmatrix}
    d_{11} (\chi)     & 0 & 0  \\
   0      & d_{22}(\chi) & 0  \\
    0 & 0 & d_{33}(\chi) \\
\end{bmatrix} \; .
\end{equation}
For $\chi = n \pi$, according to Eq. (\ref{eq:secular}) we should expect that $d_{11}(\chi)=\omega_x^2, \; d_{22}(\chi)=\omega_y^2 $ and $d_{33}(\chi)=0$. The last case occurs because, as previously mentioned, the eigenvector $\vert 2 \rangle$ becomes $\vert 1 \rangle$, and conversely $\vert 1 \rangle \rightarrow \vert 2 \rangle$

Hence, when $\alpha = \beta$,  $\chi=n \pi$ and the corresponding eigenvectors becomes
\begin{equation}
\vert 2+\beta \rangle =\frac{1}{\sqrt{2}} 
\begin{bmatrix}
(-1)^n \\ (-1)^{n+1} \\ 0\\
\end{bmatrix} \; .
\end{equation}
\begin{equation}
\vert 3\beta \rangle =\frac{1}{\sqrt{6}} 
\begin{bmatrix}
(-1)^{n+1} \\ (-1)^{n+1} \\ 2(-1)^n\\
\end{bmatrix} \; .
\end{equation}

Finally, the Hamiltonian in terms of the normal coordinates and momenta is
\begin{equation}
H=\frac{1}{2}\left(p_x^2 + p_y^2 + p_z^2 \right) + \frac{1}{2}\left(\omega_x^2 x^2 + \omega_y^2 y^2 \right) + \frac{3 \gamma}{2^{1/2}} \nu \left(\frac{1}{3} y^3- x^2 y \right) \; ,
\end{equation}
where $\nu= \lbrace -1, 1 \rbrace$.

\bibliography{mybib}

\widetext
\clearpage
\begin{center}
\textbf{\large Supplemental Materials for: Escape time, relaxation and sticky states of a softened Henon-Heiles model: low-frequency vibrational modes effects}
\end{center}
\setcounter{equation}{0}
\setcounter{figure}{0}
\setcounter{table}{0}
\setcounter{page}{1}
\makeatletter
\renewcommand{\theequation}{S\arabic{equation}}
\renewcommand{\thefigure}{S\arabic{figure}}
\renewcommand{\bibnumfmt}[1]{[S#1]}
\renewcommand{\citenumfont}[1]{S#1}
In figures \ref{fig:fit08}, \ref{fig:fit06}, \ref{fig:fit04} and \ref{fig:fit02} we show the curves of $\log N(t)/N(0)$ vs $t$ for times corresponding to the exponential decay regime and their linear fit for each of the combinations of $\beta$ and $\Delta E$ shown in figure \ref{fig:saddle}. The time intervals $t_{fit}$ used in the fitting process for each of the curves were selected following two criteria. The remaining population after $\Delta t$ should be below $50\%$, which ensures that the regime  in that time interval corresponds to the exponential decay. The fitting curve should go over the real data for the most part, such that the absolute value of the slope of the fitting curve, i.e., $\alpha$ will be smaller than or equal to the absolute value of the slope of the real data. In this sense, these $\alpha$ values correspond to lower bounds. Notice from figure \ref{fig:alfaComparing} that the $\alpha$ obtained from the fitting method almost always goes over the $\alpha$ obtained by the flux method. Thus, if one were to improve the fit by choosing a smaller time interval, the divergence between both methods would grow.

\begin{figure}[hbtp]
\centering
\includegraphics[width=3.in]{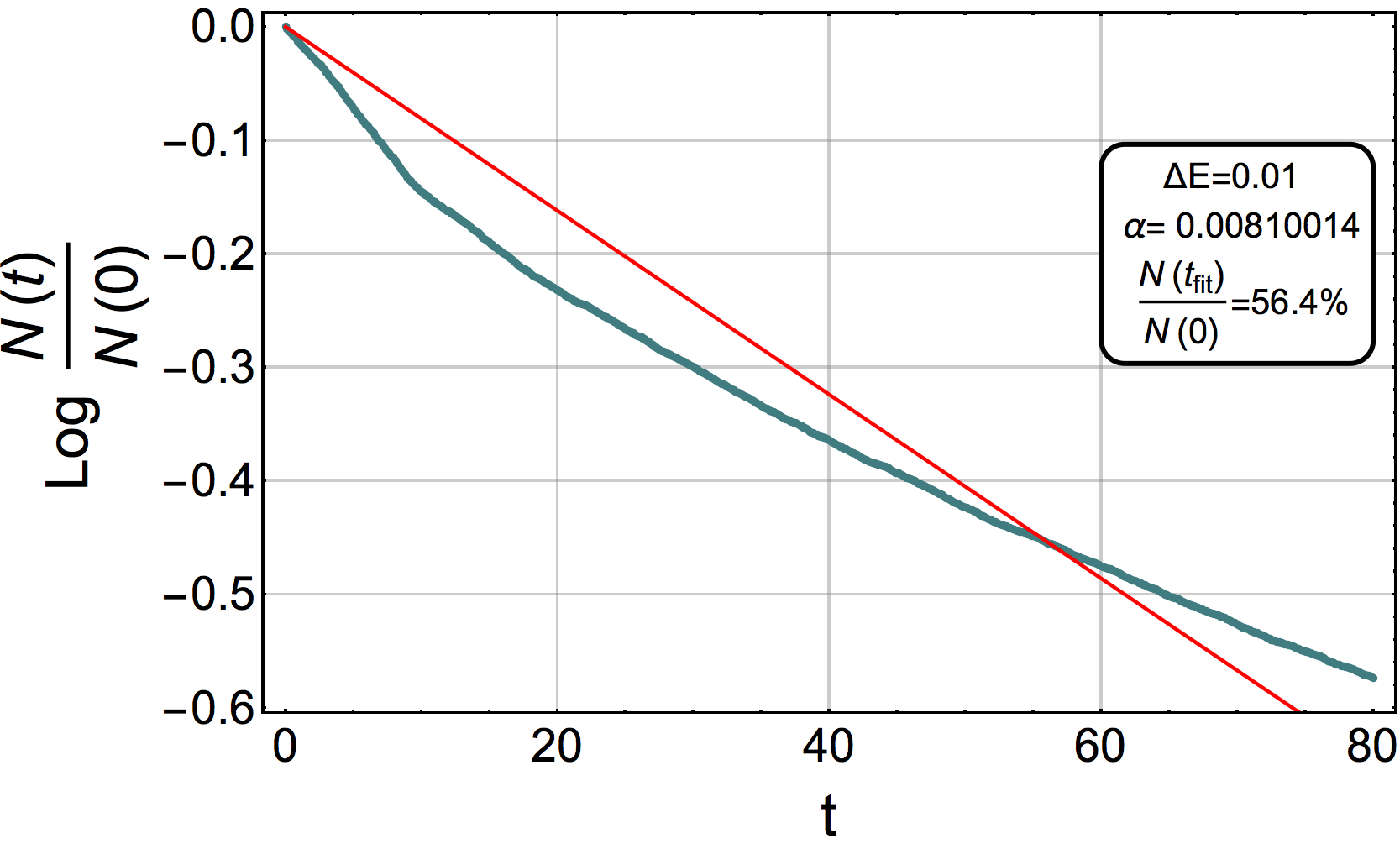}
\includegraphics[width=3.in]{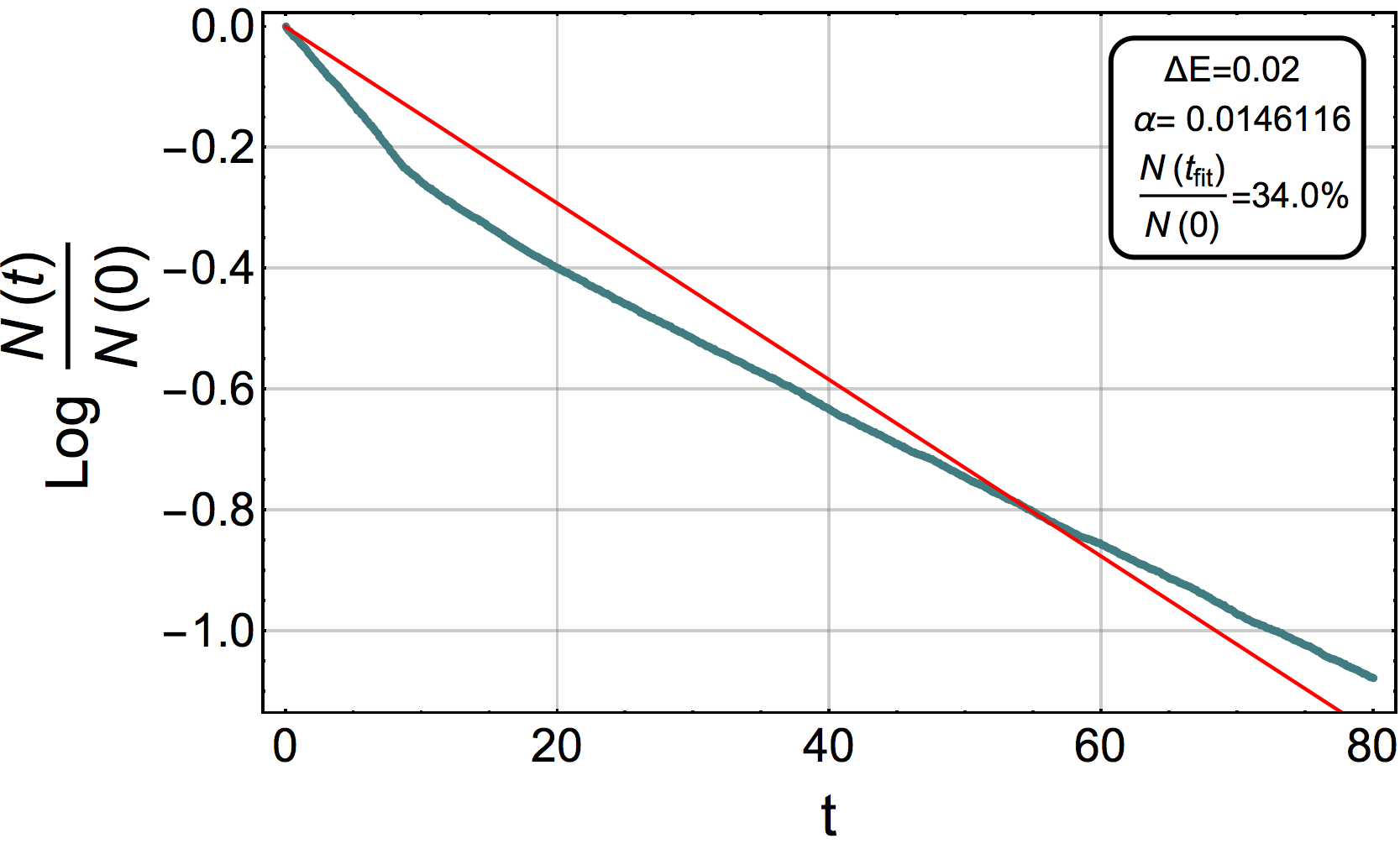}
\includegraphics[width=3.in]{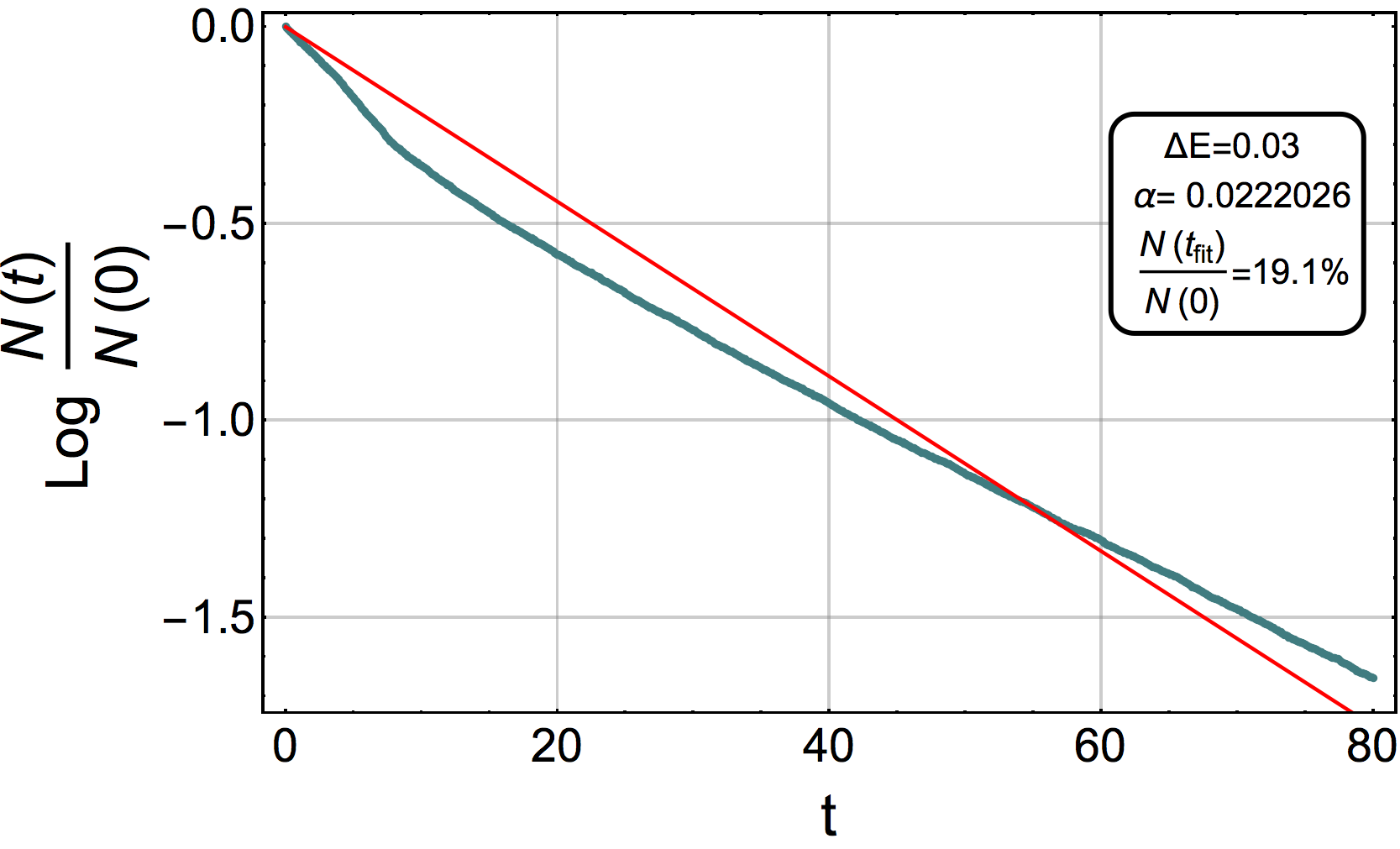}
\includegraphics[width=3.in]{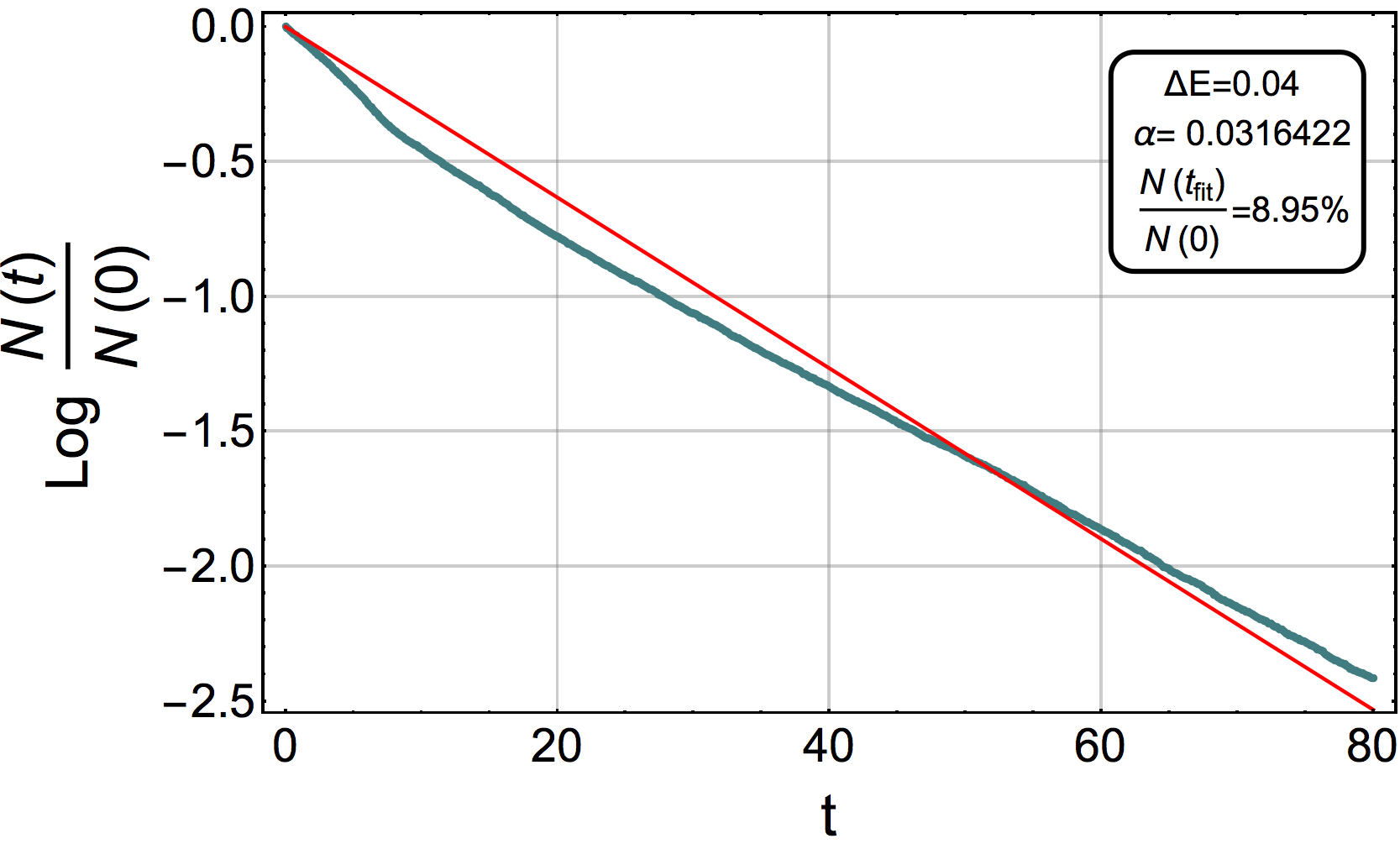}
\includegraphics[width=3.in]{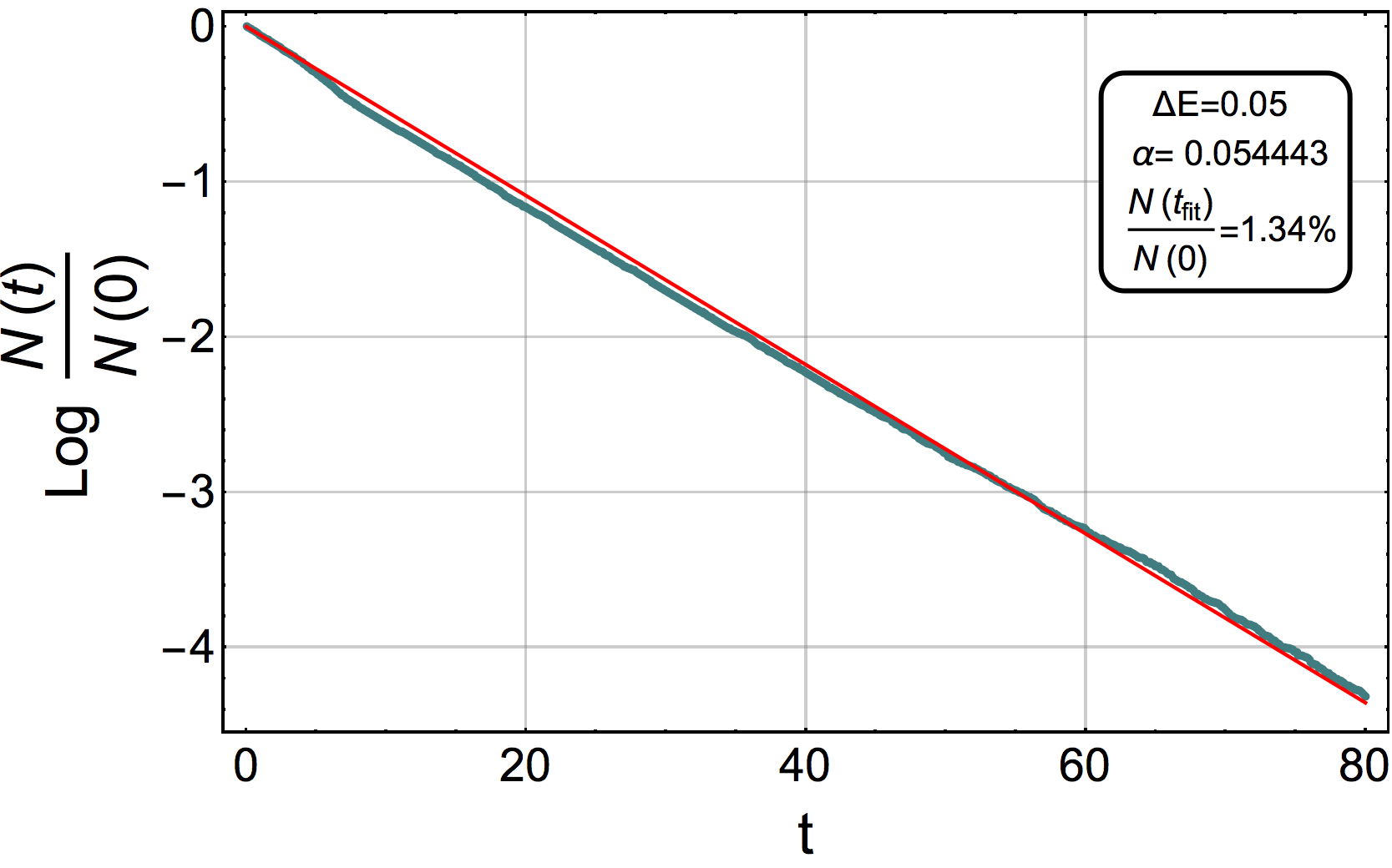}
\includegraphics[width=3.in]{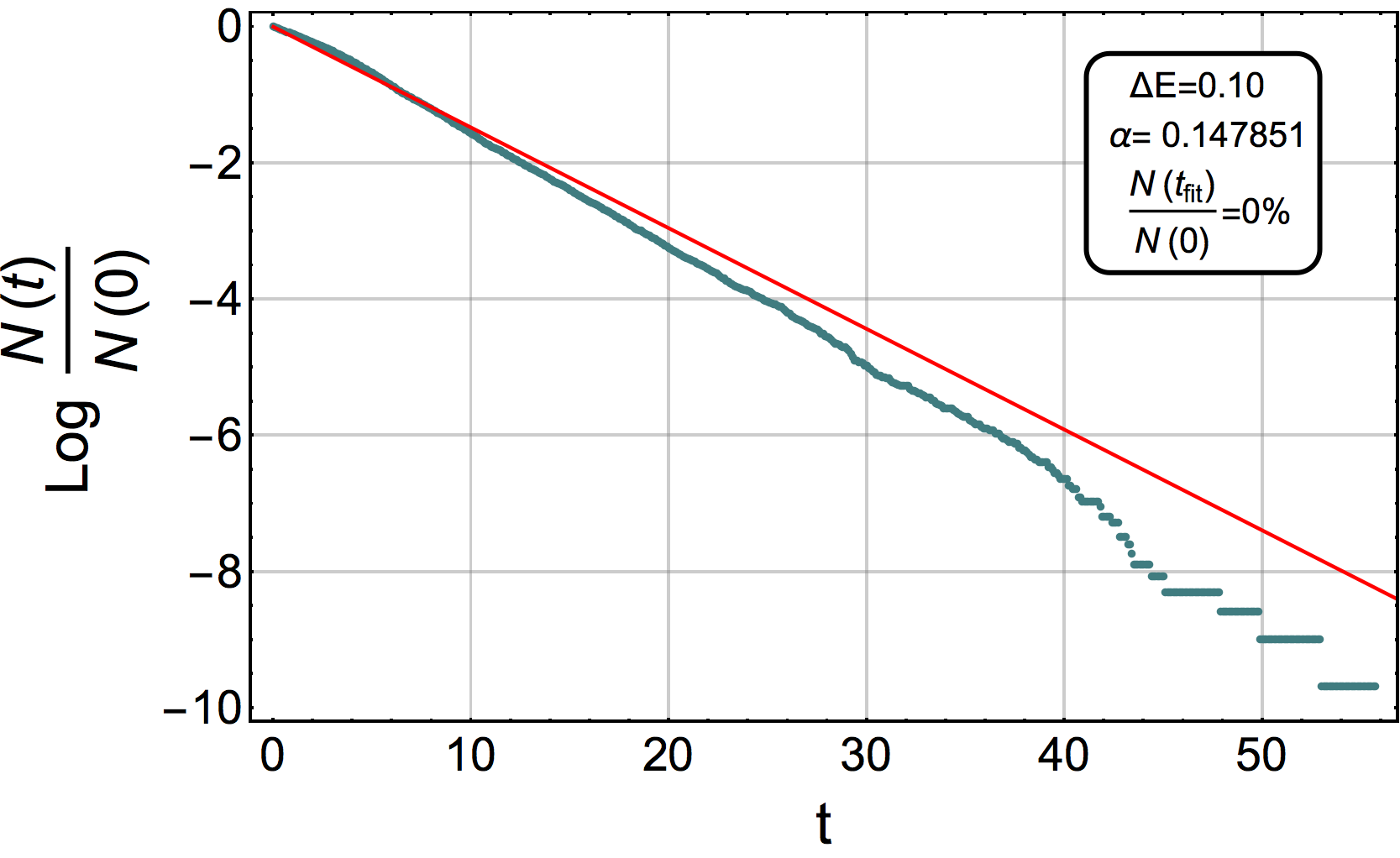}
\includegraphics[width=3.in]{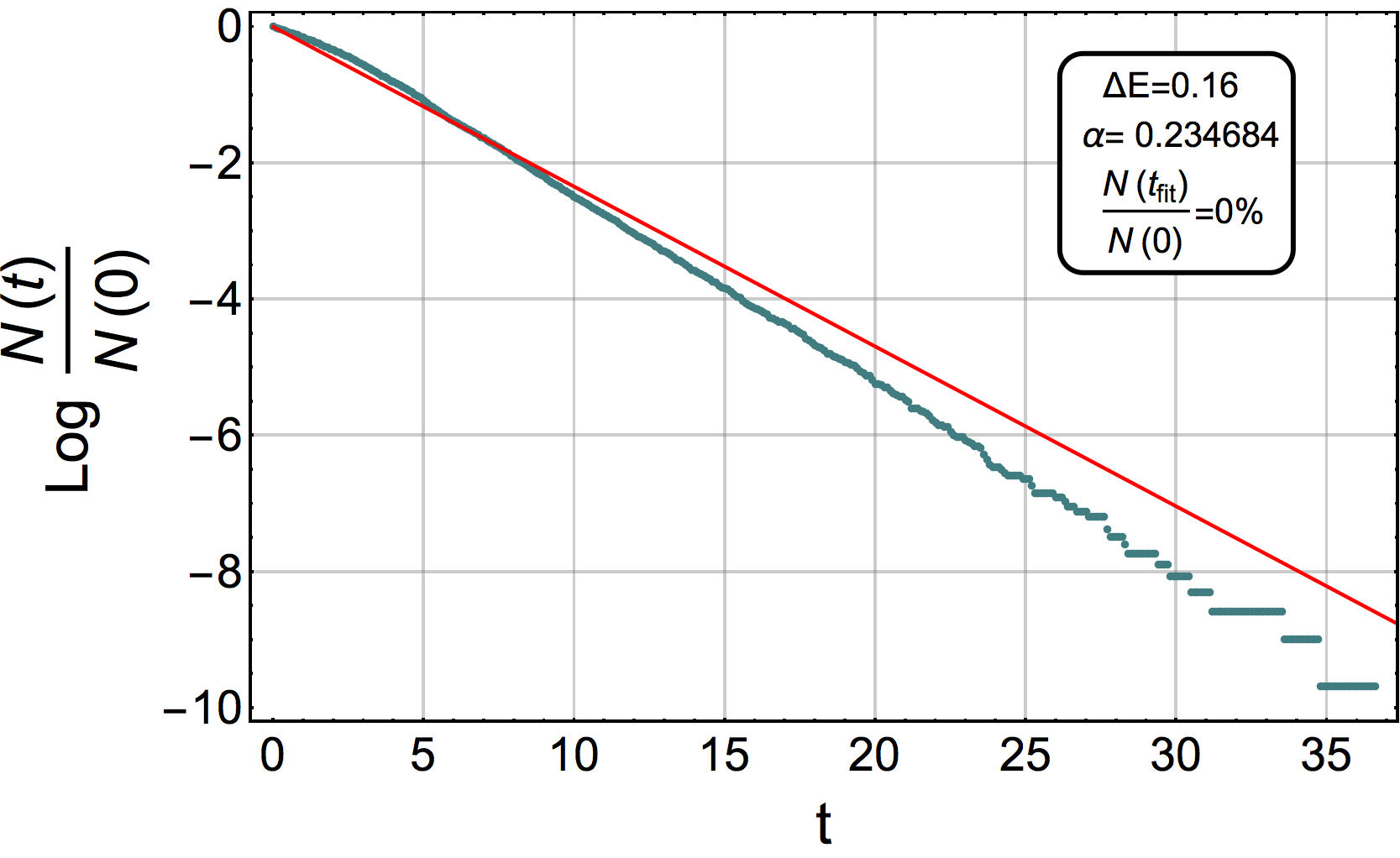}
\caption{{\footnotesize $\beta=0.8$. $\log N(t)/N(0)$ vs $t$ (blue curve) and their linear fit (red curve) for different $\Delta E$ (see insets). The $\alpha$ parameter shown in the inset is the slope absolute value of the fit. The fit was done in a time interval $t_{fit}$ such that after that time the remaining population is the one shown in the inset. }} \label{fig:fit08}
\end{figure}

\begin{figure}[hbtp]
\centering
\includegraphics[width=3.in]{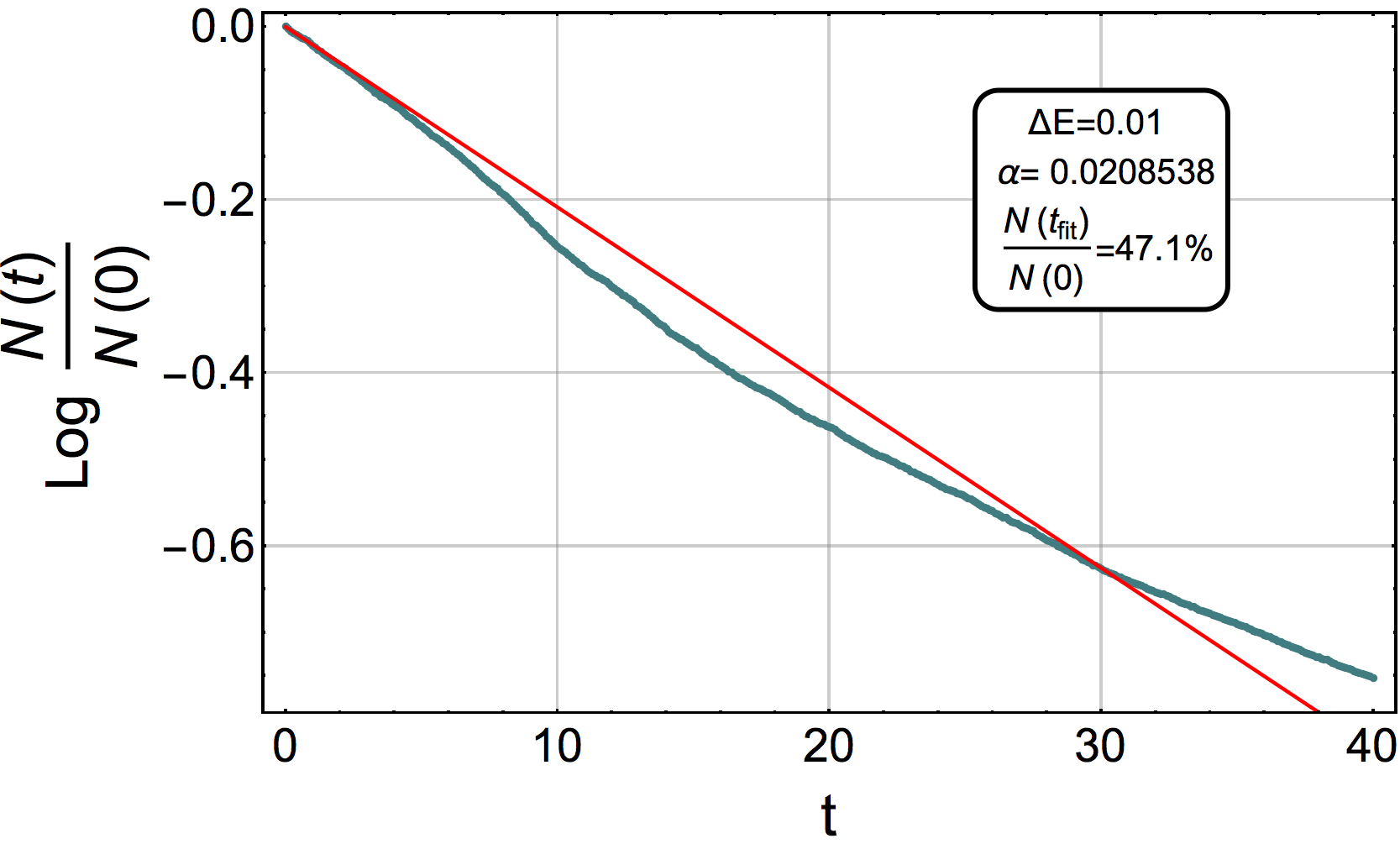}
\includegraphics[width=3.in]{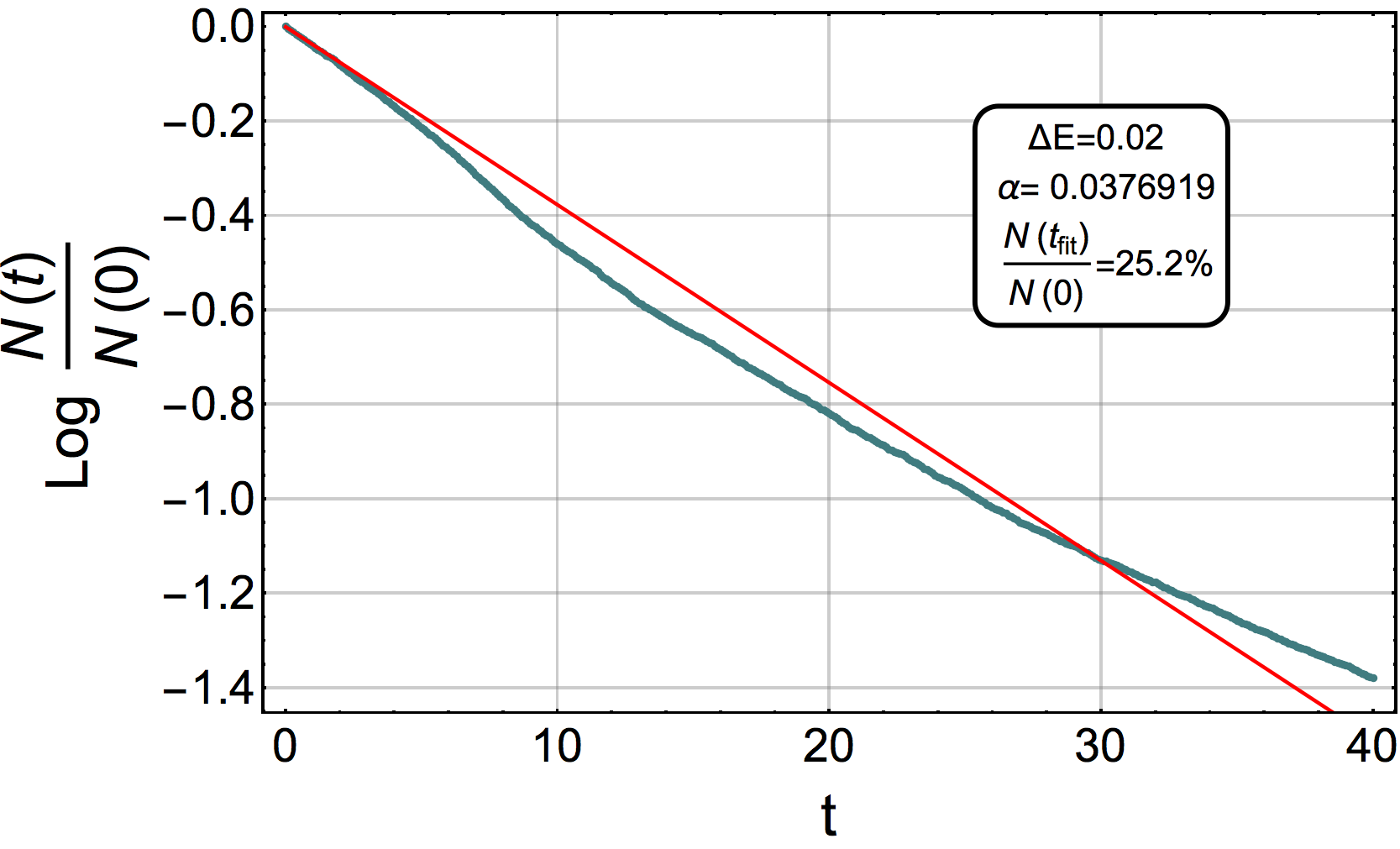}
\includegraphics[width=3.in]{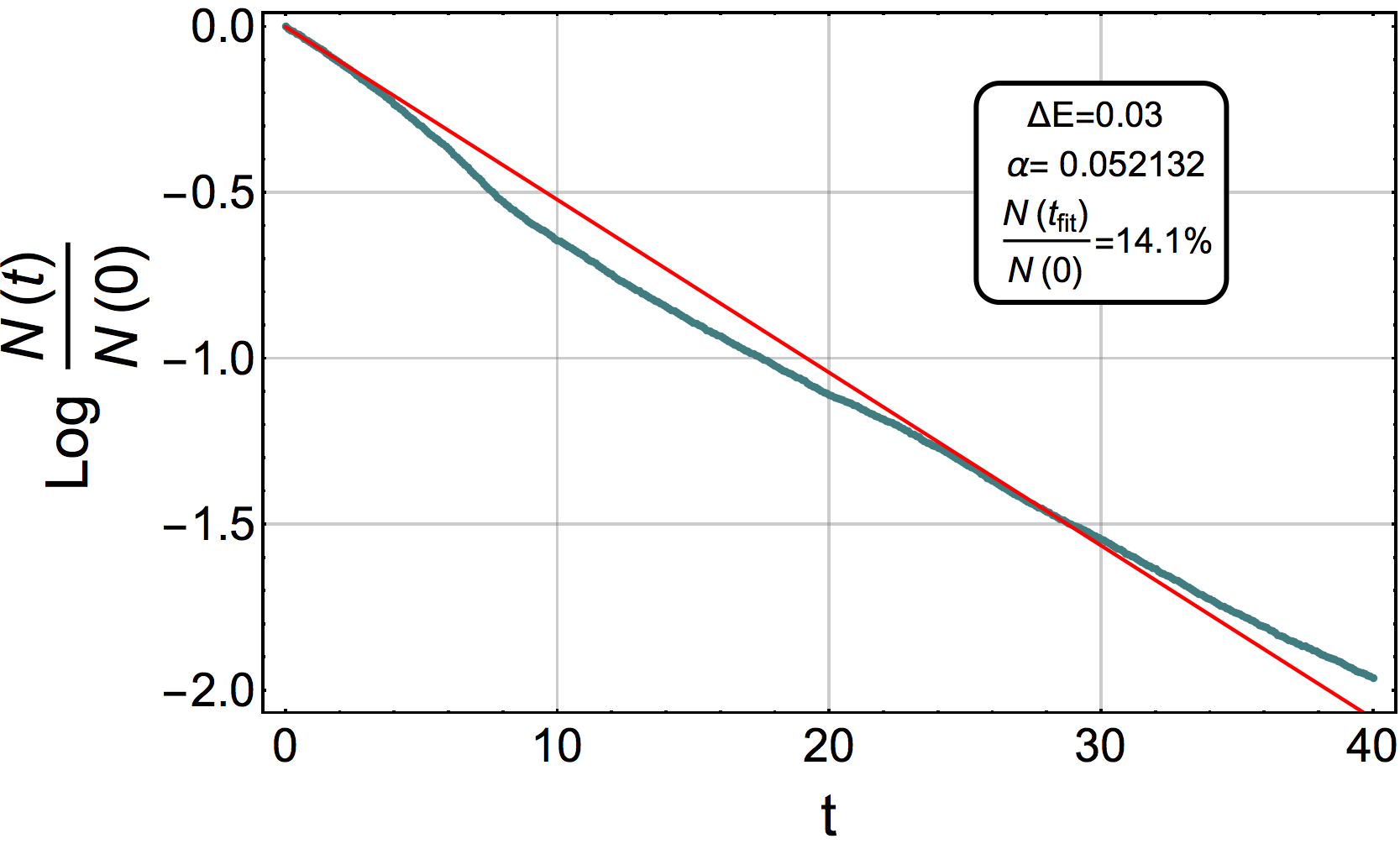}
\includegraphics[width=3.in]{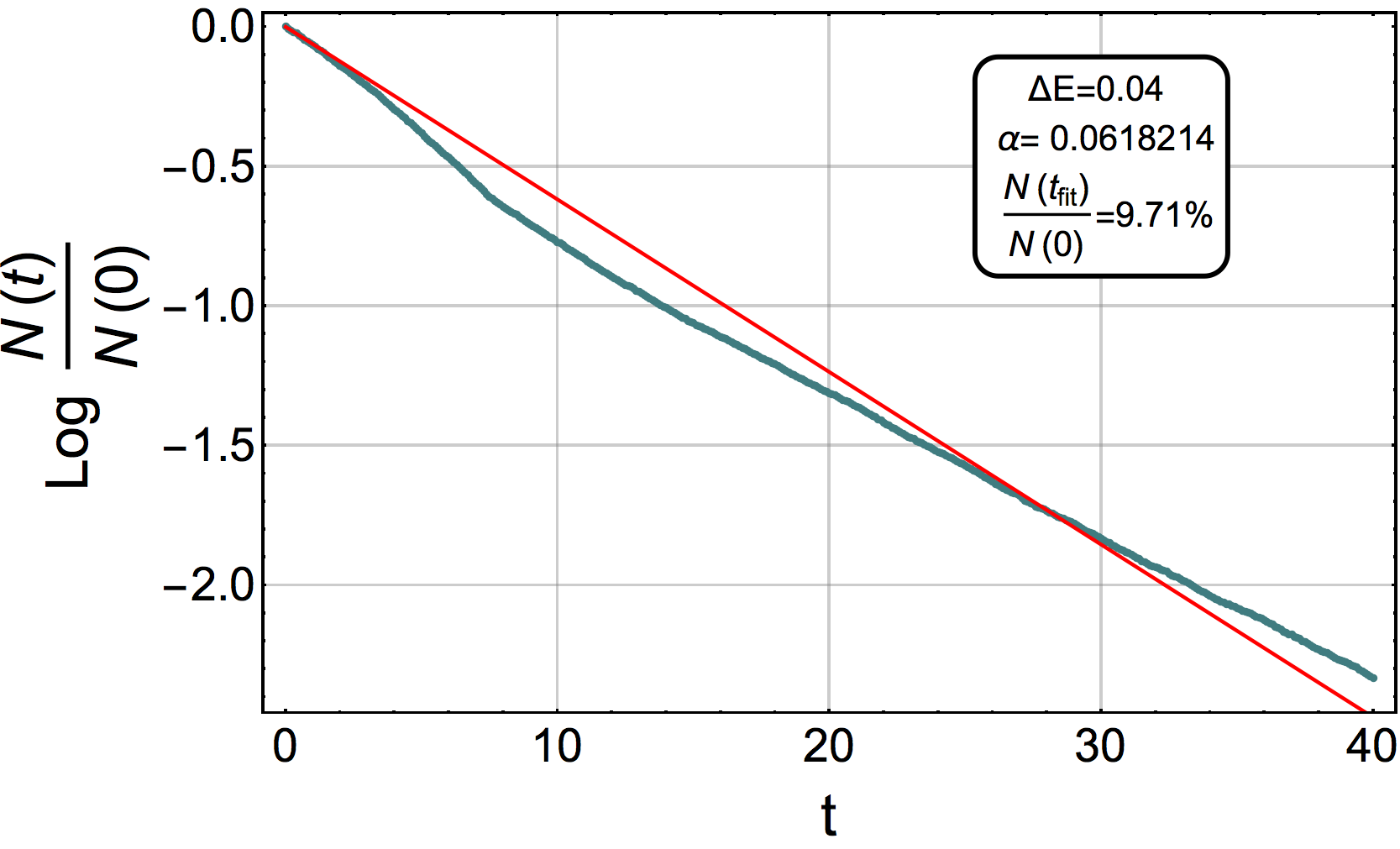}
\includegraphics[width=3.in]{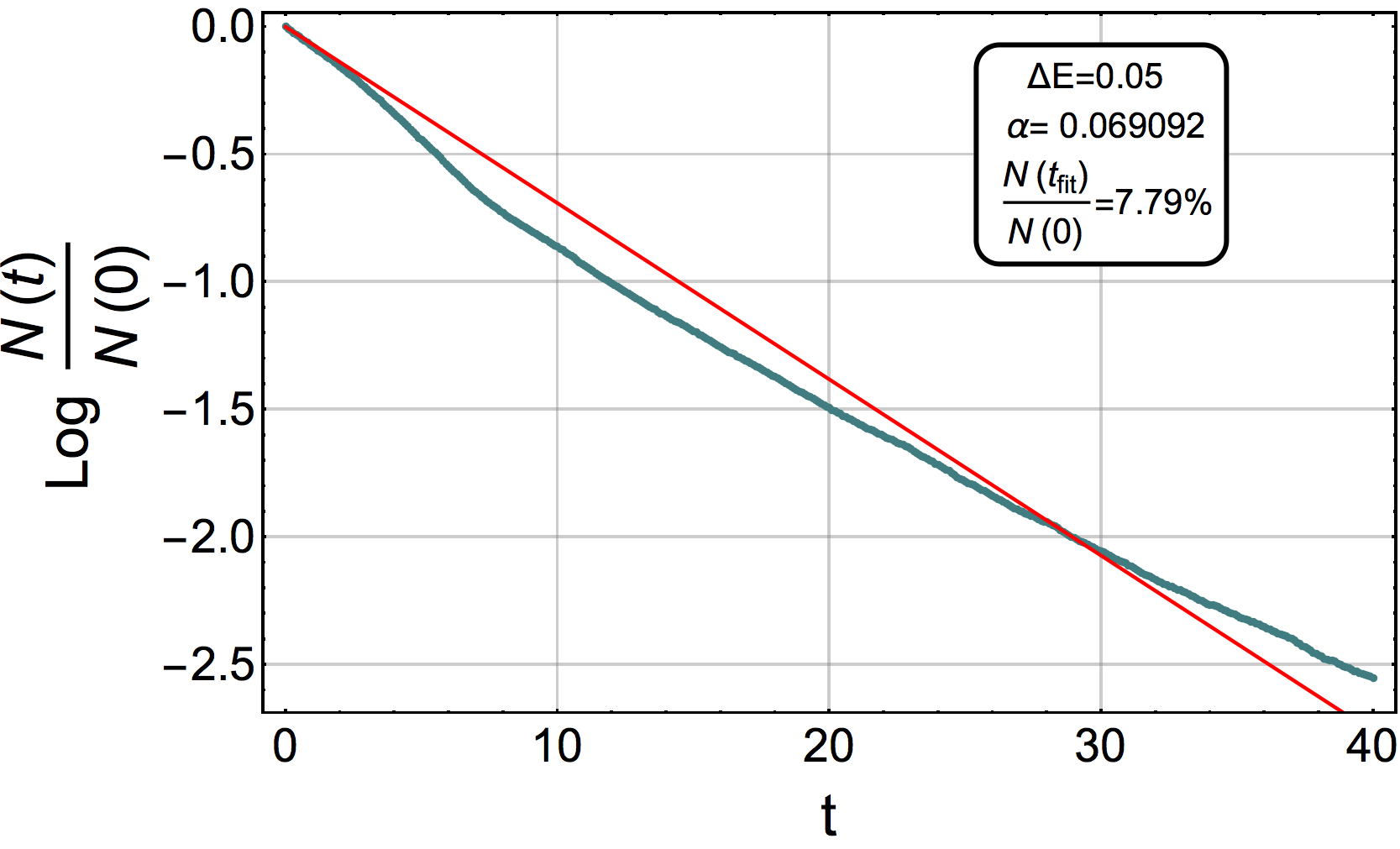}
\includegraphics[width=3.in]{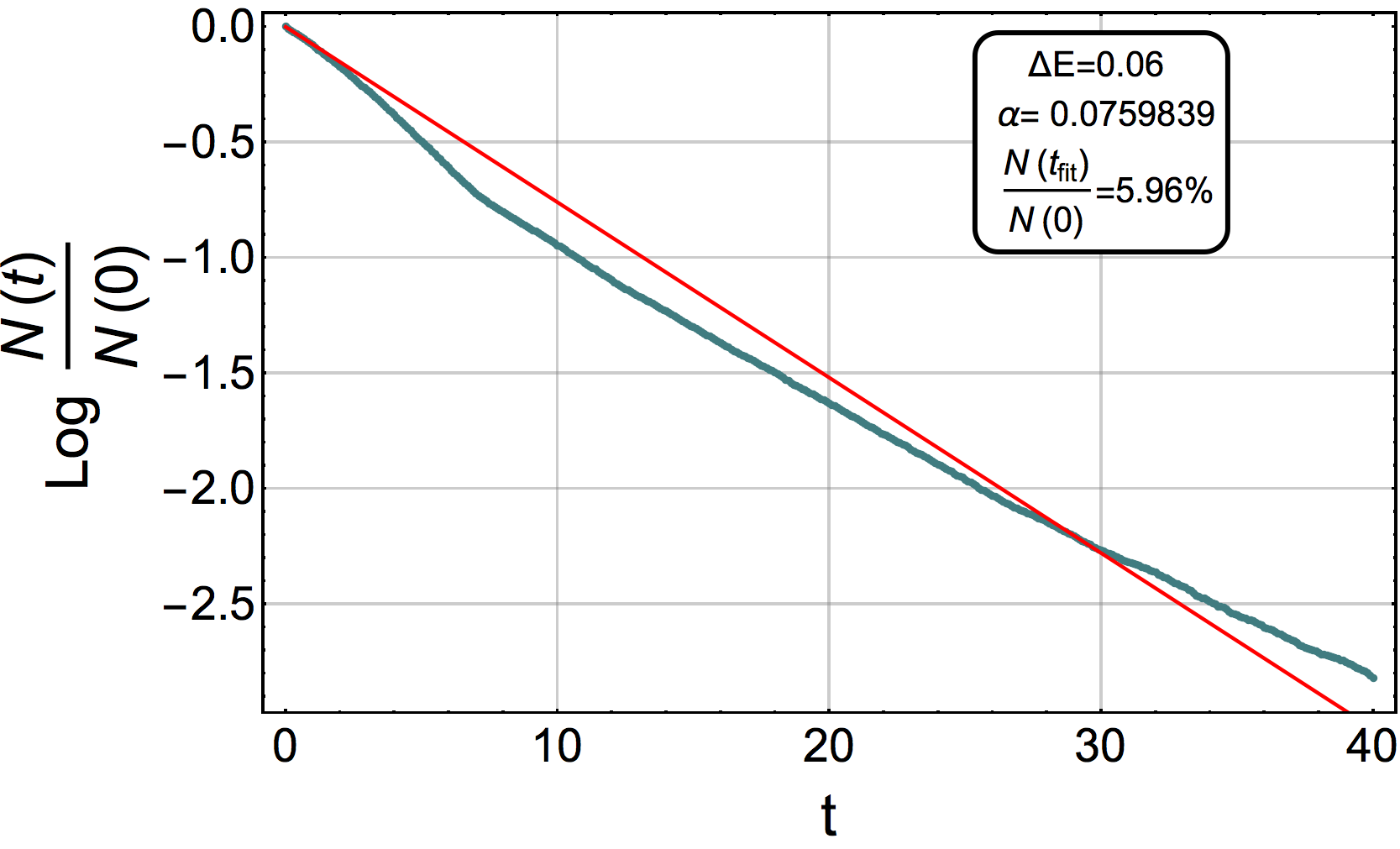}
\includegraphics[width=3.in]{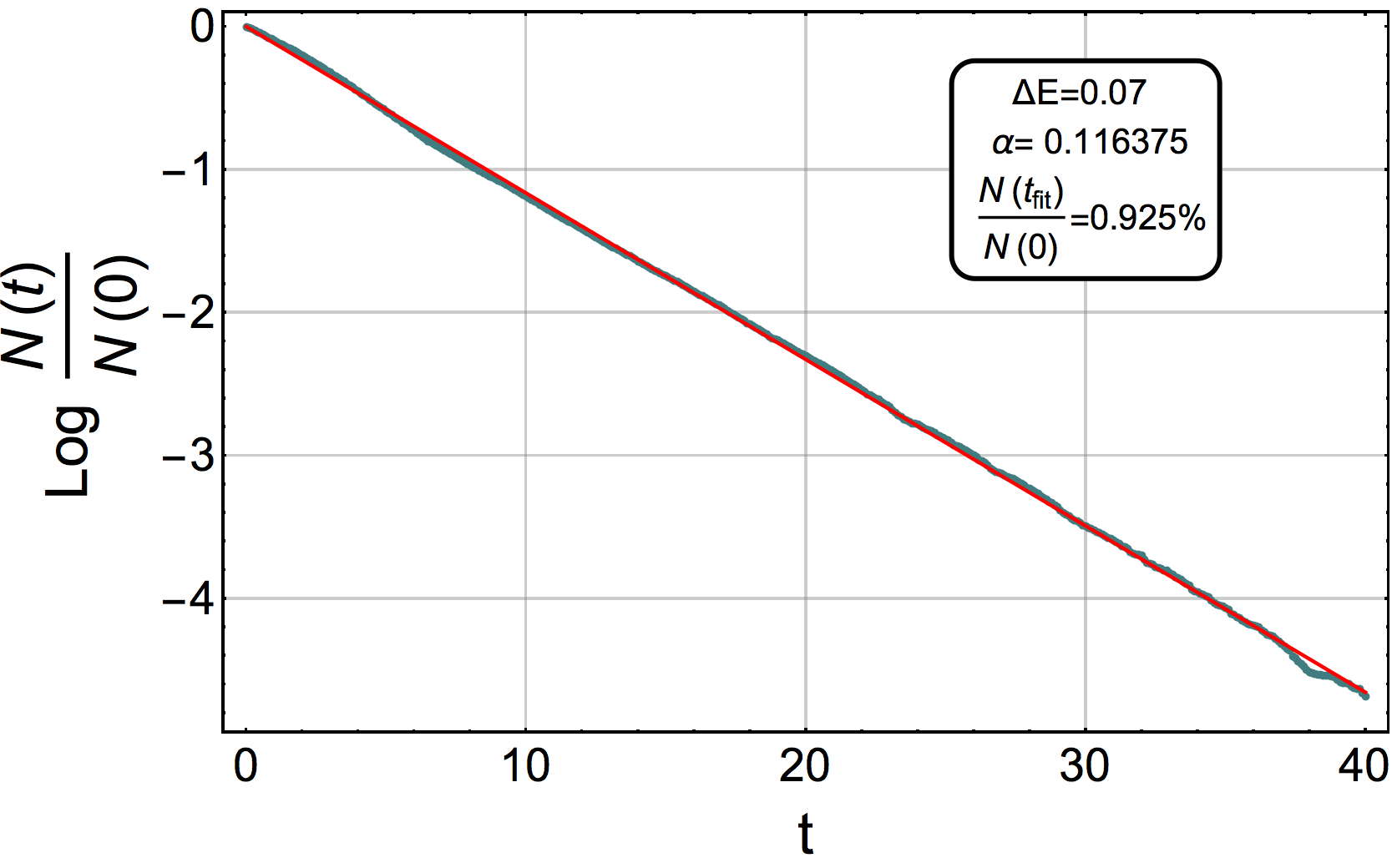}
\includegraphics[width=3.in]{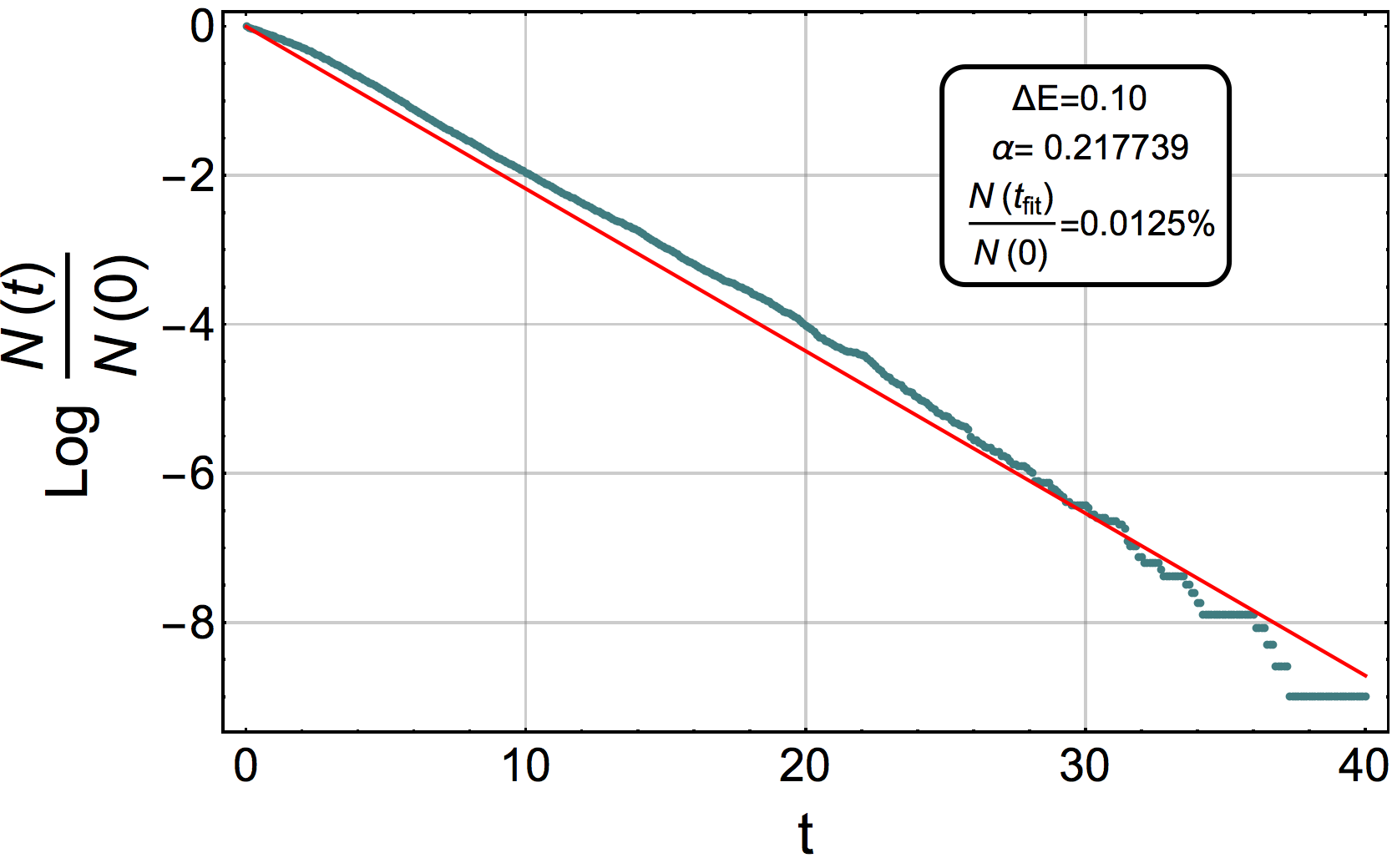}
\includegraphics[width=3.in]{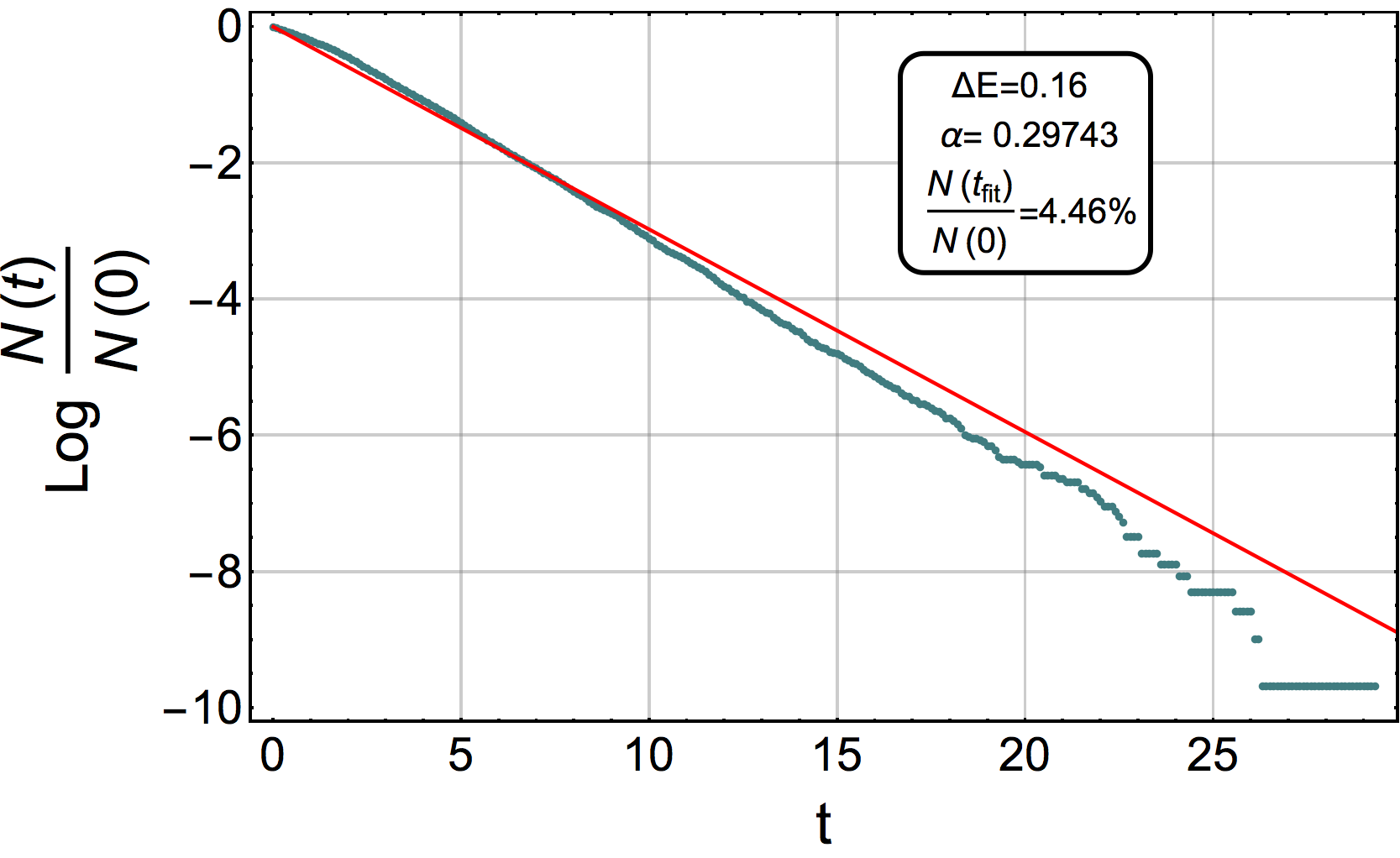}
\caption{{\footnotesize $\beta=0.6$. $\log N(t)/N(0)$ vs $t$ (blue curve) and their linear fit (red curve) for different $\Delta E$ (see insets). The $\alpha$ parameter shown in the inset is the slope absolute value of the fit. The fit was done in a time interval $t_{fit}$ such that after that time the remaining population is the one shown in the inset. }} \label{fig:fit06}
\end{figure}

\begin{figure}[hbtp]
\centering
\includegraphics[width=3.in]{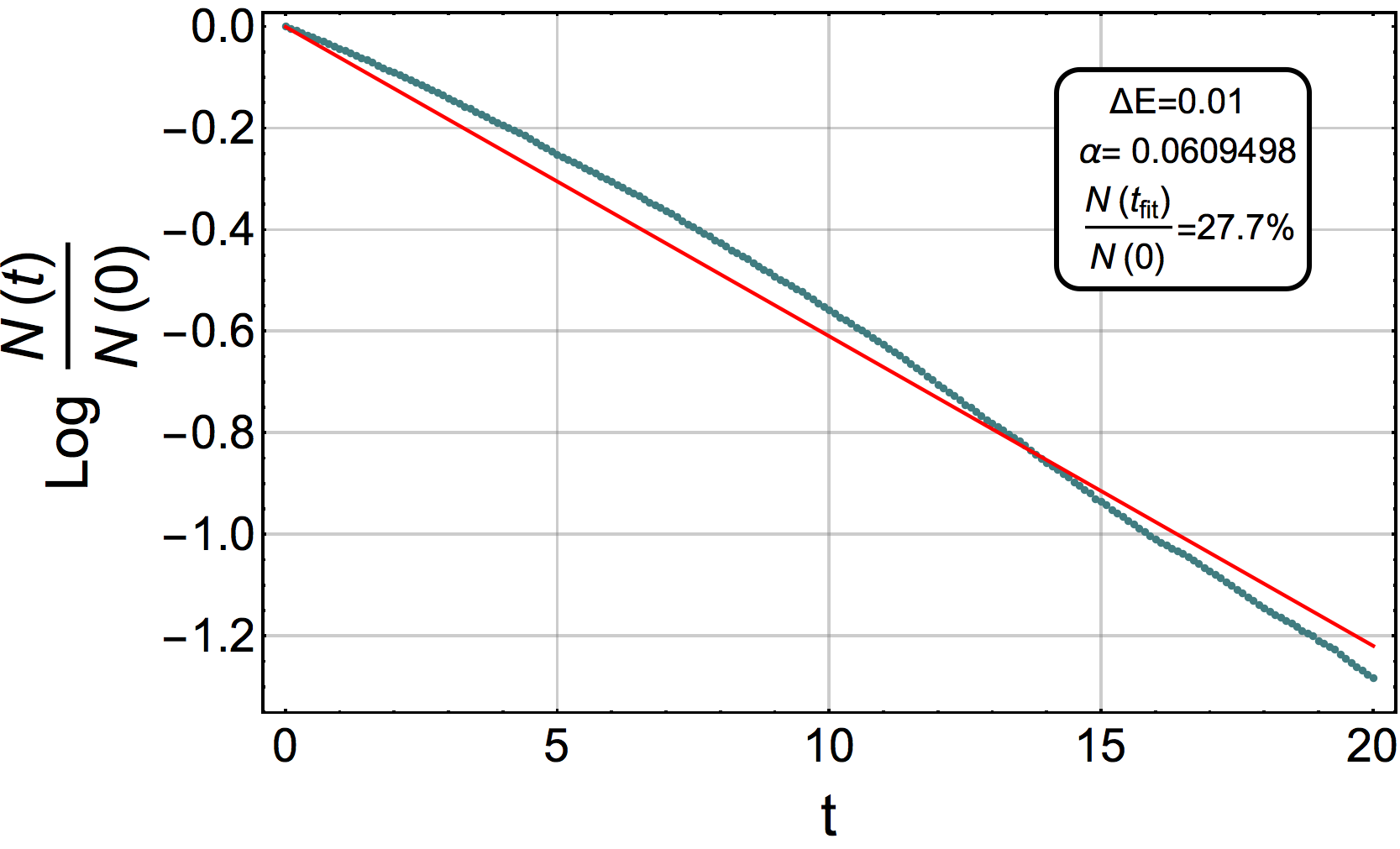}
\includegraphics[width=3.in]{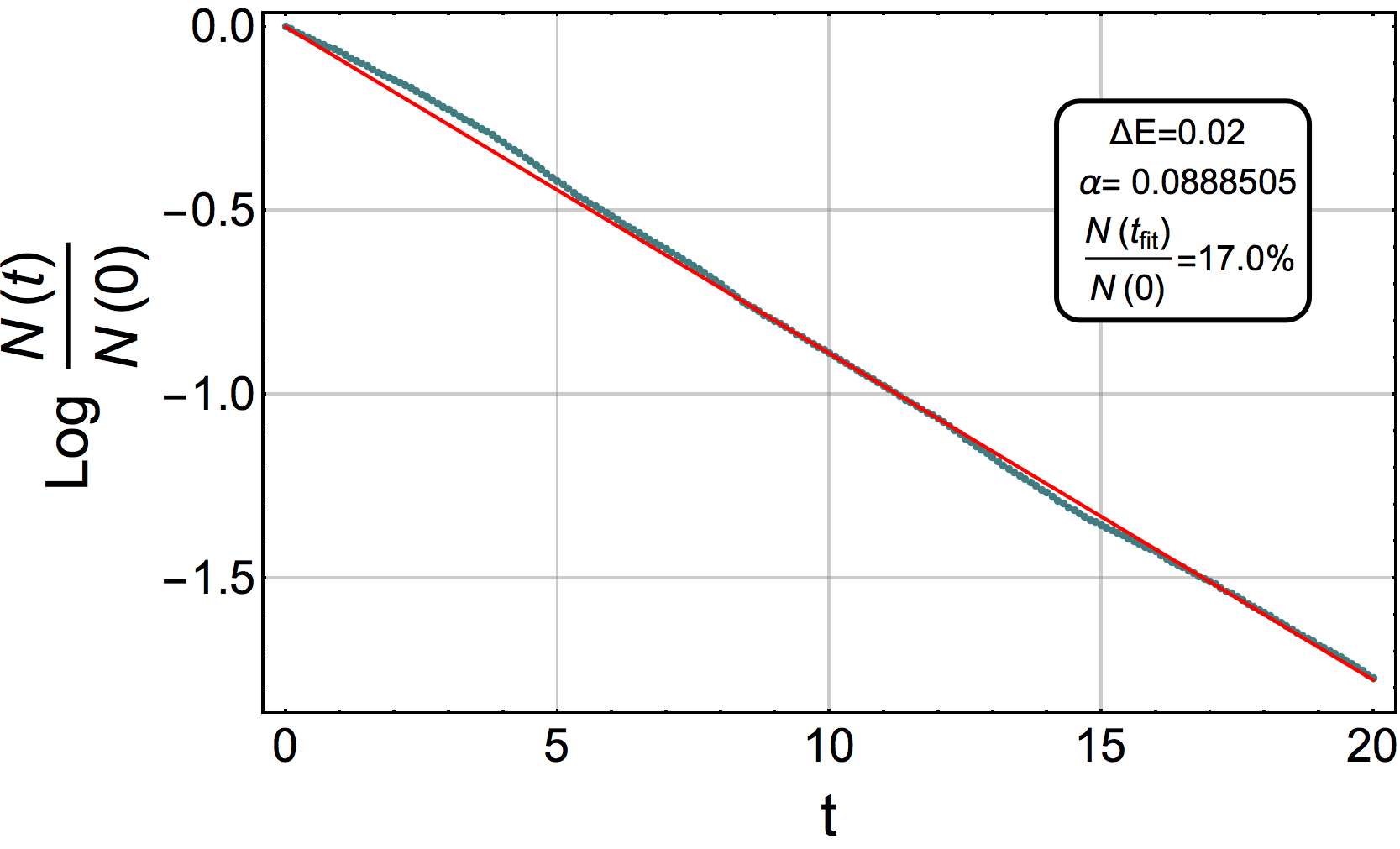}
\includegraphics[width=3.in]{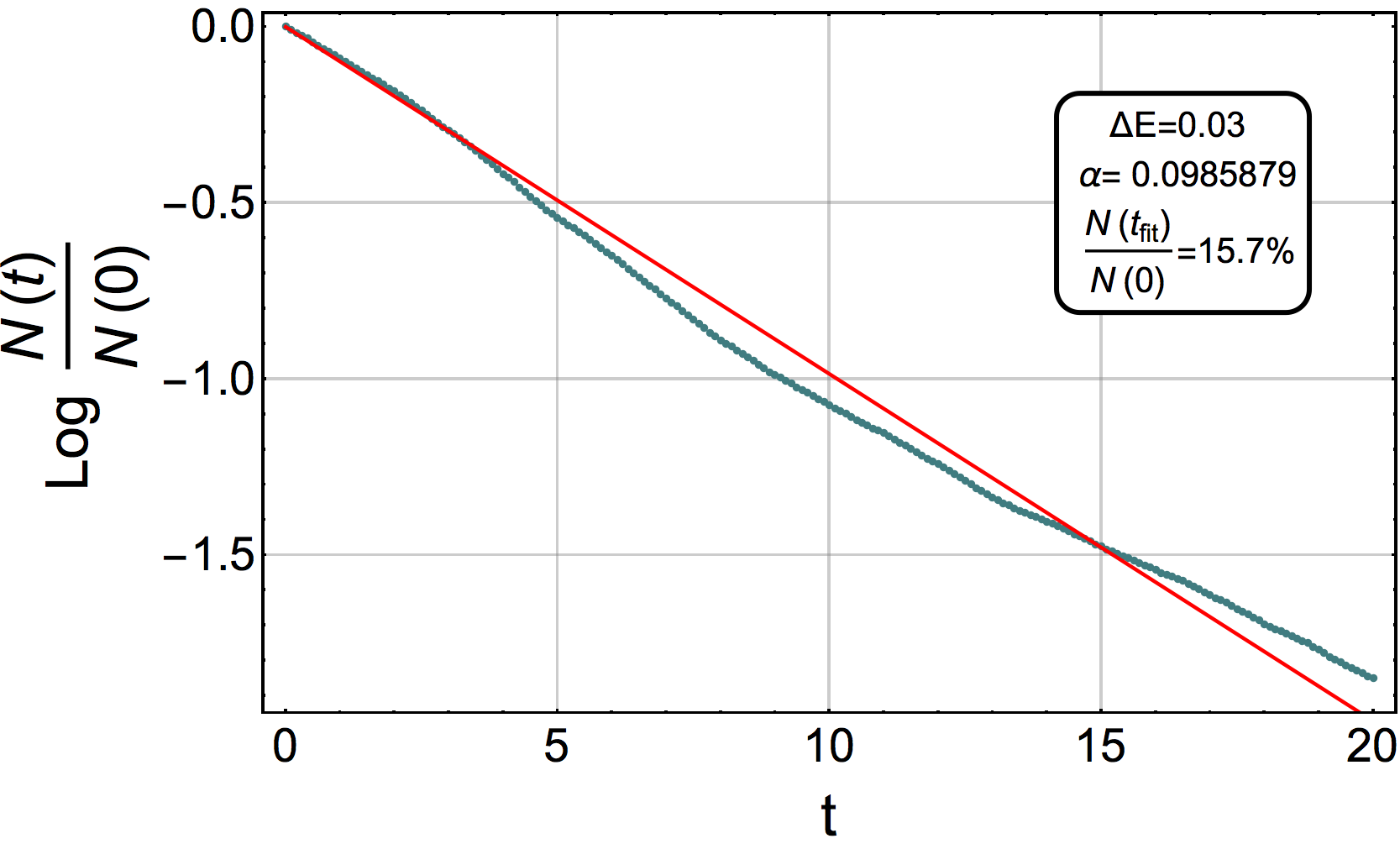}
\includegraphics[width=3.in]{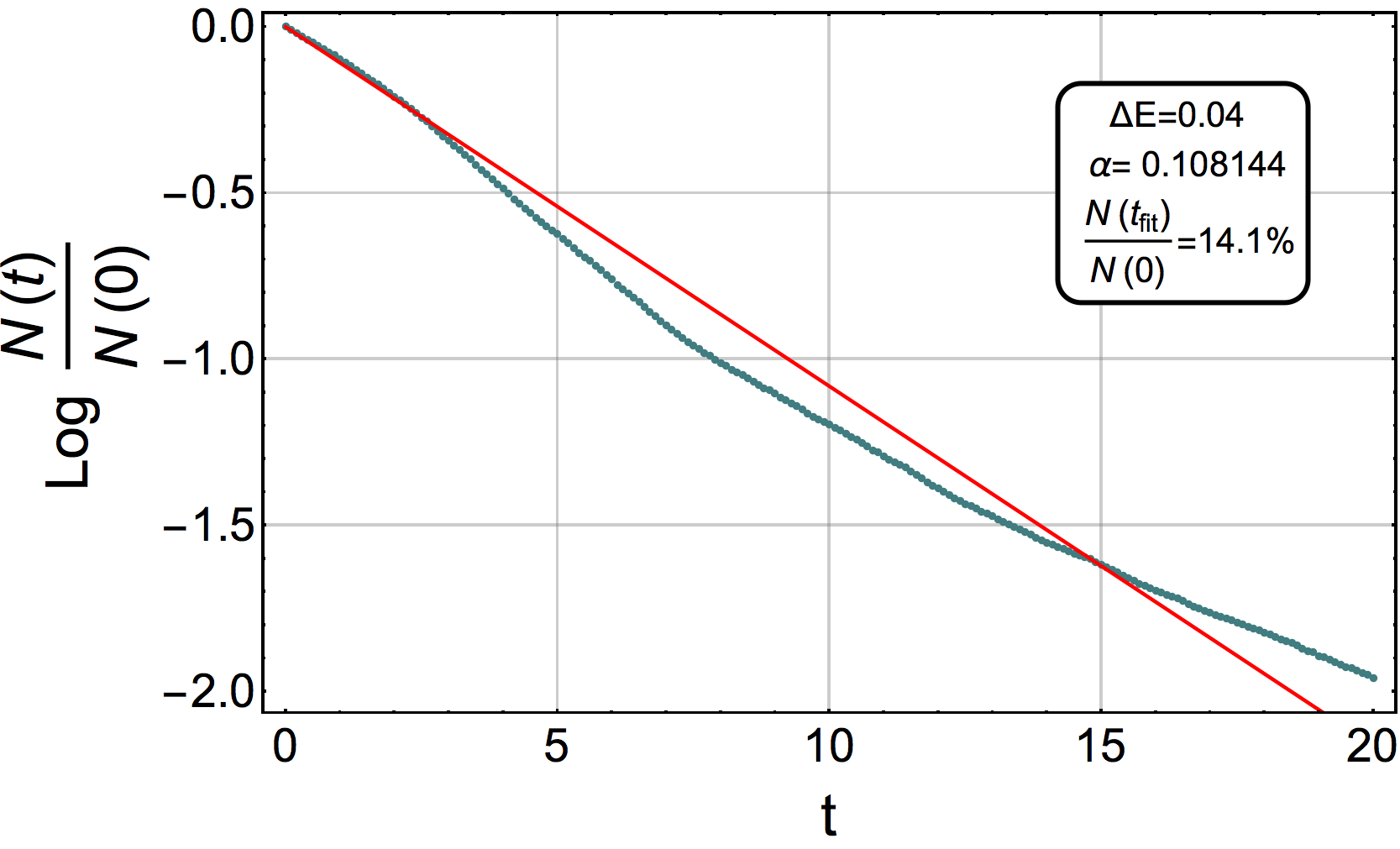}
\includegraphics[width=3.in]{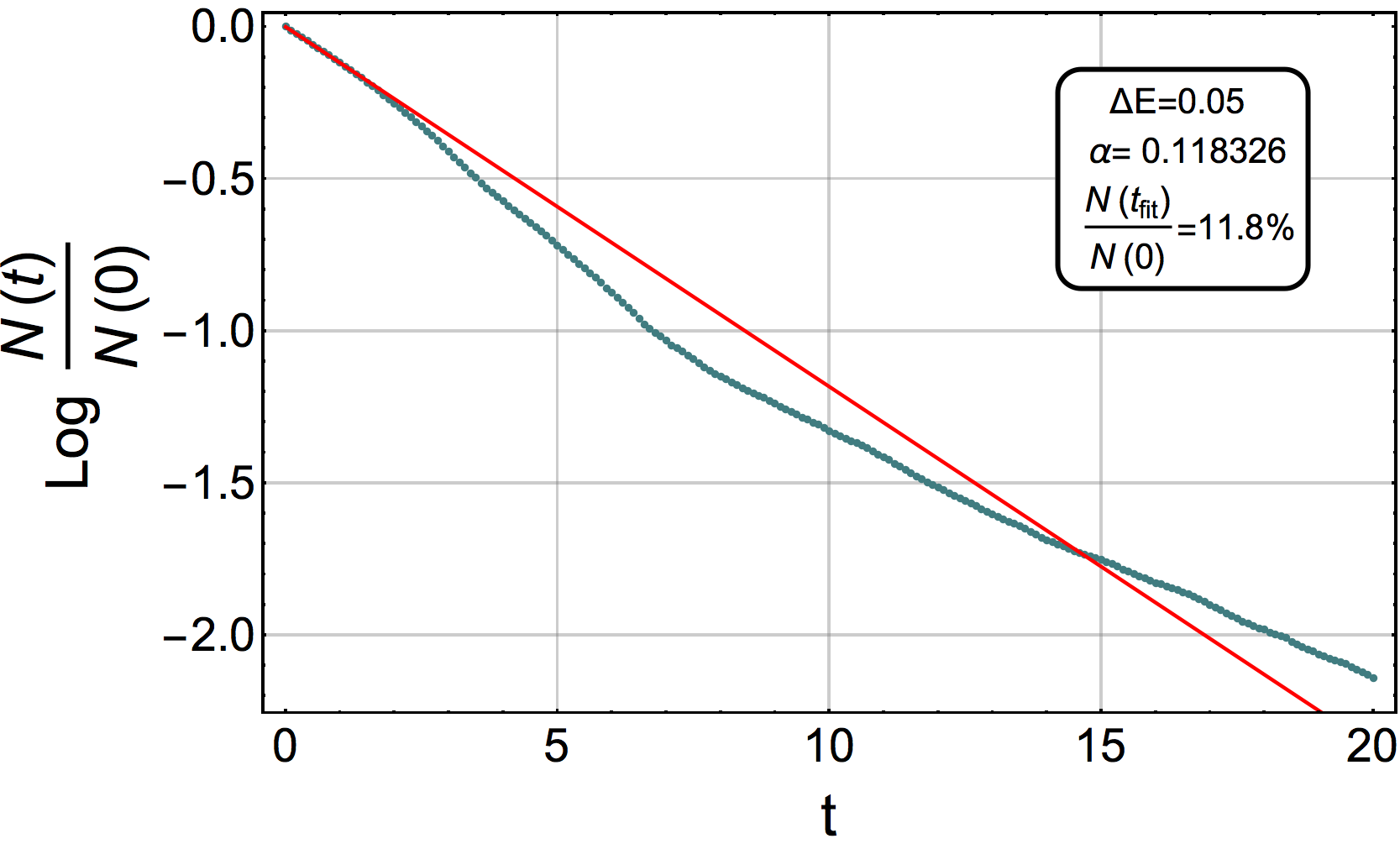}
\includegraphics[width=3.in]{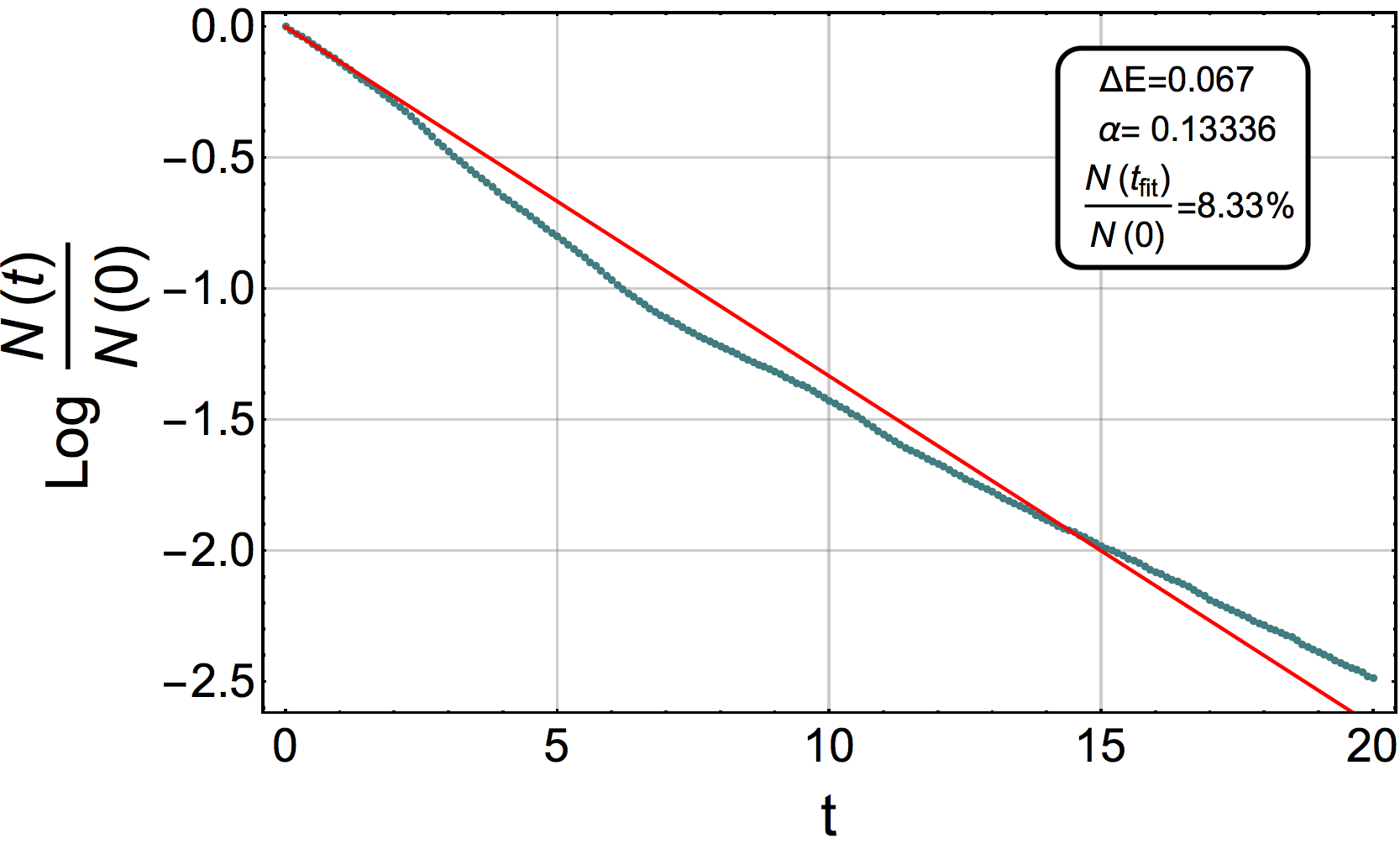}
\includegraphics[width=3.in]{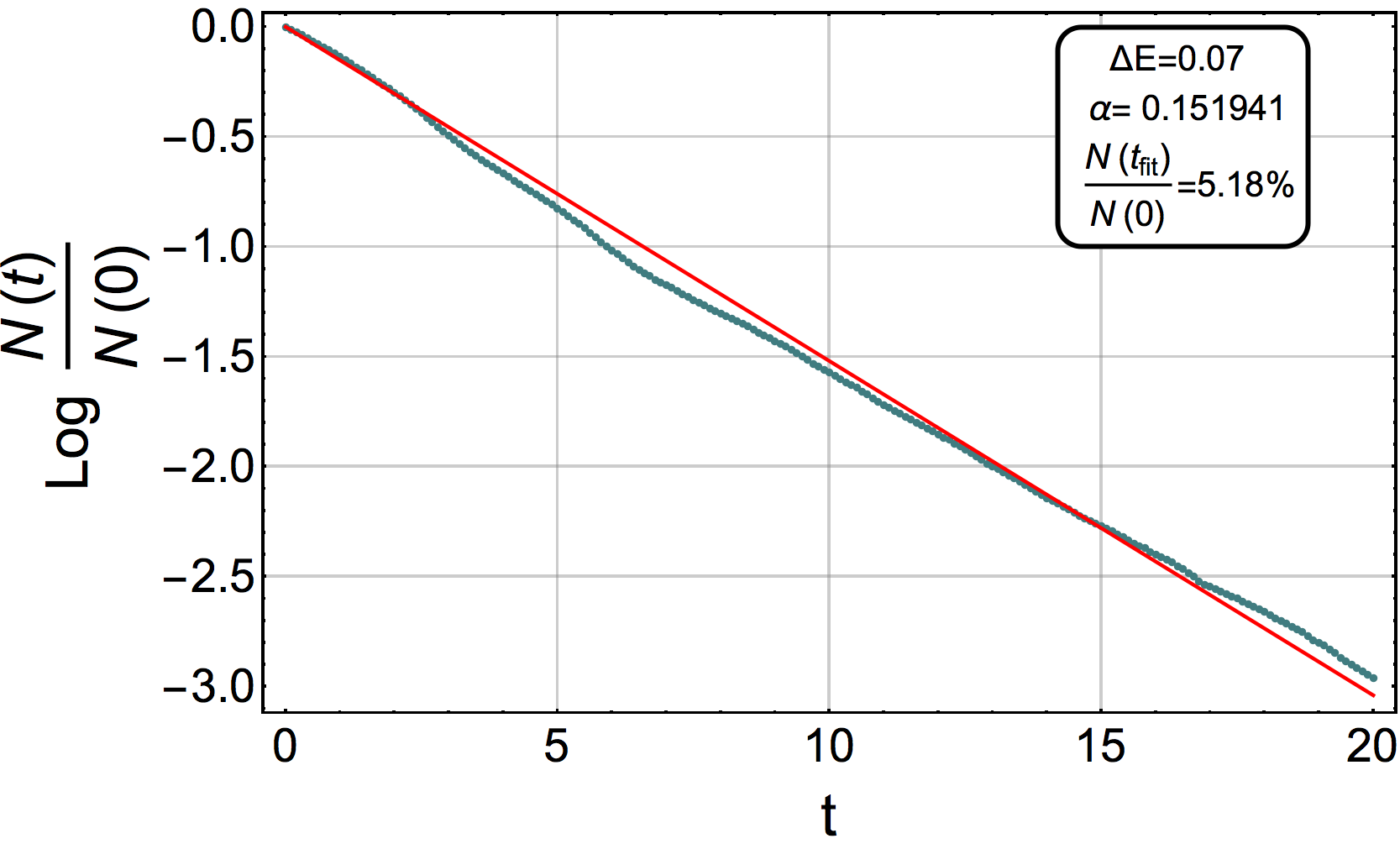}
\includegraphics[width=3.in]{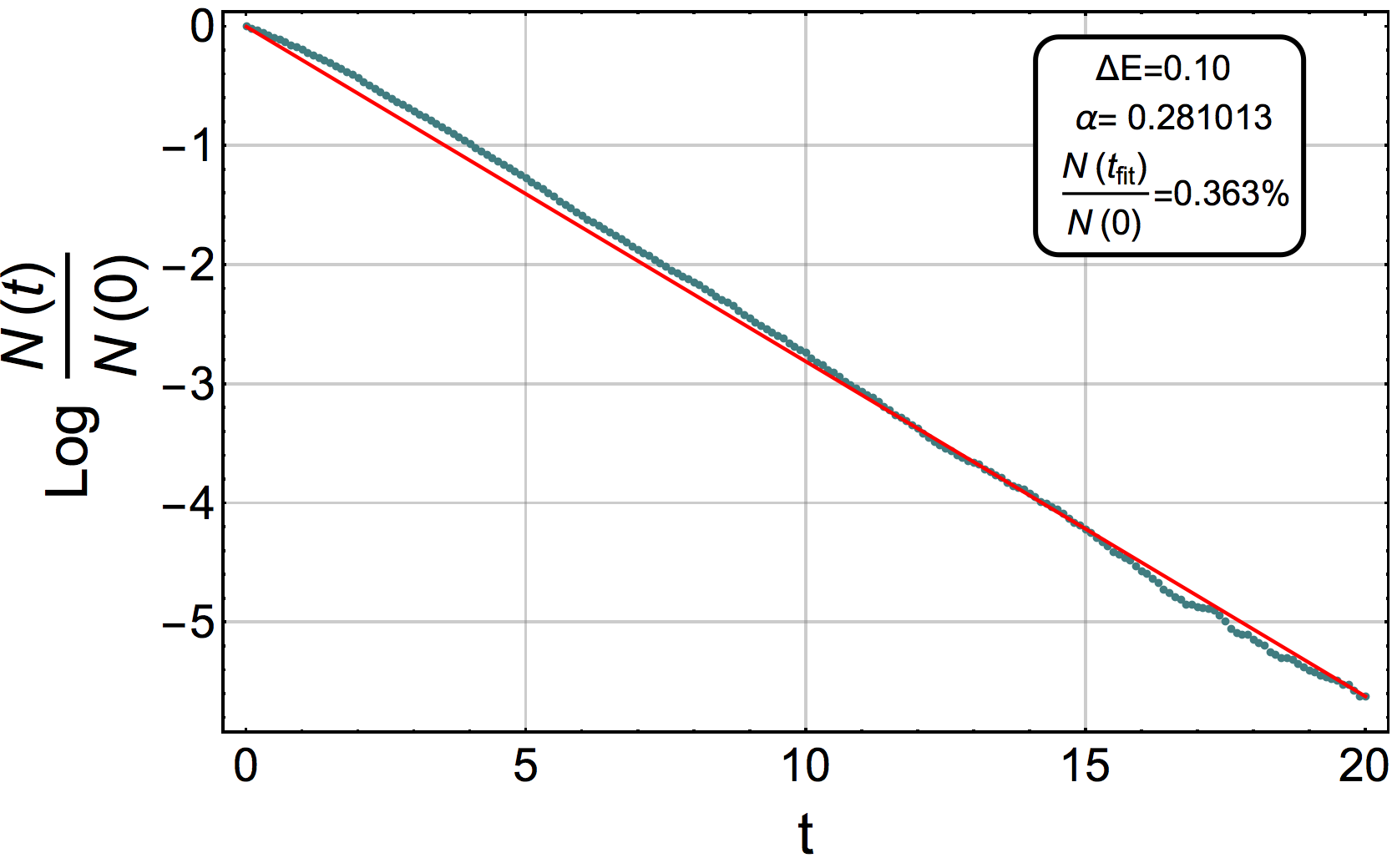}
\includegraphics[width=3.in]{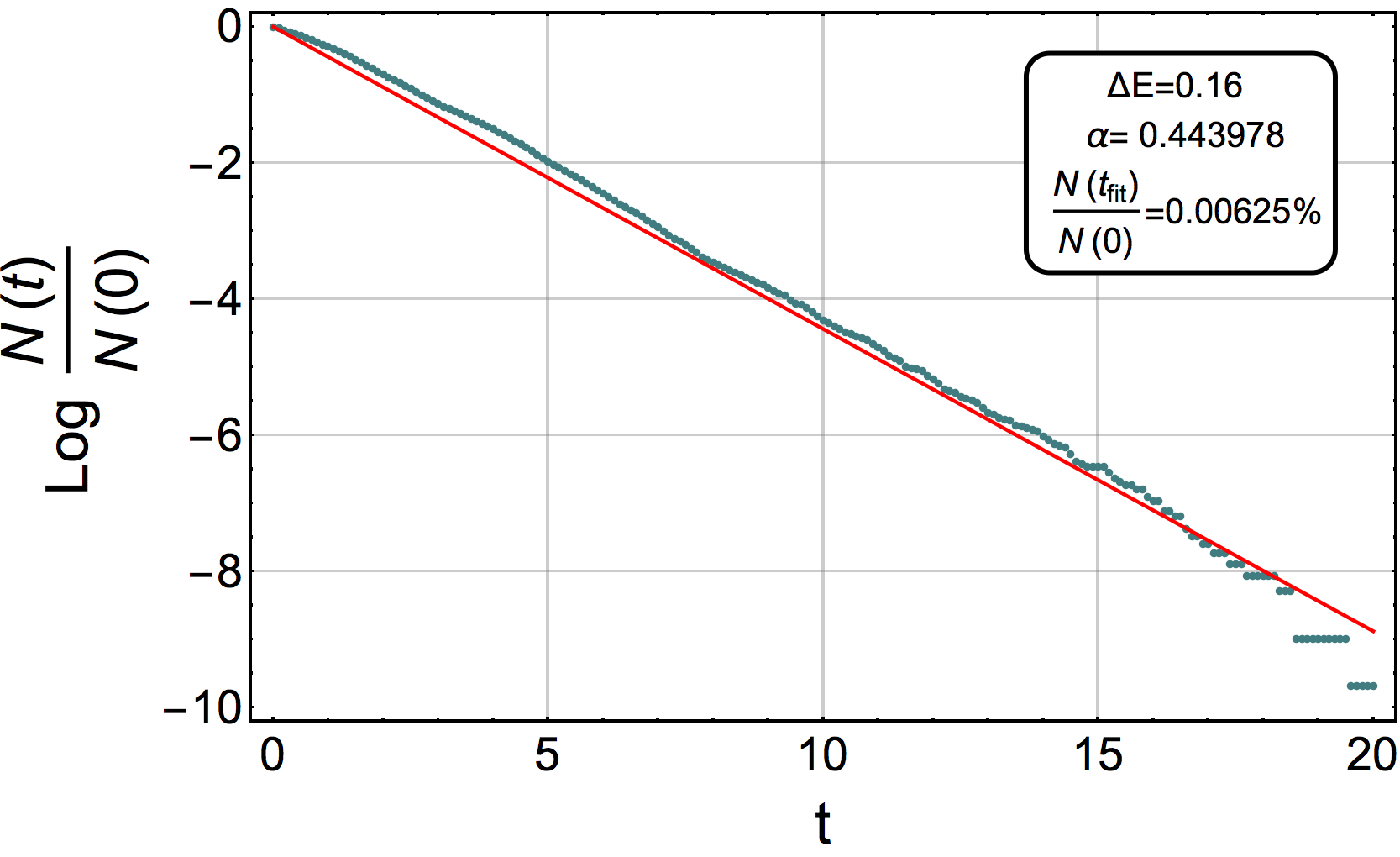}
\caption{{\footnotesize $\beta=0.4$. $\log N(t)/N(0)$ vs $t$ (blue curve) and their linear fit (red curve) for different $\Delta E$ (see insets). The $\alpha$ parameter shown in the inset is the slope absolute value of the fit. The fit was done in a time interval $t_{fit}$ such that after that time the remaining population is the one shown in the inset. }} \label{fig:fit04}
\end{figure}

\begin{figure}[hbtp]
\centering
\includegraphics[width=3.in]{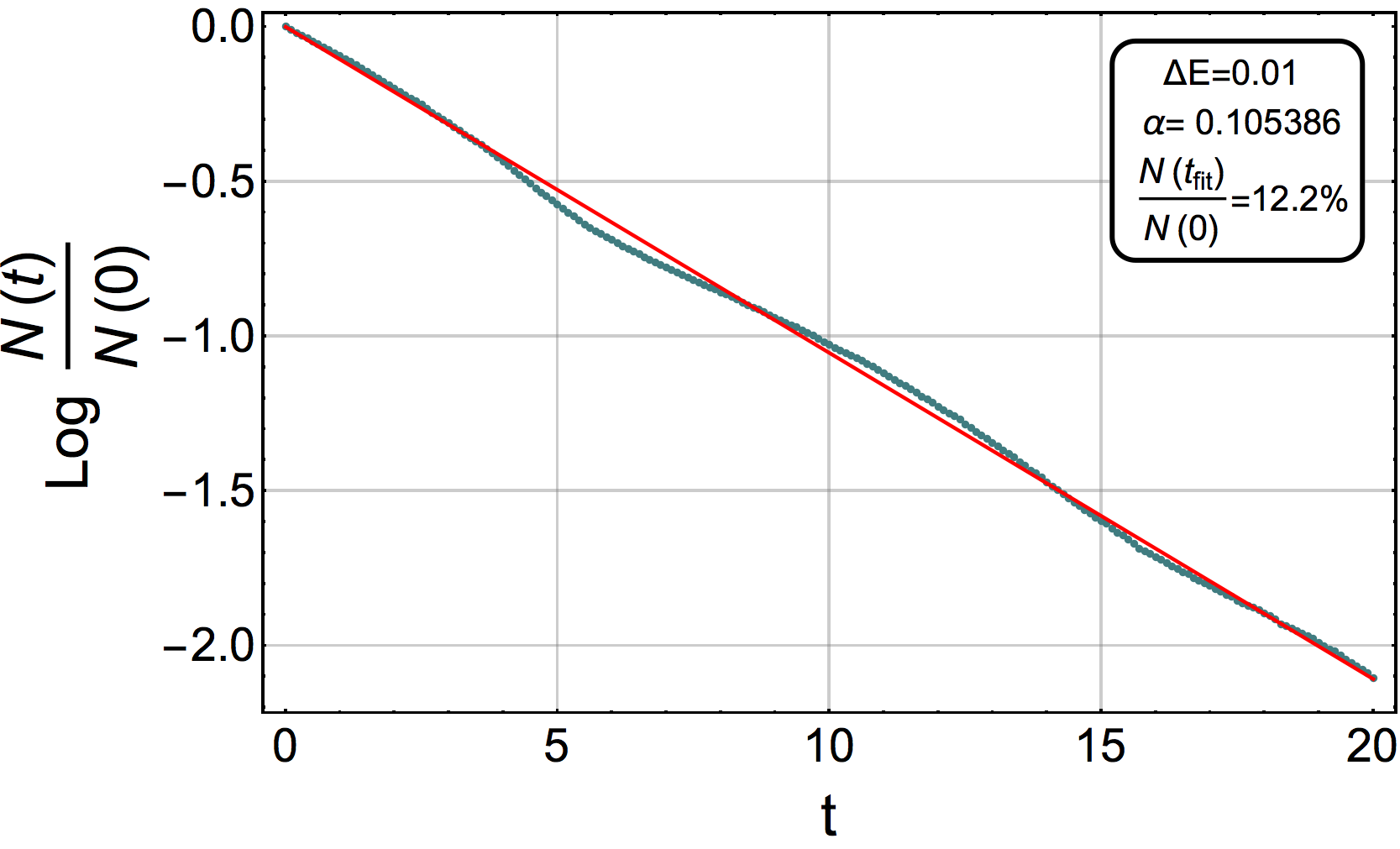}
\includegraphics[width=3.in]{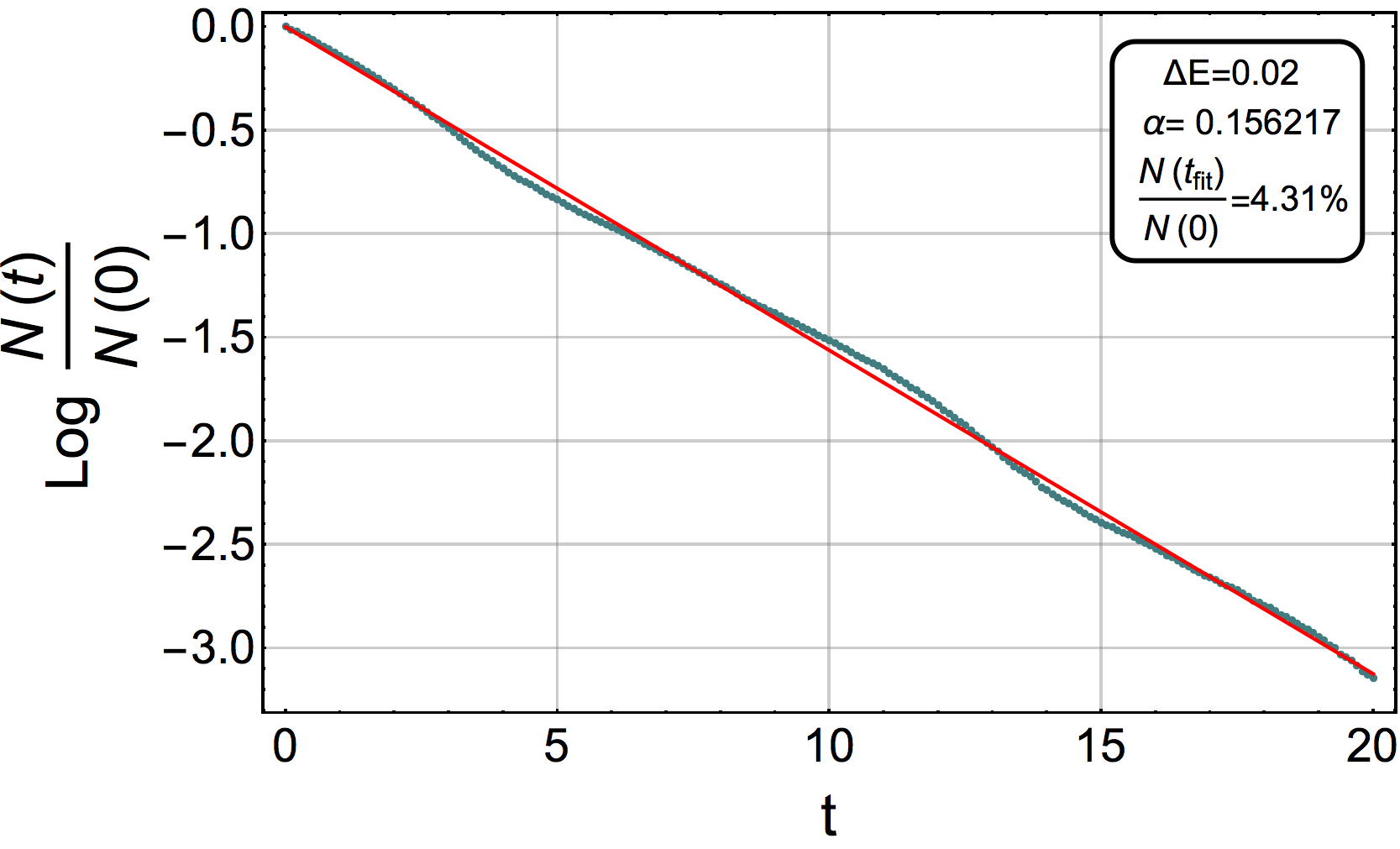}
\includegraphics[width=3.in]{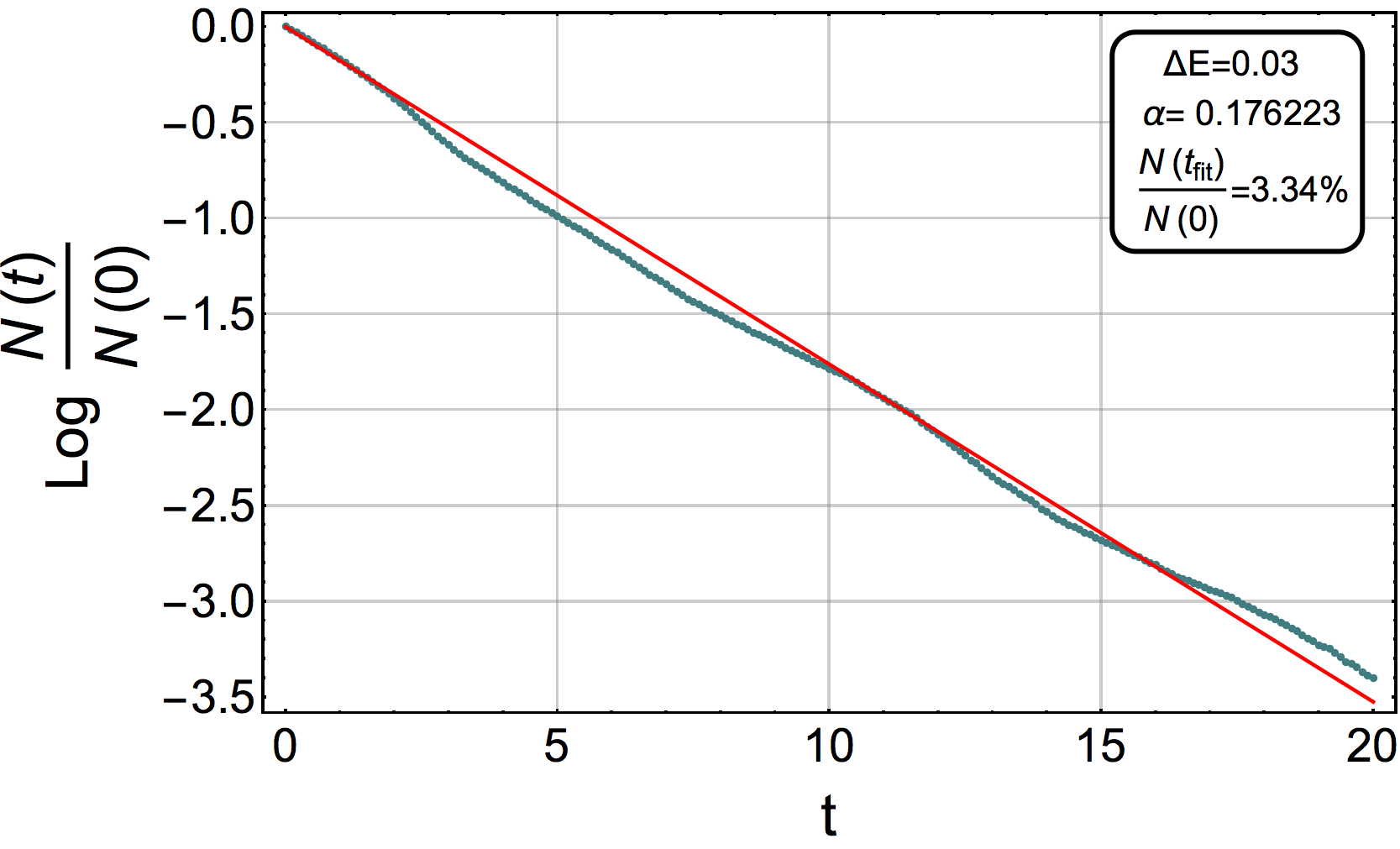}
\includegraphics[width=3.in]{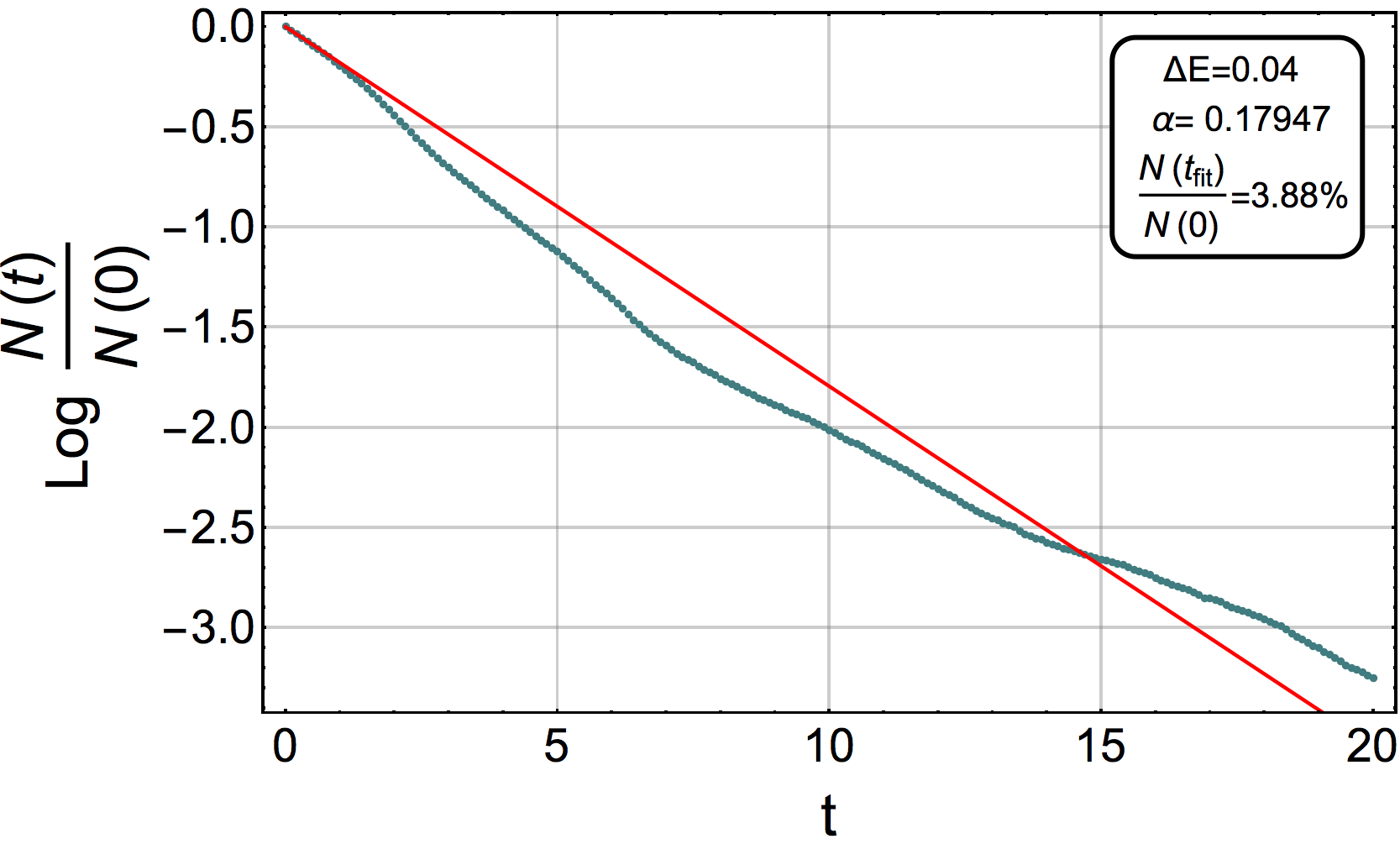}
\includegraphics[width=3.in]{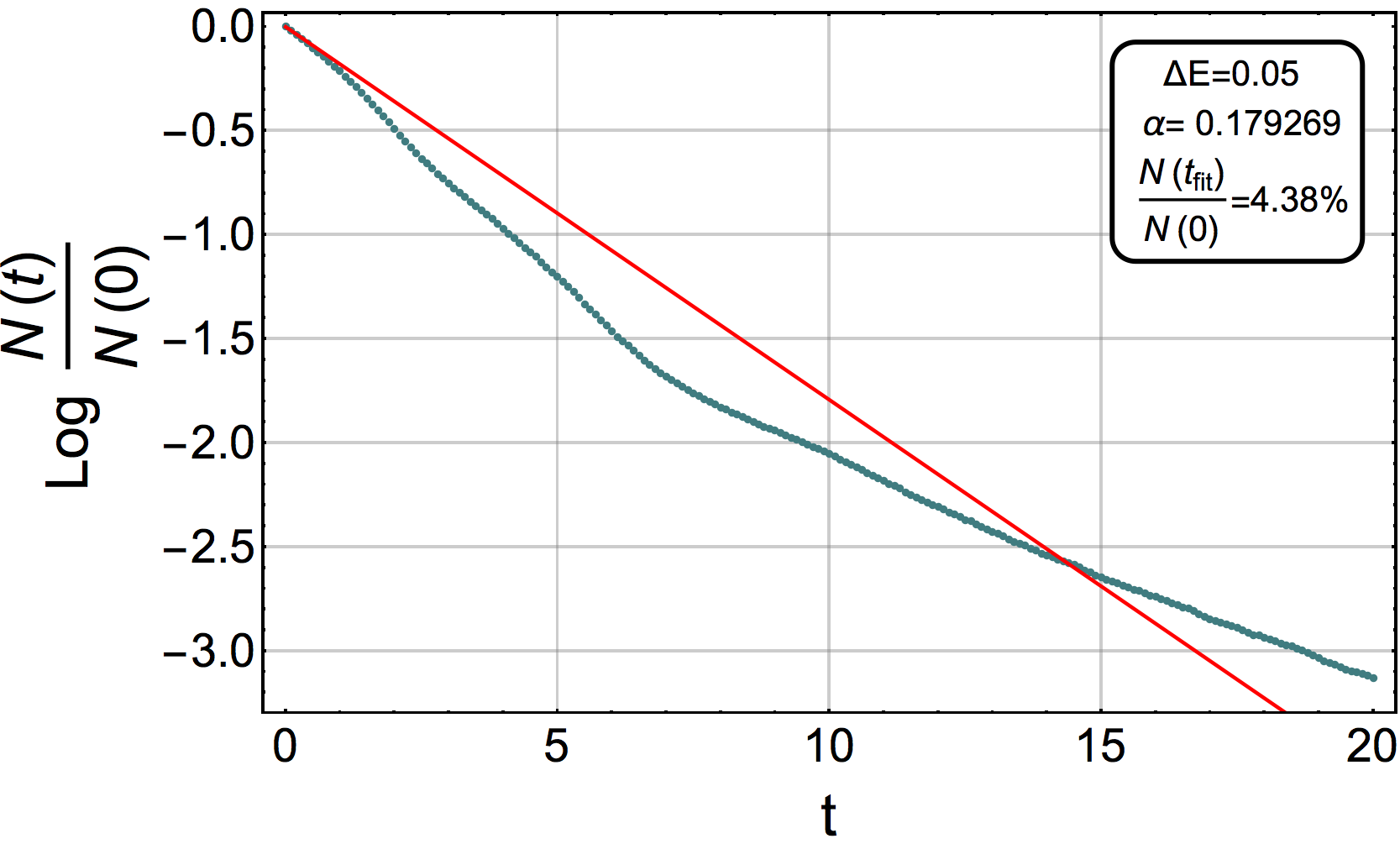}
\includegraphics[width=3.in]{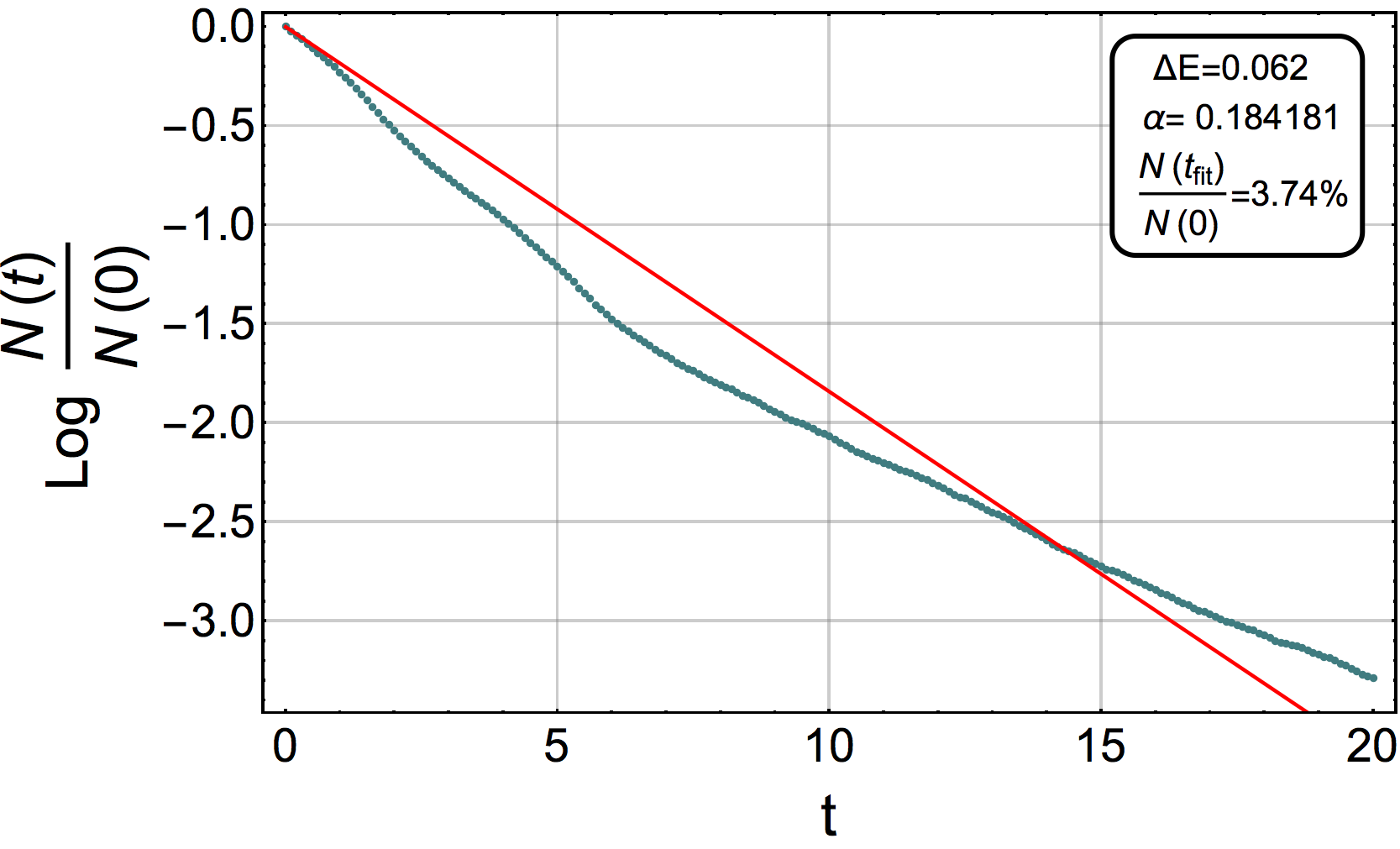}
\includegraphics[width=3.in]{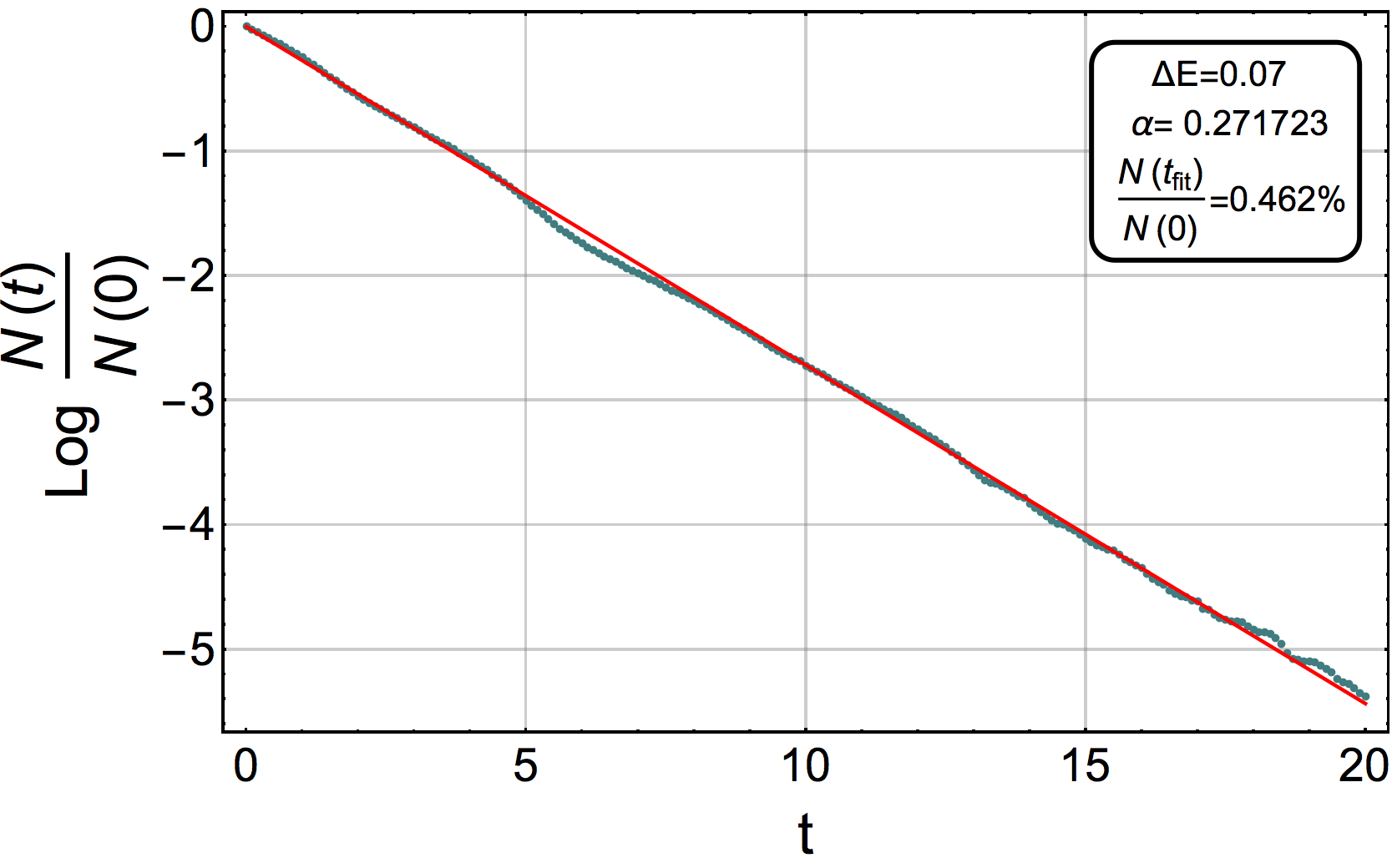}
\includegraphics[width=3.in]{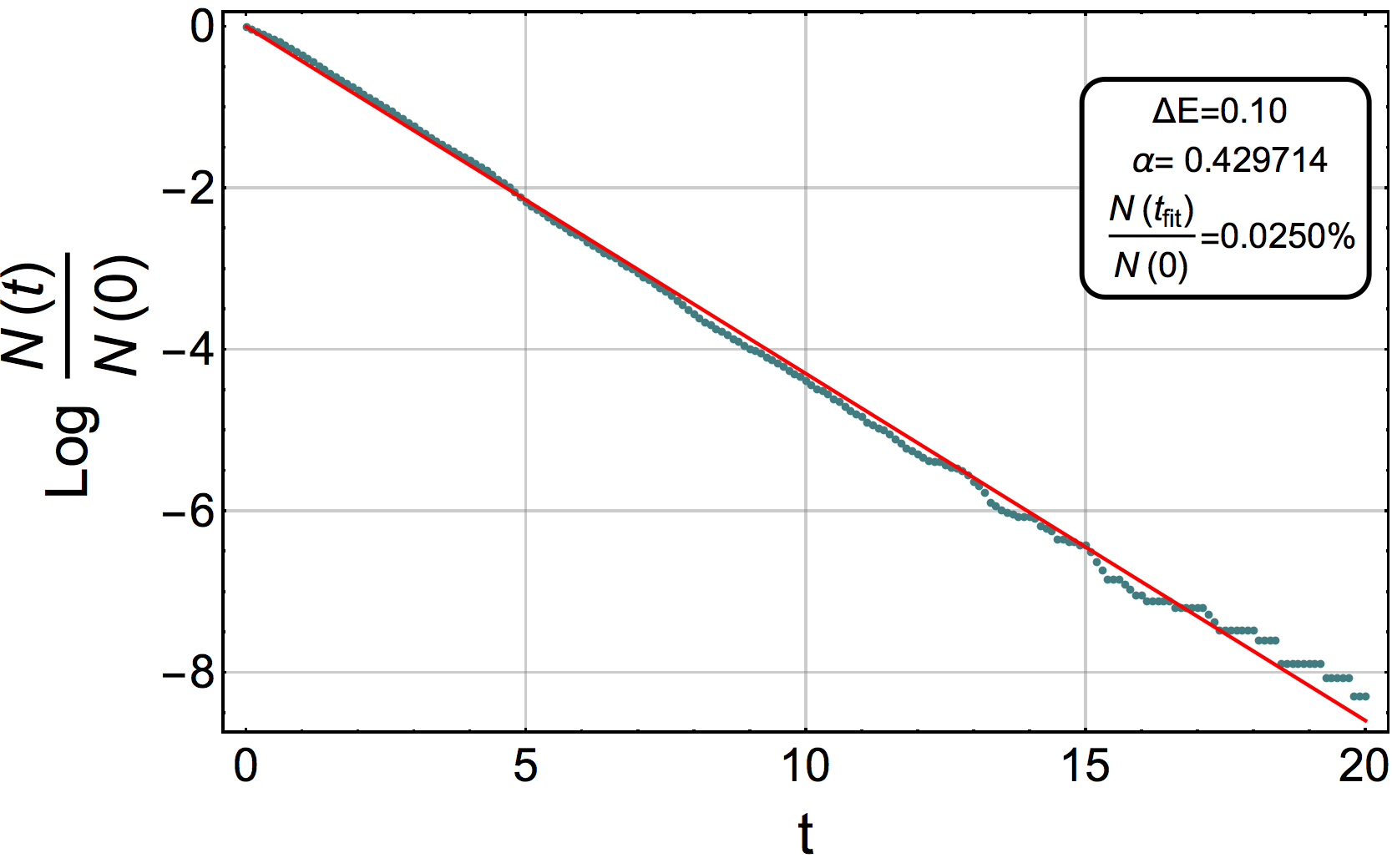}
\includegraphics[width=3.in]{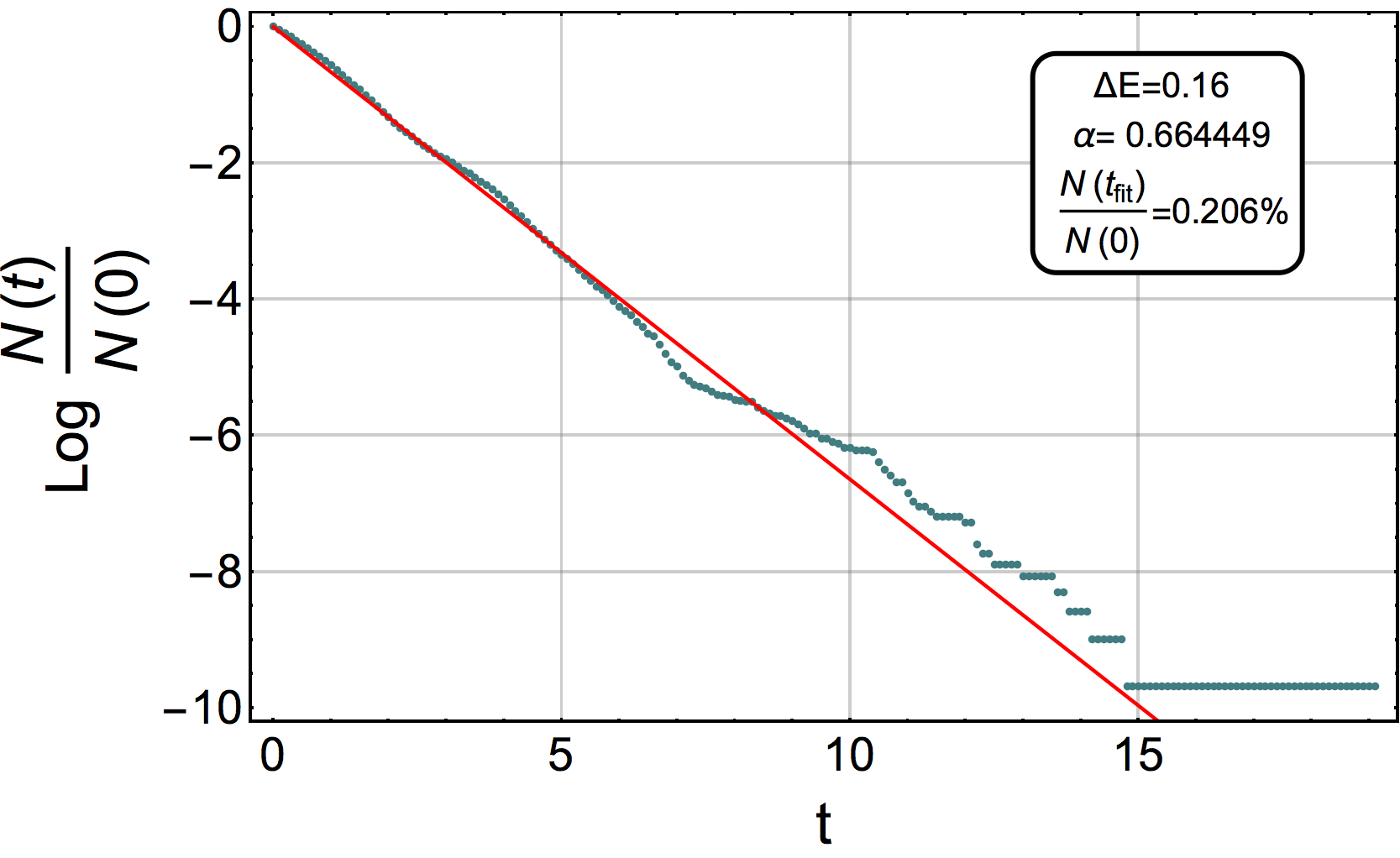}
\caption{{\footnotesize $\beta=0.2$. $\log N(t)/N(0)$ vs $t$ (blue curve) and their linear fit (red curve) for different $\Delta E$ (see insets). The $\alpha$ parameter shown in the inset is the slope absolute value of the fit. The fit was done in a time interval $t_{fit}$ such that after that time the remaining population is the one shown in the inset. }} \label{fig:fit02}
\end{figure}

\end{document}